\documentclass[
    aps,
    prx,
    reprint,
    letterpaper,
    superscriptaddress,
    twocolumn,
    longbibliography,
    floatfix
]{revtex4-2}

\usepackage{graphicx, textcomp, graphics, epsfig}
\graphicspath{ {./figures/} }
\usepackage{amsmath,amssymb,amsthm,wasysym, latexsym, braket}

\usepackage{xcolor}
\usepackage{color}
\usepackage{dsfont}
\usepackage{longtable}
\usepackage{siunitx}
\usepackage{nicematrix}
\usepackage{soul}
\usepackage{chngcntr}  
\usepackage{bm}      	

\usepackage{hyperref}
\definecolor{LinkColor}{rgb}{0.256,0.439,0.588} 
\definecolor{CiteColor}{rgb}{0.0, 0.52, 0.43}   

\hypersetup{
    colorlinks=true,
    linkcolor=LinkColor,
    citecolor=CiteColor,
    urlcolor=magenta      
}

\hypersetup{linktocpage}
\usepackage[nameinlink,capitalise]{cleveref}

\newcommand{\blank}{\,&}

\newcommand{\fig}[1]
{Fig.~\ref{fig:#1}}
\newcommand{\eq}[1]
{Eq.~\eqref{eq:#1}}
\newcommand{\sect}[1]
{Sec.~\ref{sec:#1}}
\newcommand{\app}[1]
{App.~\ref{app:#1}}

\def\ftm{\tilde{f}_0}
\def\Ut{\tilde{U}_0}
\def\maxttm{\tilde{t}_\mathrm{max}}

\def\m#1{#1^{(m)}}

\begin{document}

\title{A universal and efficient hybrid digital-analog fermionic quantum simulator}
\author{Hao-Tian Wei}
\email[]{htwei@rice.edu}
\affiliation{Department of Physics and Astronomy, Rice University, Houston, Texas 77005-1892, USA}
\author{Kaden R. A. Hazzard}
\affiliation{Department of Physics and Astronomy, Rice University, Houston, Texas 77005-1892, USA}
\affiliation{Smalley-Curl Institute, Rice University, Houston, Texas 77005-1892, USA}
\affiliation{Department of Physics, University of California, Davis, California 95616, USA}

\begin{abstract}
We present a universal framework to harness fermionic ultracold atom platforms for quantum simulation, showing how variational algorithms on existing hardware can simulate many-body systems well beyond the hardware's native Hamiltonian. Our analysis provides evidence that one can quantum simulate the ground-state properties of a broad class of gapless target Hamiltonians of local observables in a quantum evolution time that grows polynomially with the inverse relative error, $T\sim O(\mathrm{poly}(1/\epsilon))$ up to logarithmic corrections,  offering an exponential speedup over na{\"i}ve classical algorithms such as exact diagonalization. We provide numerical evidence and theoretical argument that this holds for energy density,  density-density, and spin-spin correlations in three qualitatively distinct models -- the repulsive Hubbard model; a Hubbard model augmented with nearest-neighbor attractive interactions, which introduces the phenomenon of pairing; and the Hofstadter-Hubbard model, which introduces a gauge field and fractional quantum Hall physics. This work demonstrates quantum simulation using current fermionic  platforms  far beyond the models natively implemented in the hardware.
\end{abstract}

\maketitle

\section{Introduction}

\begin{figure*}[hbt]
    \centering
    \includegraphics[width=0.8\linewidth]{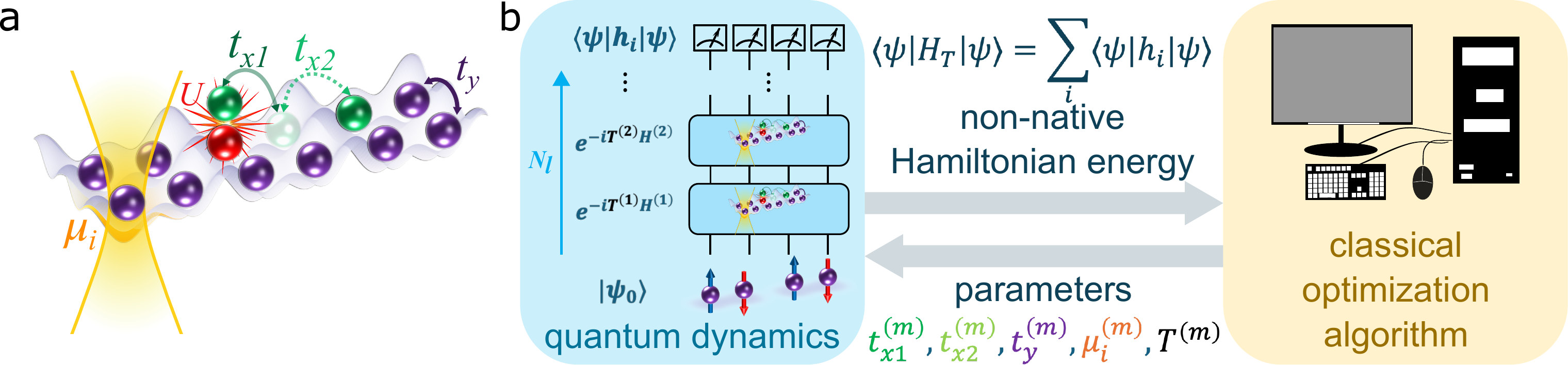}
    \caption{Programmable fermionic quantum simulator framework. (a) Experimental implementation of the fermionic quantum simulator. Neutral atoms are trapped in an optical superlattice or tweezer array, whose effective dynamics are described by the  FHM. The superlattice depth tunes staggered tunnelings, the atom-atom interaction provides an on-site repulsion, and lattice parameters -- or, for more flexibility, site-resolved tweezers -- control the on-site potentials. (b) The feedback loop of the variational quantum eigensolver (VQE). The quantum device measures terms of a potentially non-native target Hamiltonian after executing parameterized Hubbard evolution and feeds the results to a classical computer, where an optimizer uses this data to propose updated parameters for the next round of time evolution. This is iterated to convergence.}
    \label{fig:framework}
\end{figure*}


Programmable quantum simulators that realize models of  quantum systems with highly controllable quantum matter \cite{QuantSim,Feynman1982} have emerged as a powerful complement to classical computational methods, which generally suffer exponential cost with system size or uncontrolled approximations. Ultracold fermions in   optical lattices \cite{Esslingerreview,QuantSim,Tarruell_review,Bloch2012,Gross2017,altman_quantum_2021,kohlFermionicAtomsThree2005,hartObservationAntiferromagneticCorrelations2015a,shaoAntiferromagneticPhaseTransition2024,Taie2022,xuNeutralatomHubbardQuantum2025}, especially quantum gas microscopes \cite{GrossBakr_review,bohrdtAngleresolvedPhotoemissionSpectroscopy2018}, 
tunnel-coupled optical tweezer arrays \cite{altman_quantum_2021,BosonDoubleWell,FermionDoubleWell,Bergschneider2019,Becher2020,Spar2022,Florshaim2023,wei_hubbard_2024,wallEffectiveManybodyParameters2015}, and the combination of lattices and tweezers \cite{taoHighFidelityDetectionLargeScale2024,youngAtomicBosonSampler2024,youngTweezerprogrammable2DQuantum2022,parkerDirectObservationEffective2013} have become 
flexible analog quantum simulation platforms that natively realize the Fermi-Hubbard model (FHM) \cite{QuantCompChem,Bohrdt2021}.
These experiments are of broad interest because the FHM is a minimal model, yet which has resisted solution, that captures the essential interplay of interactions and kinetic energy, one of the most important ingredients of the  electronic structure problem at the center of quantum chemistry and condensed matter physics \cite{Esslingerreview,imada1998,montorsi1992,Tasaki1998,Arovas2022,Qin2022,Hubbard_model,liQuantumScienceArrays2026}.

While analog neutral-atom simulators have had numerous successes, their potential to address a broader range of fermionic many-body systems remains in its infancy. 
Many protocols to extend the range of native Hamiltonians that can be experimentally realized have been proposed, such as using Rydberg-dressed atoms for long-range interactions \cite{DefenuLRI2021,JohnsonRydbergDress2010,RydbergHubbard2015} or Floquet techniques for synthetic gauge fields \cite{jaksch_zoller_2003,aidelsburger_realization_2013,jotzuExperimentalRealizationTopological2014,kennedyObservationBoseEinstein2015,leonard_realization_2023}, but their reach extends to just a few select families of models. Recent proposals and  experimental demonstrations have described how to add ingredients to  fermionic quantum simulation platforms to make them digital and fully universal. For example Ref.~\cite{ZollerLukin2023} proposes to supplement the fermionic Hamiltonian with Rydberg gates to access the universal Bravyi-Kitaev gate set \cite{BK2002}, and experiments have demonstrated both spin- and pair-exchange gates with 99.75\% fidelity using double-well optical superlattices \cite{bojović2025highfidelitycollisionalquantumgates}.
The exciting benefits of these gate-based architectures come at the cost of significant additional experimental developments, and impose strong requirements on hardware fidelity and algorithm efficiency for implementation.

A promising approach to increase the flexibility of quantum simulators while retaining the simplicity and efficiency of analog fermionic quantum hardware is to use variational quantum eigensolver (VQE) techniques~\cite{peruzzo_variational_2014,grimsleyAdaptiveVariationalAlgorithm2019,kokail_self-verifying_2019,yuan_theory_2019,ho_efficient_2019,cerezo_variational_2021,altman_quantum_2021,nam_gs_2020,ebadiQuantumOptimizationMaximum2022,foss-feig_experimental_2023}. This method can create ground states of non-native Hamiltonians by relying on the variational principle, and generalizes to the preparation of finite-temperature states and quantum dynamics \cite{cerezo_variational_2021,yuan_theory_2019}.
The VQE is a hybrid quantum-classical algorithm that iteratively uses an analog quantum simulator to prepare parameterized quantum states, and optimizes its parameters by a classical algorithm whose input is observables experimentally measured in the prepared quantum state. The set of observables measured forms a basis for Hamiltonians that can be studied by the VQE method, and since the set of these is much larger than those appearing in the native interaction terms of the analog hardware, this allows quantum simulations to access the physics of models that cannot be realized natively.
Despite the  increased flexibility and the intuitive potential of being able to prepare classically-hard-to-simulate variational quantum states,
we still lack a clear demonstration that fermionic VQE quantum simulators can be more efficient than classical algorithms in a strong and precise sense, such as scaling with system size or  desired accuracy.
Ref.~\cite{li_fermionic_2023} proposes VQE using the aforementioned approach of augmenting fermions with Rydberg gates \cite{ZollerLukin2023} to find the ground state of a spinless FHM with density-density interaction between nearest neighbors.
Refs.~ \cite{Preiss2024} and~\cite{tabaresProgrammingOpticallatticeFermiHubbard2025} propose and demonstrate the capabilities of fermionic gate-based circuit designs without relying on Rydberg interactions, the former focusing on a a gate fabric numerically suggested to be universal, and the latter on the application of hybrid variational-adiabatic (HyVA) and variational quantum imaginary time-evolution (VarQITE) protocols to prepare ground states.
While these digital-analog frameworks are powerful, a systematic understanding of this framework's  efficiency to reach target accuracies (accounting for finite-size and variational error) remains an open question. Additionally,  
the considered models have not  included  cases with qualitatively different physics from either the FHM or very small molecules, and  little attention has been given to physically important observables beyond ground-state energy and fidelity, such as other local observables and correlations. 

In this work, we introduce an experiment-friendly protocol for implementing VQE on fermionic neutral-atom quantum simulators (fVQE), thereby realizing a universal framework for fermionic hardware that simulates a broad range of many-body systems beyond those enabled by the native interactions. Based on exact-diagonalization simulations of the protocol for systems with up to $14$ particles, we numerically demonstrate that this approach can efficiently calculate ground-state properties -- energy densities and other local observables -- for finite-range interacting gapless models. We numerically observe that the total experimental time $T$ grows polynomially with inverse algorithm error $T\sim O(\mathrm{poly}(1/\epsilon))$ up to logarithmic corrections. This achieves an exponential speedup over classical methods such as exact diagonalization (ED) and matrix product states (MPS) in dimensions greater than $1$. Rather remarkably, the fVQE seems to offer a polynomial advantage over classical MPS methods even in one dimension. We note that the number of variational parameters for a given circuit depth does not grow with the system size.
We benchmark our approach across three physically significant models, with distinct and sometimes exotic physics: the repulsive FHM, the extended FHM (EFHM) with nearest-neighbor attraction, and the Hofstadter-Hubbard model (HHM, in the hard-core boson limit for tractability of the simulations).  These not only have distinct Hamiltonians, but each feature qualitatively different physics: the EFHM introduces pair formation and the HHM features fractional quantum Hall physics. 
We demonstrate the effectiveness of the approach for observables beyond the energy, including two-point and string correlations. Finally, we provide a theoretical understanding of the observed numerical convergence, tying it to locality and finite-size scaling,
and we compare evolution-time scaling to classical baselines under explicit assumptions about the dominant error sources.

The rest of the paper is organized as follows. \sect{framework} outlines the proposed fermionic quantum simulation experimental platform and major VQE steps to realize our protocol. \sect{vqe} introduces the fVQE ansatz design and demonstrates the rapid convergence of the energy by numerical simulation of three representative models: the FHM in \sect{FHM}, the EFHM with nearest-neighbor attraction in \sect{EFH}, and the HHM with a uniform gauge flux in \sect{HH}. \sect{correlation} moves beyond energies to evaluate various correlation functions, showing that the VQE final states capture the essential physical signatures of each target state: magnetism in the FHM, binding of pairs in the EFHM, and edge currents and string-correlations characteristic of fractional quantum Hall physics in the HHM. In \sect{efficiency}, we consider the methods' computational complexity, arguing that our results suggest the VQE yields a speedup,  exponential in the targeted accuracy of local observables, over classical algorithms such as ED and MPS, and understands this via analytic considerations. Finally, \sect{summary} summarizes the main findings, discusses practical considerations for near-term implementations, and outlines future directions.

\section{Framework}
\label{sec:framework}

\begin{figure*}[htb]
    \centering
    \includegraphics[width=0.8\linewidth]{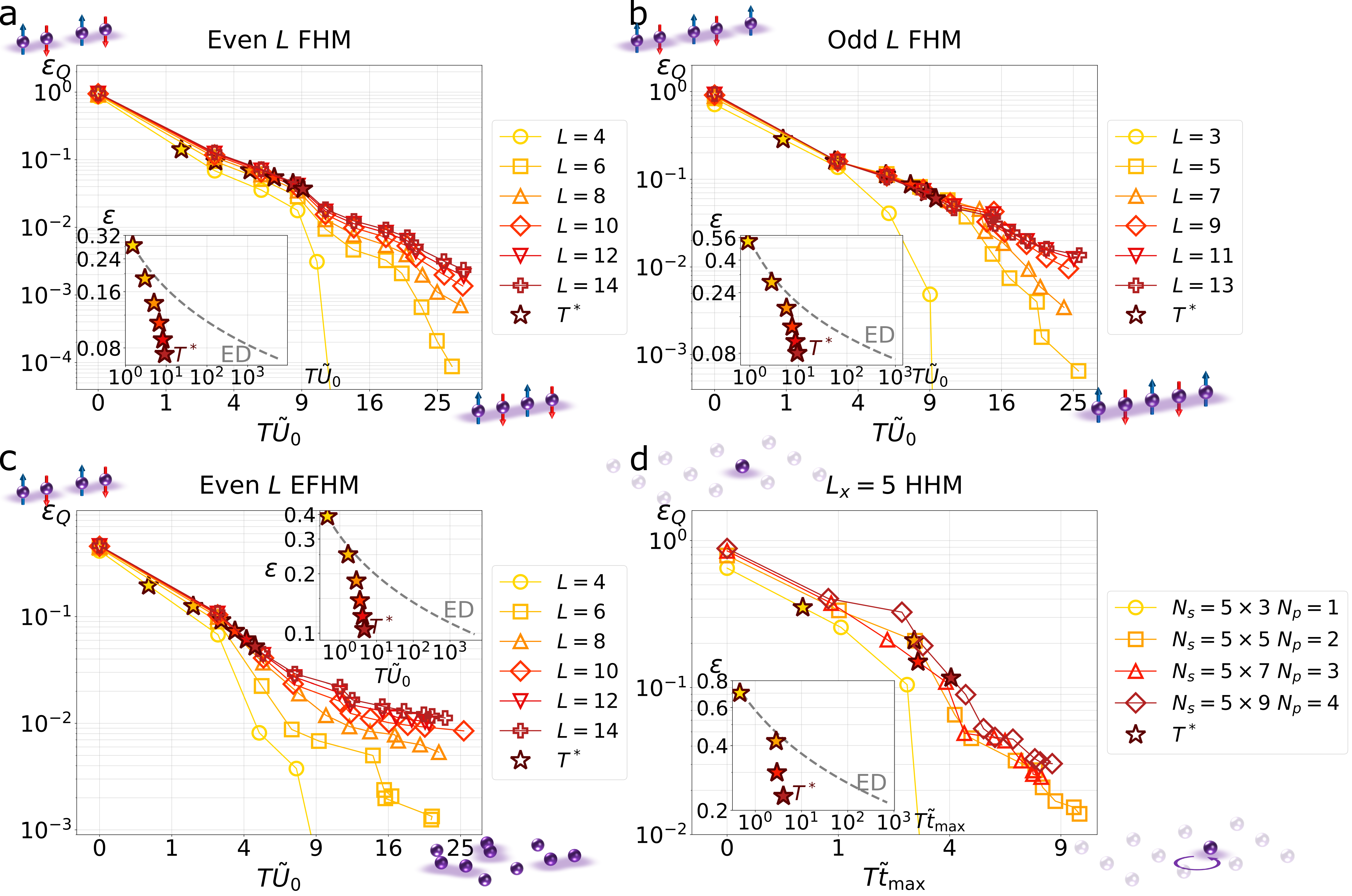}
    \caption{VQE relative energy error $\epsilon_Q$ as a function of total evolution time $T$ for various system sizes $L$ for the (a) even-$L$ FHM, (b) odd-$L$ FHM, (c) even-$L$ EFHM, and (d) $L_x=5$ square lattice bosonic HHM. The total evolution time $T$ is defined in \app{qtime}. Note the plot axes are logarithmic and square-root scales, so straight lines indicate $A e^{-B \sqrt{T}}$ behavior of $\epsilon$ for some constants $A$, $B$.  The upper-left and lower-right cartoons show the initial states  and the target final states, respectively.  The insets show the total simulation error $\epsilon$ (variational error plus finite-size error) evaluated at the optimal VQE time $T^*$ (star markers, defined in \sect{efficiency}) together with the exact-diagonalization (ED) scaling $\epsilon \sim 1/\log^{p/D} T$ (gray dashed curve), with $T$ the classical computation time. The ED computation time is exponentially longer in the target accuracy $\epsilon$ than the optimal VQE time $T^*$.}
    \label{fig:vqe}
\end{figure*}

We begin by introducing the fVQE framework, describing both the required experimental hardware and the hybrid digital-analog VQE algorithm. We envision hardware as shown in \fig{framework}a,  a quantum gas microscope with a two-site periodic optical superlattice obtained by interfering laser beams with a $2:1$ wavelength ratio.  This suffices for nearly all results in the paper, but further flexibility can be provided by superimposed optical tweezers to extend the programmability of site-resolved potential energies, or even further by realizing the lattice with programmable tunnel-coupled optical tweezer arrays  \cite{liQuantumScienceArrays2026,Yan2022,Spar2022}.
One then uses the hardware to perform a series of time evolutions of spinful fermionic atoms, each of which is governed by the dynamics of the FHM,
\begin{align}\label{eq:drive}
    \m{H} &= - \sum_{\braket{ij},\sigma} \left( 
 \m{t_{ij}} c^\dagger_{i,\sigma} c_{j,\sigma}^{\phantom \dagger} +\text{h.c.} \right)  \nonumber\\
        &{}\hspace{0.2in}+\m{U} \sum_i n_{i,\uparrow} n_{i,\downarrow} - \sum_i \m{\mu_i} n_i
\end{align}
with $c^{\phantom \dagger}_{i,\sigma}$ ($c^{\dagger}_{i,\sigma}$) fermionic annihilation (creation) operators on lattice site $i$ and spin flavor $\sigma$, $n_{i,\sigma}\equiv c_{i,\sigma}^\dagger  c_{i,\sigma}^{\phantom \dagger}$, and $n_i= n_{i,\uparrow}+n_{i,\downarrow}$, and $\sum_{\braket{ij}}$ runs over all nearest-neighbor edges of the square lattice. Other lattice geometries could be used, but we focus on the square lattice for simplicity. The tunneling rates $t_{ij}^{(m)}$ follow a strong-weak Su-Schrieffer-Heeger (SSH) modulation in the $x$ direction and are uniform in the $y$-direction, and can be arbitrarily set at each evolution step $m$.
The native on-site Hubbard interaction denoted by $\Ut$ is not treated as a layer-dependent variational control parameter. Instead, we use $\m{U}=\Ut$ as the energy (or angular-frequency by setting $\hbar=1$) scale, so the optimized parameters are the dimensionless quantities $\m{t}_{ij}/\Ut$, $\m{\mu_i}/\Ut$, and $\m{T}\Ut$. Keeping constant $U$ avoids layer-by-layer tuning of the atom-atom interaction, which can be challenging on the required experiment time scales.
While the site-resolved potential $\m{\mu_i}$ is in principle arbitrarily tunable with tweezer beams, in this work we utilize only parameterizations whose number of parameters per layer are $L$-independent -- in some schemes requiring $\m{\mu_i}$ implementable with only the superlattice itself -- making the experiment implementation and optimization of the variational ansatz easier and more scalable.

The hybrid classical-quantum digital-analog VQE toggles between quantum dynamics and classical optimization, as shown in \fig{framework}b. 
To create the ground state of the target Hamiltonian $H_T$, it iterates through the following steps:

\begin{enumerate}
    \item \textbf{State Preparation}: Initialize a high-fidelity product state $\ket{\psi_0}$. Choices of $\ket{\psi_0}$ are straightforward and discussed later in the main text and \app{initState}.
    \item \textbf{Parameterized Evolution}: Apply time-sliced dynamics $U^\text{VQE}=\prod^{N_l}_{m=1} e^{-i\m{H} \m{T}}$ (setting $\hbar=1$ throughout) generated by $N_l$ parameterized $\m{H}$'s defined in Eq.~\eqref{eq:drive} with time parameter $\m{T}$.
    \item \textbf{Energy Measurement}: Estimate $\braket{H_T}=\sum_i\braket{h_i}$ by measuring all local terms $\braket{h_i}$ on the final state $\ket{\psi} = U^{\text{VQE}}\ket{\psi_0}$.
    \item \textbf{Classical Optimization}: Update parameters to minimize $\braket{H_T }$ and repeat the whole loop until convergence.
\end{enumerate}

This algorithm is highly flexible and independent of the specific quantum hardware and the classical optimization method~\footnote{As a side remark, we note that the method is robust to the experimental calibration error of the Hamiltonian parameters during the time evolution since the time evolution is used only to generate variational quantum states.}.
The target Hamiltonian $H_T$ can be either native as $\m{H}$ or non-native, i.e. containing terms $h_i$ that are not directly implemented on the quantum hardware. 
Rather, any $H_T$ whose all terms $h_i$ can be measured on the quantum device can be simulated, because the value $\braket{H_T}=\sum_i\braket{h_i}$ is available to the classical optimizer.
This enables the study of a wide range of Hamiltonians that are not natively available in analog quantum simulators. In \app{measurement}, we discuss the full range of Hamiltonians that can be simulated and the measuring techniques.
In this work for simplicity, we use the gradient-based Sequential Least Squares Programming (SLSQP) method for classical optimization~\cite{nocedal2006numerical}, with techniques introduced in \app{optim} that accelerate optimization and analysis of convergence.


\section{Ansatz designs \& energy convergence}
\label{sec:vqe}

We demonstrate this fVQE's versatility and efficiency by adapting and using it to find ground states of various target Hamiltonians $H_T$ on lattices with open boundary conditions, simulating the protocol on a classical computer. \fig{vqe} summarizes results for three representative models. 
A key quantity is the VQE error in the energy density relative to the exact finite-size ground state energy density $E_0$, 
\begin{equation}
    \epsilon_Q=\left|\frac{\braket{H_T}/N_s-E_0}{E_\infty}\right|,
\end{equation}
where $E_\infty$ denotes the thermodynamic-limit energy density (estimated by finite-size extrapolation of $E_0(L)$ to $1/L^p\to 0$, where $p$ is obtained by fitting $E_0(L)$).  We observe that as the total evolution time $T=\sum_m \m{T}$ is increased (see \app{qtime} for the dimensionless form of $T$ used in each model), $\epsilon_Q$ decreases exponentially with $\sqrt{T}$.
Importantly, $\epsilon_Q$ is not the only source of algorithmic error: there is also a finite-size error
\begin{equation}\label{eq:fse}
    \epsilon_L=\left|\frac{E_0-E_\infty}{E_\infty}\right|.
\end{equation}
The full relative error of the finite-size, finite-layer VQE to the true thermodynamic limit result is
\begin{equation}
\epsilon=\left|\frac{\braket{H_T}/N_s-E_\infty}{E_\infty}\right|\leq \epsilon_Q + \epsilon_L,
\end{equation}
by the triangle inequality.

As analyzed in \sect{efficiency}, for each desired accuracy $\epsilon$,   there is an ``optimal $T^*$'' such that $\epsilon_Q(T^*)=\epsilon_L(N_s)$. Using much larger $T^*$ would improve $\epsilon$ negligibly since the error would be dominated by finite-size effects, while using much smaller $T^*$ would not take advantage of the rapid fVQE convergence. This choice of $T^*$ ensures a rapid convergence with target error, shown in Fig.~\ref{fig:vqe}, at least as rapidly as a power law with increasing $T^*$ there.
This scaling emphasizes the quantum hardware's exponential advantage over classical methods like ED, which have an error scaling as $\epsilon \sim 1/\log^{p/D} T$ for $D$-dimensional gapless systems for $p$ a constant, as we will show and discuss in Sec.~\ref{sec:efficiency}. Two of the models shown (EFHM, HHM) are non-native to the hardware, with each introducing qualitatively new physics not present in the native Hamiltonian.

\subsection{Fermi-Hubbard model}
\label{sec:FHM}

We first study the simple case where the target Hamiltonian $H_T$ is the uniform on-site-repulsive FHM, the same the hardware realizes natively, 
\begin{equation}
\label{eq:FHM}
    H_\mathrm{FH} = - t\sum_{\braket{ij},\sigma} \left( c^\dagger_{i,\sigma} c_{j,\sigma}^{\phantom \dagger} + \text{h.c.} \right) + U \sum_i n_{i,\uparrow} n_{i,\downarrow}.
\end{equation}
We consider $t=0.2U$ at half-filling in a one-dimensional (1D) chain. Its thermodynamic-limit ground state is a Mott insulator with a vanishing spin gap \cite{1dHubbard}.

Our ansatz initial state is arrays of dimers, each in a one-particle-per-site SU(2) singlet, which then evolves  under piecewise time-dependent two-site unit cell SSH tunnelings $\m{t}_1$ and $\m{t}_2$. This ansatz has $3N_l$ parameters, independent of $L$. This initial state can be prepared with high fidelity by several methods: in \app{initState}, we briefly propose one preparation method starting from a band insulator state, a low-entropy state that the experiment can prepare with $>99\%$ fidelity \cite{xuNeutralatomHubbardQuantum2025}.

This choice follows principles that guide the design of ansatzes for each of our target Hamiltonians, and can be applied generally. First, our initial state should have the same symmetry as the desired final state. Here, this is an SU(2) symmetry. This is crucial, since we included no  magnetic field terms to change the SU(2) quantum numbers and local magnetic field terms are difficult to engineer. We note that convenient SU(2)-symmetric initial states in the correct particle number sector required breaking the translation symmetry. However, this causes no issue because spatially varying tunnelings during the time evolution can create states that restore the translation invariance if beneficial to the VQE energy. Therefore, we  choose the simplest product states compatible with the SU(2) symmetry~\footnote{One may question the effectiveness of this procedure in the presence of spontaneous symmetry breaking. However, strictly speaking, in finite systems there is never spontaneous symmetry breaking. Moreover, even as the system size approaches infinity, local observables cannot distinguish between a symmetry-broken state and the large-but-finite-system-size ground state which is a superposition of all symmetry-equivalent configurations.}.

To use this ansatz, in addition to implementing the Hamiltonian dynamics, experiments must be able to measure $\braket{H_{\text{FH}}}$. This can be done by measuring each term in the Hamiltonian individually. Here these are the on-site density correlations $\braket{n_{i,\uparrow} n_{i,\downarrow}}$, which follow immediately from quantum gas microscopy, and the nearest-neighbor coherence $\braket{c^\dagger_{i,\sigma} c_{j,\sigma}^{\phantom \dagger}}$, which can be measured by isolating pairs of adjacent  wells using the superlattice and measuring the amplitude of tunneling oscillations \cite{impertroLocalReadoutControl2024,kesslerSinglesiteresolvedMeasurementCurrent2014}, as discussed in \app{measurement}.

\fig{vqe}a and b show the exponential decay of the VQE relative energy error $\epsilon_Q$ with  $\sqrt{T}$ for the even-$L$ and odd-$L$ FHM, respectively. It further shows the power law (or faster) convergence of the total (VQE plus finite-size) error $\epsilon$ of both models for an appropriately chosen  time $T^*$~\footnote{Since for odd system size, the ground state is not in the total spin $S=0$ sector, we fix the dynamics to be in the total $S^z=\frac{1}{2}$ sector by setting the last spin in the initial state to be in the spin-up state.}.
This error decay is exponentially faster than the classical $\epsilon \sim 1/\log^p T$ scaling (see Sec.~\ref{sec:efficiency}).
In real experiments, $\epsilon\sim 1\%$ would require a total evolution time $T\sim 15/U\sim 3/t$, which is $ 2\text{ms}$ for tunneling  $t=2\pi\times200\text{Hz}$ (taking $\hbar=1$). This evolution time is well below the coherence time, which is usually longer than $100\text{ms}$ for optical lattices \cite{shaoAntiferromagneticPhaseTransition2024,xuNeutralatomHubbardQuantum2025}.

We also observe that for $L=3$ and $4$, the VQE energy error decreases much faster than exponentially in $\sqrt{T}$, reaching machine precision after only a few layers. We interpret this as a small-system over-parameterization effect: in such small Hilbert spaces, the ansatz has sufficiently many continuously tunable parameters that at low depth any state in the Hilbert space can be represented exactly -- note that the depth grows exponentially in $N_s$. Similar behavior has been observed in previous ansatz constructions and described as ``numerical universality''~\cite{QNP2021}. As we will see, the same small-$L$ behavior appears in all three examples for our ansatzes.

\subsection{Extended Fermi-Hubbard model (nearest-neighbor attractions)}
\label{sec:EFH}

With our framework's success in simulating the analog simulator's native Hamiltonian, we now study its application to non-native Hamiltonians. First, we apply our framework to  $H_T$ being an extended Fermi-Hubbard model (EFHM) in the itinerant regime. This is the FHM of a 1D chain at half-filling, augmented with a nearest-neighbor attraction, 
\begin{align}\label{eq:EFH}
    H_\mathrm{EFHM} = \blank - t \sum_{\braket{ij},\sigma} \left(c^\dagger_{i,\sigma} c_{j,\sigma}^{\phantom \dagger} + \text{h.c.} \right) \nonumber \\
    \blank \hspace{0.1in} {}+ U \sum_i n_{i,\uparrow} n_{i,\downarrow} - V \sum_{\braket{ij}} n_i n_j
\end{align}
with $t=U=V>0$. Its thermodynamic-limit ground state features unconventional triplet-pairing \cite{qu_efh_2022} with a spin gap (the one-dimensional analog of a spin-triplet superconductor), in stark contrast with the original FHM ground state.

It is challenging for neutral-atom analog quantum simulations to simulate this model due to its unusual combination of on-site repulsion and nearest-neighbor attraction. 
Our fVQE framework, however, is able to simulate the target ground state without needing to natively realize such interactions. Our framework evolves the state under the native time-dependent FHM Hamiltonian while minimizing the variational energy determined by summing the measured expectation values of the EFHM terms. The kinetic energy and on-site interactions are measured as they were for the usual FHM discussed in the last subsection, while the nearest-neighbor interaction is obtained by measuring density-density correlations $\braket{n_in_{i+1}}$, which are measured directly in a quantum gas microscope.

Following the same guiding principle as in the FHM case, we design our EFHM's ansatz starting from an initial state that respects the symmetry of our target Hamiltonian, again choosing the one-particle-per-site SU(2) singlet product state. 
We also use the same staggered tunnelings as in the FHM scenario. However, we now allow spatially varying on-site potential, expanded in the lowest $2$ reflection-symmetric, long-wavelength modes,
\begin{equation}
    \mu_i= \sum_{r=1,2}\mu_{F,r}\cos \left( \frac{2r\pi (i-1)}{L-1} \right ),
\end{equation}
with $i=1,\dots,L$ the site indices, to capture the spatial redistribution that arises from the attractive density-density interaction. 
Intuitively, the $r=1$ component penalizes particle occupation near the open boundary sites. This encourages particles to cluster in the lattice center, which is energetically favored by the attractive interaction. The $r=2$ component then supplies finer control to the density distribution.

\fig{vqe}c demonstrates the relative energy error's rapid decay and weak dependence of $L$. This highlights our framework’s potential for non-native quantum simulation.

\subsection{Hubbard model with a gauge field (Hofstadter-Hubbard model)}
\label{sec:HH}

Now we show that our framework extends straightforwardly to a higher-dimensional system, with qualitatively new terms in the Hamiltonian and exotic new physics, namely the HHM on a two-dimensional (2D) number of sites $N_s=L_x\times L_y$ square lattice with a gauge field,
\begin{equation}\label{eq:HH}
    H_\mathrm{HH} = - t \sum_{\braket{ij}} \left( e^{i\phi_{ij}} a^\dagger_i a_{j}^{\phantom \dagger} + \text{h.c.} \right) + U \sum_i n_i(n_i-1),
\end{equation}
with $N_p=\frac{\phi N_\Box}{2\pi m}$ particles and a flux $\phi=2\pi/4$ through  each of the $N_\Box=(L_x-1)\times(L_y-1)$ square plaquettes. Its thermodynamic-limit ground state is expected to be the $1/m$ fractional quantum Hall (FQH) state~\cite{repellin_fractional_2020,heRealizingAdiabaticallyPreparing2017a,motrukPhaseTransitionsAdiabatic2017,wangMeasurableSignaturesBosonic2022a,leonard_realization_2023,FQH2005,rosson_bosonic_2019,pauw_detecting_2024} (also see \app{fqh} for evidence). Given the classical computing constraints, we focus on the hardcore boson version of this model, replacing fermionic operators with bosonic ones and setting $U\to\infty$; this allows $m=2$ to realize FQH behavior (instead of the $m\ge 3$ required by fermions) and thus requires fewer sites than the fermionic model. 
We emphasize that in real experiments, our framework applies equally to fermions and bosons, and this choice is made for classical computational simulability. 

Quantum simulation of this model has been a long-standing challenge for neutral atom analog quantum simulators. Existing methods to engineer gauge fields
\cite{jaksch_zoller_2003,linSyntheticMagneticFields2009,aidelsburger_realization_2013,miyakeRealizingHarperHamiltonian2013a,jotzuExperimentalRealizationTopological2014,kennedyObservationBoseEinstein2015,leonard_realization_2023}
show significant heating in the strongly interacting regime, and have limited ability to spatially vary the magnetic flux to create more general models. Our framework circumvents these difficulties by avoiding engineering the gauge field. Instead, it approximates the target state's properties by minimizing the variational energy of its final state evolved by an ansatz without  complex tunnelings.
As elaborated in \app{measurement}, the local gauge flux can be encoded through the classical post-processing that obtains the variational energy by multiplying the measured nearest-neighbor coherence expectation value $\braket{a^\dagger_i a_j}$
with the 
complex phase $e^{i \phi_{ij}}$.

To achieve the right filling for the $m=2$ Laughlin state, we consider $L_x\times L_y = (1+4n_x)\times (1+2n_y)$ site lattices ($n_x$, $n_y$ integers), resulting in $N_\Box=8n_xn_y$ plaquettes. Accordingly, in our variational ansatz, we use an initial state that is a product state with one particle localized at a single site in each $4\times 2$ cell, as illustrated in \fig{HHInitState}.
For the ansatz's Hamiltonian, the primary factor to consider is that the HHM breaks time-reversal symmetry. 
A na\"ive uniform hardcore Bose-Hubbard Hamiltonian containing only real-valued tunnelings lacks the phase structure needed to generate the ground state of a time-reversal-breaking target Hamiltonian. A non-uniform on-site potential varying with the layer of the ansatz (``time-dependent"), however, can create states with broken time-reversal symmetry. Inspired by the stroboscopic proposal to implement synthetic gauge fields in Ref.~\cite{FQH2005}, we introduce a layer-dependent quadrupolar potential $\mu_{\bm{i}}=\mu_Q x_i y_i$, 
where $\bm{i}=(x_i,y_i)$ is the coordinate vector of the $i$-th site with the origin defined at the lattice center. This quadrupolar potential, controlled by a single parameter $\mu_Q$, can be implemented in the experiment in several ways, for example using two pairs of wide, equal-waist Gaussian laser beams or an SLM to imprint the desired potential.
We also apply a staggered single-parameter potential $(-1)^{x_i + y_i}\mu_S$ on top of the quadrupolar potential.
Since in the hardcore limit $U\to\infty$, $U$ is no longer  a useful reference energy, we absorb the time parameter into the drive Hamiltonian parameters $\m{t_{x1}}\m{T},\,\m{t_{x2}}\m{T},\,\m{t_y}\m{T}$, $\m{\mu_Q}\m{T}$ and $\m{\mu_S}\m{T}$, making the ansatz five parameters per layer.

Similar to the results for the models above, \fig{vqe}d shows the variational energy error's exponential decay with $\sqrt{T}$~\footnote{The definition of $T$ for the HHM differs slightly from the one we used earlier to work naturally in the hard core limit, see \app{qtime}.} for every system size $N_s$,
and depends weakly on $L_y$. This shows that our quadrupolar potential ansatz can efficiently generate time-reversal symmetry broken states arising from models with gauge fields.

\section{Correlations}
\label{sec:correlation}

\begin{figure}
    \centering
    \includegraphics[width=\linewidth]{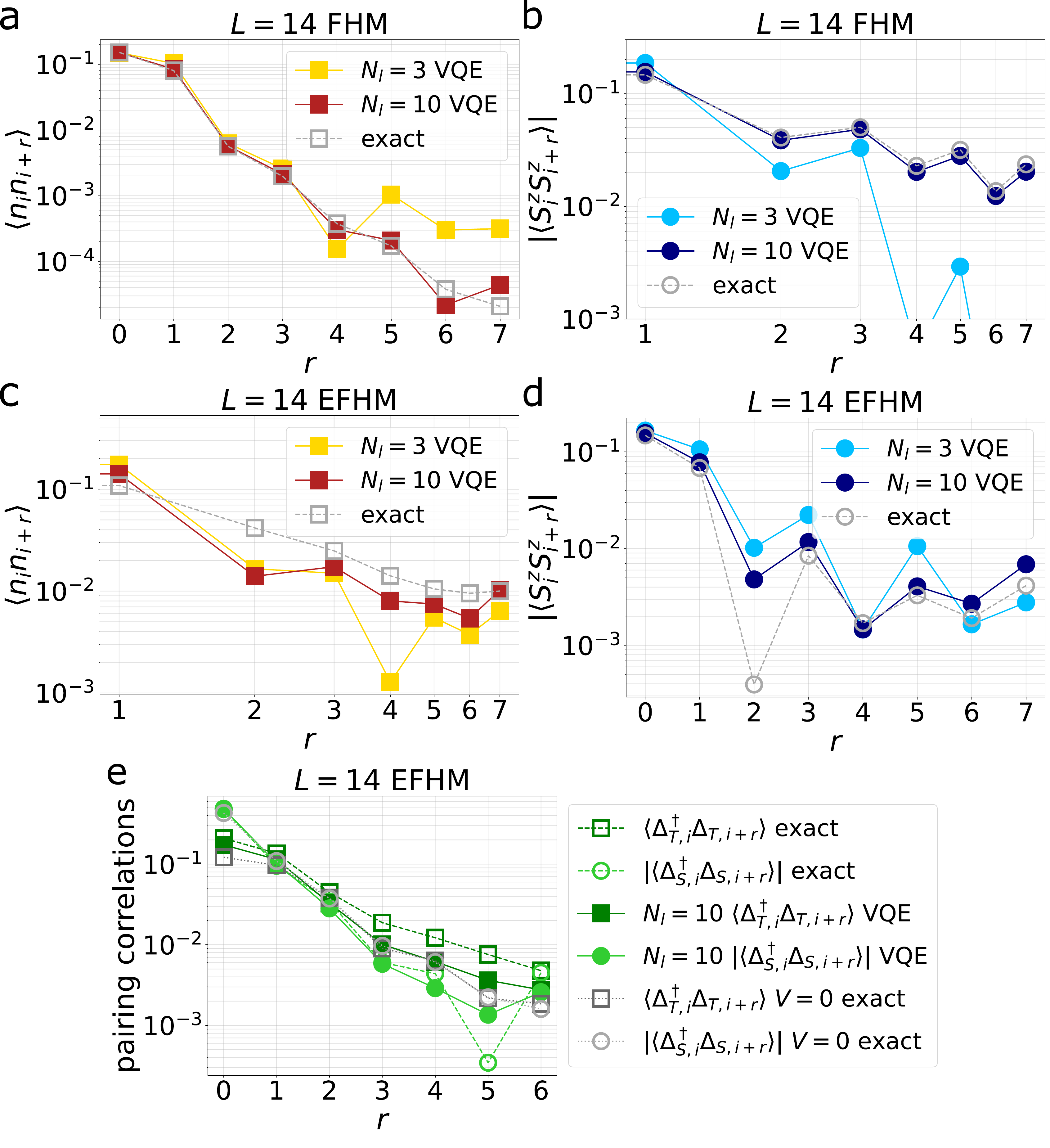}
    \caption{Correlations as a function of separation $r$ for $N_l=3$ and $N_l=10$ VQE final states (solid markers) and ED results (open gray markers). For the $L=14$ half-filled FHM, the (a) density-density and (b)  absolute spin-spin correlations show that the $N_l=10$ VQE captures the expected exponential and power-law decay, respectively, reflecting the FHM ground state's finite charge gap and the vanishing spin gap. Similarly, for the $L=14$ EFHM, the (c) density-density and (d)  absolute  spin-spin correlations show that $N_l=10$ VQE captures the expected behavior of the EFHM ground state, with gapless charge and gapped spin sectors. (e) Finally, for the $L=14$ EFHM, the $N_l=10$ VQE reproduces the dominance of triplet over singlet pairing correlations at large $r$.  The splitting between the triplet and singlet pairing correlations is a qualitatively new effect of the attractive interactions, as shown by contrast with the ED results for the $V=0$, $t=|U|$ FHM ground state.}
    \label{fig:correlation}
\end{figure}


The previous section  demonstrated the efficiency and expressiveness of the fVQE through the convergence of the  energy. While this  is commonly used to characterize VQE convergence to the ground state, it does not directly characterize other physically important observables,
such as spin correlations, particle number densities, or local order parameters. Although it is difficult to establish scalings as cleanly as for the case of the energy, here we show that such local observables of VQE final states in all three models converge rapidly,
similar to the energy, and obtain the correct qualitative structure.

We start with the charge and spin correlations of the FHM. As mentioned in \sect{vqe}, in the thermodynamic limit the FHM's ground state has a finite charge gap but a vanishing spin gap, so the density-density correlations $C_n(r)=\braket{n_{i}n_{i+r}}-\braket{n_{i}}\braket{n_{i+r}}$ decay exponentially and the absolute values of spin-spin correlations $|C_{S^z}(r)|=|\braket{S^z_{i}S^z_{i+r}}|$ decay as a power law with separation $r$, where $S^z_i=(n_{i,\uparrow} - n_{i,\downarrow})/2$.
These observables are directly measurable in quantum gas microscope or tweezer experiments. 
Figs.~\ref{fig:correlation}a and b show the $L=14$ FHM's VQE final state and confirm that  $C_n(r)$ and $|C_{S^z}(r)|$ match the true ground state behaviors. Here, we set $i=L/2$ (the center site) to reduce the boundary effects. Significant quantitative improvement from $N_l=3$ to $N_l=10$ (corresponding to increasing $T$) can be seen as well. Generally, the range of correlations captured by the VQE increases with increasing $T$.

In contrast to the FHM ground state, the EFHM ground state in the thermodynamic limit is gapped in the spin sector but remains gapless in the charge sector. Figs.~\ref{fig:correlation}c and d illustrate how   $C_n$ and $C_{S^z}$ approach their expected decay laws as we increase $N_l$ from $3$ to $10$.  We also consider the triplet pairing correlation $C_T(r)=\braket{\Delta^\dagger_{T,i}\Delta_{T,i+r}^{\phantom \dagger}}$ and singlet pairing correlation $C_S(r)=\braket{\Delta^\dagger_{S,i}\Delta_{S,i+r}^{\phantom \dagger}}$, where
\begin{align}
    \Delta^\dagger_{T,i}=\frac{1}{\sqrt{2}}(c^\dagger_{i,\uparrow}c^\dagger_{i+1,\downarrow} - c^\dagger_{i+1,\uparrow}c^\dagger_{i,\downarrow}),  \\
    \Delta^\dagger_{S,i}=\frac{1}{\sqrt{2}}(c^\dagger_{i,\uparrow}c^\dagger_{i+1,\downarrow} + c^\dagger_{i+1,\uparrow}c^\dagger_{i,\downarrow}).
\end{align}
To measure the pairing correlations in the experiment, in \app{measurement} we propose a strategy that uses local two-site basis rotations similar to the basis rotation protocol used for measuring the two-site coherence $\braket{c^\dagger_{i,\sigma} c_{j,\sigma}^{\phantom \dagger}}$, which could be combined with atom shuttling by optical tweezers to measure longer-range correlations.
In the true ground state, the triplet pairing dominates, which \fig{correlation}e  shows is reproduced by the fVQE. This is in notable contrast to the initial singlet product state -- which has nonzero singlet pairing only at $r=0$, and a vanishing triplet pairing everywhere -- as well as the $V=0$ FHM ground state, which has an identical singlet and triplet pairing at large $r$~\footnote{This can be understood by bosonization. The triplet and singlet pairing correlations are in the form of $C_T(r)\sim\braket{\sin(\sqrt{2}\phi_\sigma(r))\sin(\sqrt{2}\phi_\sigma(0))}$ and $C_S(r)\sim\braket{\cos(\sqrt{2}\phi_\sigma(r))\cos(\sqrt{2}\phi_\sigma(0))}$ with $\phi_\sigma$ the spin field \cite{giamarchi2003quantum}. If the phase has a gapless spin sector as $V=0$ FHM ground state, the two pairing correlations have the same asymptotic behavior; but when the spin field $\phi_\sigma$ is pinned by the $\cos(4\phi_\sigma)$ term \cite{seidelLutherEmeryLiquidSpin2005} arising from the attractive interaction in the EFHM, the two correlations differ.}.

\begin{figure}
    \centering
    \includegraphics[width=0.95\linewidth]{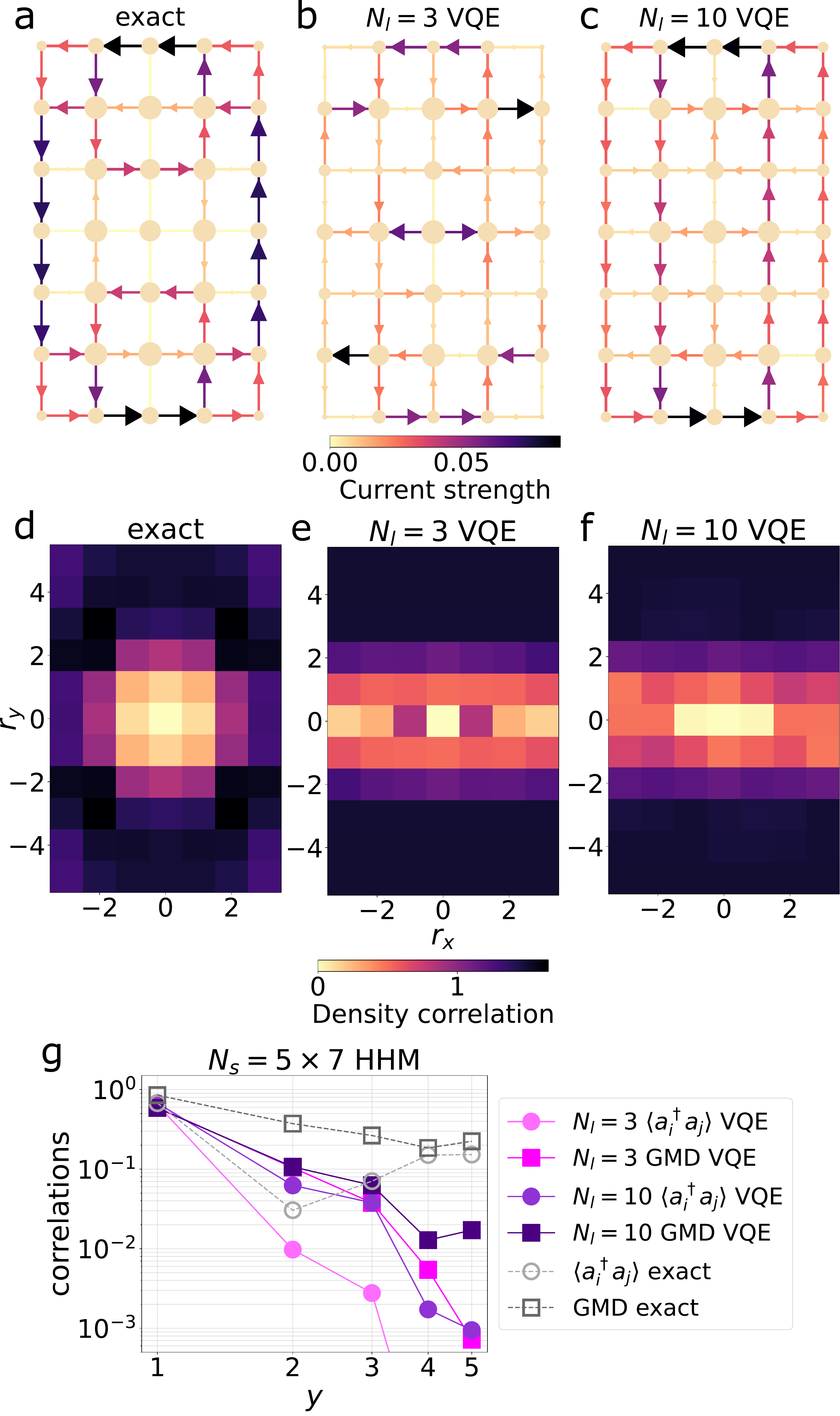}
    \caption{The particle current of an $N_s=L_x\times L_y=5\times 7$ square lattice for the (a) ground state, (b) $N_l=3$ VQE final state, and (c) $N_l=10$ VQE final state. The bond color and arrow size indicate the current strength, and the vertex radius is proportional to the local particle density. The reduced bulk density-density correlations in the same $5\times 7$ square lattice for the (d) ground state, (e) $N_l=3$ VQE final state, and (f) $N_l=10$ VQE final state. (g) The ED, $N_l=3$, and $N_l=10$ fVQE results for the Girvin-MacDonald order parameter $C_{\bm{ij}}^{\text{GMD}}$ and bare two-point correlator $C^{\text{bare}}_{\bm{ij}}$ of the $5\times 7$ lattice as functions of $y$ where $\bm{i}=(0,-2)$ and $\bm{j}=\bm{i}+(0,y)$ where the origin is the lattice center. For all quantities, the VQE final state reasonably reproduces features of the ED ground state, with increasing accuracy as the ansatz depth $N_l$ (evolution time $T$) increases.}
    \label{fig:current}
\end{figure}

Finally, the HHM ground state displays qualitatively new features connected to the targeted Hamiltonian's nontrivial topological order, and the fVQE framework still captures key observables as illustrated in Fig.~\ref{fig:current}. 
Figs.~\ref{fig:current}a--c illustrate the particle currents $J_{\bm{ij}}=ie^{i\phi_{\bm{ij}}}\braket{a^\dagger_{\bm{i}}a_{\bm{j}}}+\text{h.c.}$, which can be measured by a method similar to measuring the nearest-neighbor coherence detailed in \app
{measurement}. The key feature observed is the presence of chiral edge currents. Figs.~\ref{fig:current}d--f  plot the reduced density-density correlations summing over site pairs $(\bm{i},\bm{i+r})$ such that either $\bm{i}$ or $\bm{i+r}$ is in the $N_\text{bulk}=(L_x-2)\times(L_y-2)$ bulk sites, defined as in  Ref.~\cite{leonard_realization_2023},
\begin{equation} \label{eq:FQH-dd-corre}
    C^\text{bulk}_{n}(\bm{r})
    =\frac{N_p}{N_\text{bulk}(N_p-1)}
    \sum_{\bm{i}\,\text{or}\,\bm{i+r}\in\text{bulk}}
    \frac{\braket{a^\dagger_{\bm{i}}a^\dagger_{\bm{i+r}}a_{\bm{i+r}}a_{\bm{i}}}}{\braket{n_{\bm{i}}}\braket{n_{\bm{i+r}}}}.
\end{equation}
The chiral edge states and the anti-correlation of particle positions are qualitatively captured, becoming more accurate as the ansatz depth $N_l$ (hence total evolution time $T$) increases.

A particularly non-trivial feature captured by the fVQE is the  non-local string correlations directly characterizing the topological order. For Laughlin liquids, the relevant correlator is  the Girvin-MacDonald (GMD) order parameter, defined as~\cite{GMD1987,pauw_detecting_2024,weitenbergProtocolsManybodyPhase2026}
\begin{equation}
    C^{\text{GMD}}_{\bm{ij}} = \braket{\prod_{\bm{k}\neq \bm{i},\bm{j}} e^{im (\varphi_{\bm{ik}} - \varphi_{\bm{jk}}) n_{\bm{k}}} a^\dagger_{\bm{i}} a^{\phantom\dagger}_{\bm{j}}}. \label{eq:GMD-def}
\end{equation}
Here, $\varphi_{\bm{ij}}\equiv \arg(x_i - x_j + i(y_i - y_j))$ is the phase angle of the complex number representing the spatial vector $(x_i - x_j, y_i - y_j)$.   \fig{current}g  shows that, at fixed system size, increasing the ansatz depth $N_l$ (and the total evolution time $T$) systematically improves the accuracy of both the non-local GMD string correlation and the bare two-point correlation $C^{\text{bare}}_{\bm{ij}}=\braket{a^\dagger_{\bm{i}} a^{\phantom\dagger}_{\bm{j}}}$ in the VQE final state. While the agreement between the fVQE and exact results is quantitative only at short distances, the VQE final state captures that non-trivial GMD correlation that persists to long-range even as the bare two-point correlation  decays rapidly with distance. Moreover the fVQE result improves as $N_l$ is increased.
To measure GMD in experiments, one can combine a double-well basis rotation on $i$ and $j$ with site-resolved density readout over the entire lattice. We propose a detailed scheme in \app{measurement}. The success of VQE in simulating the GMD correlations associated with topological order further demonstrates that our framework can faithfully capture the  physical properties of systems whose physics is significantly different than the native Hamiltonian's.

\section{Quantum universality \& efficiency}
\label{sec:efficiency}

The above results numerically demonstrate that the fVQE framework can allow experiments to simulate models beyond the analog system's native Hamiltonian, and the fVQE's total error (VQE plus finite-size) seems to decrease exponentially more rapidly with run-time $T$ than na\"ive classical methods.
as we argue in this section, these numerical observations are well-grounded in our understanding of the method. We first establish the universality of our framework, argue for its expected scaling of error with system size  and experimental time $T$ for an optimized ansatz, and compare this to the corresponding results for na\"ive classical algorithms.

We begin by considering universality. Despite using only capabilities that are currently routine in quantum gas microscope or tweezer experiments, the framework is universal for particle-number-conserving fermionic quantum computation upon the addition of on-site spin rotations. 
This follows from Ref.~\cite{oszmaniec_universal_2017} -- universality is obtained once one can generate arbitrary passive fermionic linear-optics transformations (i.e. the evolution of arbitrary non-interacting fermionic Hamiltonians) together with at least one non-quadratic interaction gate in the relevant particle-number sector~\footnote{Technically, this result assumes that an $M$-mode system has a number of particles $N_p\notin\{0,1,M,M-1\}$. This presents no significant obstacle, as these cases involve either exactly zero or one particle, or zero or one hole. Therefore, their state space dimensions are no larger than $M$ and can be treated efficiently classically.}.
In our case, the fermionic modes are labeled site and spin indices $(i,\sigma)$, and  
as we argue in \app{universality}, the proposed framework, augmented with single-site spin rotations, can implement arbitrary  
two-mode tunneling gates using our fVQE ansatz $\m{H}$ defined in Eq.~\eqref{eq:drive} even with a uniform, nonzero Hubbard $U$ interaction, which suffices to generate the universal fermionic linear-optics operations. $\m{H}$'s Hubbard interaction term provides the gate outside of the linear fermionic optics operations required for universality.  
This universality shows in principle that the fVQE framework can generate ground states of arbitrary non-native Hamiltonians beyond those natively realized by the analog quantum simulation capability, in accord with the results above showing effective simulation of ground states of the EFHM and HHM.

While universality suffices to show that the fVQE scheme can create arbitrary ground states, it leaves unanswered why the fVQE ansatz does so efficiently in interesting cases, as our numerical results above indicate, and when this is possible. In particular, we would like to understand the rapid convergence of the VQE error with increasing total evolution time $T$ and relatively weak dependence on $L$, especially the associated scalings. 

We now argue that the VQE can obtain local observables with a target accuracy $\epsilon$ for the thermodynamic limit ground states of a broad class of local Hamiltonians, at a time exponentially faster than naive classical algorithms, such as exact diagonalization (ED).
This characterization of error decay in the thermodynamic limit is similar to Ref.~\cite{CiracAnalogAdvantage2024}. 

For many classical algorithms -- such as ED -- that obtain finite-size quantum ground states, their major error source is the finite-size error. This error can be estimated. 
The FHM and EFHM exhibit gapless phases in the thermodynamic limit.
We expect that the finite-size error of local observables in such gapless phases decays as a power law $\epsilon_L\sim L^{-p}$ \cite{Sachdev_2011,hastings_qac_2005} with $p$ a model-specific constant.
For the HHM geometry studied here, although gapped in the bulk, its spatially averaged local observables, such as the energy density, also exhibit the same $\epsilon_L\sim L^{-p}$ finite-size scaling due to contributions from critical edge states~\footnote{In this work we keep $L_x$ fixed and increase $L_y$, so the two edges, each with finite-size scaling $\sim L^{-p}$, contribute $O(L_y)$ sites, i.e. an $O(1)$ fraction of the system volume. In the isotropic 2D limit $L_x,L_y\to\infty$, the edge contribution is instead suppressed as $\sim 1/L_{x,y}$, so the finite-size error of the energy density decays slightly faster, $\epsilon_L\sim L^{-(p+1)}$. In both cases, the finite-size error remains power-law in the linear system size.}.
Also, because $L_x$ is kept fixed while $L_y$ is varied, the total number of HHM lattice sites $N_s=L_xL_y\propto L$; effectively $D=D_{\mathrm{eff}}=1$ for the scaling analysis hereafter.

For ED, the run-time $T$ needed to compute the target ground state scales exponentially $T\sim d^{N_s} \sim d^{L^D}$ where $d$ is the local Hilbert space dimension. This means that  in a computational time $T$ one can solve  a system size  $L\sim(\log T)^{1/D}$, and therefore, the ED algorithm error shows an inverse logarithmic decay with run-time
\begin{equation}\label{eq:EDerror}
    \epsilon=\epsilon_L\sim \frac{1}{L^p}\sim \frac{1}{\log^{p/D} T}.
\end{equation}
This derives the classical error scaling represented by the gray dashed curves in the insets of \fig{vqe}. The constant $p=1$ is estimated using the finite-size extrapolation of the exact ground state energy densities.

One can similarly try to understand our observed results for the fVQE.
We will show that the qualitative scaling observed in the fVQE matches what  would  be achieved by finding an adiabatic path (that is guaranteed to exist under mild assumptions), and then evolving this trajectory using known quantum simulation techniques. This provides some understanding of the fVQE scaling, while also showcasing the fVQE's ability to achieve this scaling without having to directly find the adiabatic path or compile the somewhat complex quantum simulation routines.

One strategy to  prepare a target quantum state $\ket{\psi_{1}}$ is to find a gapped trajectory of local Hamiltonians $H(s)$, $s\in[0,1]$, that connects an easy-to-prepare initial ground state $\ket{\psi_{s=0}}$ to the target ground state $\ket{\psi_{1}}$, where $\ket{\psi_s}$ denotes the ground state of $H(s)$.
If we denote by
\begin{equation}
    V_\tau=\mathcal{T}_s \exp\!\left[-i\tau\int_0^1 ds\, H(s)\right]
\end{equation}
the evolution over total adiabatic time $\tau$, where $\mathcal{T}_s$ is the time-ordering operator, then many-body adiabatic theorems imply that, for a local gapped Hamiltonian path, the error in any local observable is bounded independently of the total system size $N_s$ \cite{bachmannAdiabaticTheoremQuantum2019,bachmannAdiabaticTheoremQuantum2017,teufelQuantumAdiabaticTheorem}. 
Additionally, for a smooth trajectory $H(s)$ whose $s$-derivatives vanish to all orders at the endpoints, the error of a local observable $A$ is expected to decay in a stretched-exponential with $\tau$, consistent with our numerical observations:
\begin{equation}\label{eq:adiabatic}
    \epsilon_A=|\bra{\psi_0}V^\dagger_\tau AV^{\phantom\dagger}_\tau\ket{\psi_0} - \bra{\psi_1}A\ket{\psi_1}| \leq C_A e^{-c_1(\Delta\tau)^{1/D}},
\end{equation}
where $\Delta$ is the minimum many-body gap along the interpolation, and $C_A,c_1$ are constants independent of $N_s$. 
In the case of preparing the ground state of critical systems where the spectral gap closes with $L$, which are the models studied in the numerics, we assume that the minimum many-body gap along the path is controlled by the finite-size gap of the target Hamiltonian $\Delta_{\min}(L)\sim L^{-z}$, with $z$ the dynamical critical exponent. We note that our scaling framework also recovers the results of the gapped system by setting $z=0$ (see \app{epsScaling}).

In the above setting, preparing $\ket{\psi_1}$ to local-observable accuracy $\epsilon_A$ can be reduced to simulating evolution for a fictitious time $\tau$ under the local Hamiltonian $H(s=t/\tau)$.
This is a task for which the quantum simulator has an exponential advantage over ED-like exact classical algorithms. The HHKL quantum algorithm simulates time-$\tau$ local quantum dynamics of a size-$N_s$ system with a linear (up to logarithmic factors)
number of (spatially local) two-mode gates
\begin{equation}\label{eq:Ngate}
    N_\text{gate}\sim O(N_s\tau\log(N_s\tau/\epsilon_S)),
\end{equation}
where $\epsilon_S$ is the quantum simulation error of the local observable~\cite{haah_qalgo_2023,wolfQuantumComputingLecture2023}.
As shown in \app{universality}, due to the universality of our framework, each of such two-mode gates can be simulated using a sequence of our drive Hamiltonians. 
The number of fVQE layers to compile $N_{\text{gate}}$  two-mode local gates can then be characterized by the scaling $N_l \sim N_s^q N_{\text{gate}}$, where $q=0$ if one encodes a single gate per finite number of fVQE layers, and $q=-1$ if $O(N_s)$ local gates can be executed in parallel per layer. The $q=-1$ limit is saturated when each layer of circuit perfectly simulates $O(N_s)$ local gates in parallel, and even the most pessimistic estimate gives $q\geq 0$. The total circuit depth is thus
\begin{equation}
    N_l\sim N_s^{q+1}\tau\log(N_s\tau/\epsilon_S). \label{eq:Nl-depth-sim-plus-ovhd}
\end{equation}

Solving Eq.~\eqref{eq:adiabatic} for the required adiabatic time $\tau$ to reach local-observable error $\epsilon_A$, and then balancing $\epsilon_A=\epsilon_S=\epsilon_Q/2$, where $\epsilon_Q$ is the total quantum simulation error, we obtain, to leading order,
\begin{equation}
    N_l\sim \frac{N_s^{q+1}}{\Delta}\log^{D+1}(1/\epsilon_Q).
\end{equation}
In our framework, we set a maximum time $T_{\rm max}$ for each layer of circuit (see \app{optim}), so the total evolution time $T\leq N_lT_{\rm max}$ depends at worst linearly on $N_l$.
Then the  quantum simulation error scales with $T$ as
\begin{equation}
    \epsilon_Q\sim \exp\left[-c'_2\left(\frac{T\Delta}{N^{q+1}_s}\right)^\frac{1}{D+1}\right],
\end{equation}
with $c'_2$ an $N_s$-independent constant. 
Using $N_s\sim L^D$, and $\Delta=g_0L^{-z}$, 
we obtain the unified leading-order scaling
\begin{equation}\label{eq:vqeScaling}
    \epsilon_Q\sim \exp\left[-c_2\left(\frac{T}{L^{b}}\right)^\frac{1}{D+1}\right],
\end{equation}
with $b=D(q+1)+z$ and $c_2$ another $L$-independent constant.
A more careful asymptotic derivation in \app{epsScaling} yields the prefactor, which depends only algebraically on $L$ and $T$.
This prefactor modifies only subleading logarithmic corrections o the asymptotic scalings of the total error \eq{VQEerror} derived hereafter.

We note that, as discussed in detail in \app{epsScaling}, this asymptotic theory assumes $T\gg L^b$. Thus, it applies to the large-$T$ scaling with a fixed finite $L$ or that $L$ scales no faster than $T^{1/b}$. For $D=1$, Eq.~\eqref{eq:vqeScaling} implies at fixed $L$ that the error decays as $\epsilon_Q\sim e^{-c\sqrt{T}}$, consistent with our numerical observation in \fig{vqe} via ansatzes (despite using neither explicit smooth trajectory design nor near-optimal circuit compilation). However, this asymptotic bound is inapplicable to study the error convergence when $L\to\infty$ at fixed $T$ as this violates the $T\gg L^b$ assumption. In \app{epsScaling}, we also show numerically that the fixed-$T$ $\epsilon_Q$ converges as $L\to\infty$, and that the value it converges to decreases rapidly with $T$. 

The above discussion on quantum algorithm errors for fixed-$L$ systems does not include the finite-size error $\epsilon_L$, which for a gapless system has the same power-law decay $\epsilon_L\sim L^{-p}$ as mentioned for classical algorithms. 
The total error of the quantum approach $\epsilon$ is the sum of the quantum simulation error $\epsilon_Q(L;T)$ and the finite-size error $\epsilon_L(L)$ relative to the thermodynamic limit. For each linear system size $L$, there is an ``optimal evolution time $T^*(L)$'' at which these two contributions are comparable,
\begin{equation}\label{eq:TStar}
    \epsilon_Q(L;T^*)=\epsilon_L(L).
\end{equation}
When $T\ll T^*$, the total error is controlled by $\epsilon_Q$ and decreases with increasing $T$. When $T\gg T^*$, it is controlled by $\epsilon_L$, so increasing $T$ further does not change the leading scaling of the total error, so one should choose $T\sim T^*$ to avoid spending computational time  minimizing a negligible error.
Consequently, for large $T^*$, solving \eq{TStar} asymptotically using the result Eq.~\eqref{eq:vqeScaling} gives
\begin{equation}
    L(T^*)\sim \left(\frac{T^*}{ \log^{D+1}T^*}\right)^\frac{1}{b},
\end{equation}
up to subleading logarithmic corrections from the prefactor in Eq.~\eqref{eq:vqeScalingPref}. (As shown in detail in App.~\ref{app:epsScaling}, Eq.~\eqref{eq:vqeScaling} is valid in the required regime.)
Substituting the optimal size $L(T^*)$ back into $\epsilon_L\sim L^{-p}$ (equivalently, into the balanced error $\epsilon_Q\sim\epsilon_L$), we obtain the total error scaling up to a logarithmic factor,
\begin{equation}\label{eq:VQEerror}
    \epsilon(T^*) = \epsilon_Q + \epsilon_L\sim \left(T^*\right)^{-\frac{p}{b}}.
\end{equation}
This concludes the claim we made that the quantum evolution time to simulate ground state local observables at the thermodynamic limit grows polynomially with the inverse relative error $T\sim O(\mathrm{poly}(1/\epsilon))$ up to logarithmic corrections. It further shows that, to converge to accuracy $\epsilon$, the quantum evolution time is exponentially shorter than that of classical exact diagonalization given by \eq{EDerror}.

The above theoretical framework is based on the assumption of a power-law finite-size error $\epsilon_L$. This applies to the critical systems we numerically demonstrate in this work (the HHM is critical on the edge). In a fully gapped, locally interacting quantum system with finite-dimensional on-site Hilbert space, the finite-size error of a bulk local observable decays exponentially with its distance to the boundary, typically at the scale of the system size, $\epsilon_L\sim e^{-L/L_0}$ \cite{zywang_bounding_2021}, where $L_0$ is the correlation length. In this case, a similar analysis to that in \app{epsScaling} shows that VQE still has an exponential efficiency advantage over ED.

The theoretical framework presented here is not intended to prove a worst-case complexity theorem for our VQE runs. Rather, it shows that the observed rapid convergence is consistent with   locality-based state-preparation scalings established from near-optimal quantum simulation of adiabatic preparation.
Notably, in this sense, the numerically observed VQE performance behaves as if the fVQE dynamics were saturating these highly favorable bounds, despite using a much more straightforward, experiment-friendly ansatz and without explicitly constructing such optimized protocols. This convergence of theory and numerics provides strong evidence for VQE's utility for scalable, robust quantum state preparation. 

In addition to the exponentially shorter computation time, the quantum simulation also requires exponentially fewer spatial resources than classical algorithms. Similar to what we show for the computation time $T$, 
the quantum algorithm's ``space complexity'' ($M$,  the quantum system size $M=N_s$) is related to the finite-size error by
\begin{equation}
    \epsilon_L\sim L^{-p}\sim M^{-p/D}.
\end{equation}
This is, again, an exponential advantage over the classical ED-like algorithms, whose spatial resource to compute size-$L$ system grows exponentially as $M\sim d^{L^D}$, and thus its error decay with classical memory 
$M$ is another inverse logarithm $\epsilon\sim 1/\log^{p/D} M$.

While the above claims about quantum advantage are based on comparisons to ED -- relevant as a broadly-used, generally applicable classical algorithm with  well-understood asymptotic scalings -- they also imply, somewhat surprisingly, that our framework is even competitive with tensor network algorithms in 1D systems, where the latter are extremely efficient. In 1D, tensor network algorithms like MPS are significantly more efficient than ED even for gapless systems. The MPS algorithm has a time complexity $T\sim L\chi^3$ for each sweep with $\chi$ its maximal bond dimension. For ground states of critical systems, the   entanglement entropy $S_{E}$ grows with system size at most logarithmically, so MPS can represent the state with bond dimension growing only polynomially in $L$~\cite{pollmannTheoryFiniteEntanglementScaling2009,pirvuMatrixProductStates2012}. As is shown in \app{epsScaling}, to resolve a spatially averaged local observable whose finite-size error scales as $\epsilon_L\sim L^{-p}$, the required bond dimension scales as $\chi\sim L^{p/(2\kappa)}$, which then gives
\begin{equation}
    T\sim L\chi^3 \sim L^{1+\frac{3p}{2\kappa}}
\end{equation}
and hence
\begin{equation}
    \epsilon\sim T^{-\frac{p}{1+3p/(2\kappa)}},
\end{equation}
with $\kappa$ a model-specific constant, a scaling similar to the VQE error.
\fig{vqeMPS} numerically confirms this similar or even faster reduction of fVQE error with time compared to MPS across all three models, showing our quantum framework is competitive with tensor network algorithms even in their most efficient regime. 
fVQE's efficiency over MPS also holds for simulating local observables in the gapped bulk system, as we summarize in \app{epsScaling}.
For MPS representations of $D>1$-dimensional systems, the favorable 1D scaling of the MPS is lost, since the bond dimension needed to capture area-law quantum entanglement $\chi\sim d^{L^{D-1}}$ suffers from exponential growth with the transverse width of the system. Hence, the inverse logarithmic error scaling $\epsilon\sim \frac{1}{\log^{p/(D-1)} T}$ applies. This inverse-log scaling also applies to space complexity, simply by replacing the run-time $T$ with the memory resource $M$.
Because the VQE algorithm's time and spatial complexity depend much more mildly on spatial dimensions, while the MPS-like tensor network algorithm is significantly harder in such scenarios, our framework is expected to have a significant quantum advantage over  MPS-based methods in higher dimensions.

\section{Summary \& discussion: classical optimization and current experimental outlook}
\label{sec:summary}

We have  proposed an experiment-friendly framework implementing VQE on  fermionic neutral-atom quantum simulators, showing they are already a potent tool in systems such as quantum gas microscopes. The fVQE framework's universality enables it to simulate quantum many-body systems beyond its hardware's native interactions, and we presented evidence that the algorithm is efficient for simulating the ground-state properties of many locally interacting models. We present the ansatz designs for near-term experiments to simulate three physically important models. 
Through classical numerical simulations of the quantum platform, we also show that our designs achieve an exponential speedup over classical methods such as ED. Our ansatz families use a number of parameters per layer that is independent of  the linear system size $L$ to benefit the classical optimization. In the end, we present theoretical arguments on our framework's versatility, scalability, and efficiency.

The presented fVQE framework is implementable in current experiments. Unlike gate-based quantum circuit designs, local control is unnecessary for many models, or minimal (slowly spatially varying on-site potentials) for the HHM. Moreover, following the principles outlined in \sect{vqe} for our ansatz design, our framework's $L$-independent parameterization requires minimal prior knowledge of the target Hamiltonian, yet  maintains its efficiency. In \sect{efficiency}, it was seen that the fVQE matches the scaling of a quantum simulation algorithm based on smooth, regularized, and optimized adiabatic trajectories with dynamics implemented to circumvent Trotter error, requiring complicated quantum circuits such as linear combinations of unitaries (LCU)~\cite{wolfQuantumComputingLecture2023}. The presented VQE framework, on the other hand, is free from such complicated circuit designs. We also note that it could be applied -- in addition to the fermionic architecture -- to qubit or boson-based architectures.

For the concrete fVQE protocols demonstrated in this work, we use only native Hubbard dynamics together with experimentally accessible controls of tunneling patterns and on-site potentials. Separately, App.~\ref{app:universality} shows that, upon augmenting these controls with local spin rotations, the platform can synthesize a universal particle-number-conserving fermionic gate set. This universality result explains the reach of the method  numerically demonstrated in the main text, and shows the framework is  in principle capable of reaching broad ranges of ground states well-beyond those natively implemented by the hardware or the specific Hamiltonians explored here. A rigorous characterization of the expressive power of the symmetry-restricted ansatz families used here is to be published in future work \cite{sun_inprep}.
As hardware further develops, the fVQE can naturally accommodate additional native operations that arise.

We close our discussion with a brief analysis of two important issues:  the expected capabilities of the fVQE algorithm executed on existing and future hardware (and the contrast with more standard qubit-based hardware), and the cost of classical optimization. We find that existing hardware can already study interesting physics, and may already be capable of treating some classically challenging questions,  including the cost of classical optimization and accounting for experimental fidelities.

Existing experiments are capable of running the proposed fVQE algorithms to obtain high-quality ground-state expectation values, with accuracy that  may already achieve quantum advantage over classical simulations in some cases. To see this, we consider a crude estimate of the requirements for current experiments that can simulate local observables in a 2D lattice with a roughly $2-3\%$ target accuracy, including all errors --  fVQE, optimization, and finite-size error. This level is classically challenging for many Hubbard-like problems~\cite{wuVariationalBenchmarksQuantum2024}.

To estimate operation fidelities, we note a recent experiment has realized double-well fermionic collisional entangling gates of gate time $\tau_h=0.25\pi/J$ with a fidelity as high as $99.75\%$ \cite{bojović2025highfidelitycollisionalquantumgates}, where $J\sim t$ is the superexchange energy. To reach a $1\%$ VQE algorithm error, the total gate time is on the order of $T=10\pi/U=2\pi/t$, if we assume the VQE error remains similar in 2D lattices as is suggested by the theory scaling analysis. This suggests that the experimental time evolution would introduce $\sim2\%$ error. The error from the measurement shot noise can also be controlled to the same $1\%$ level by repeating sufficiently many experimental shots. In \app{run-time}, we estimate that the total run-time, including all in-loop optimization using the most na\"ive optimization algorithms, would be around 8000 experimental shots and $20$ hours, in a quantum gas microscope.

This estimate shows the current potential of the fVQE. Furthermore, rapid progress on multiple fronts is likely to allow more accurate calculations in shorter times: higher operation fidelities, shorter one-shot experimental cycle times such as that achievable with tunnel-coupled tweezers,    error mitigation techniques, and more advanced optimization strategies.
As these improvements push fVQE toward smaller target errors, it becomes important to account for overheads beyond the quantum evolution time. In \app{optim} and \app{run-time}, we analyze the optimization and measurement overheads, respectively, and argue that both add only polynomial factors, thereby preserving the main polynomial-in-$1/\epsilon$ scaling conclusion.

Finally, it is useful to compare the capabilities of the fVQE to similar quantum algorithms implemented on a traditional qubit-based quantum computer. There are a few distinctions. First, the fermionic hardware such as quantum gas microscopes already operate with a thousand or more atoms, and have demonstrated the ability to create rich many-body equilibrium states of Hamiltonians already in a beyond-classically-regime that could be used as initial states in a fVQE scheme. Second, they avoid the overhead of mapping fermions to qubits, although this does not change the expected scalings of fVQE error.
Fermion-to-qubit mappings, such as Jordan-Wigner and ancilla-free Bravyi-Kitaev transformations, require non-local Pauli strings to encode fermionic statistics \cite{mcardle_qchem_2020,BK2002}.  
Topological compact encodings for local fermionic operators proposed \cite{chienSimulatingFermionsDigital2026,BK2002,setiaSuperfastEncodingsFermionic2019,chien_optimizing_2022,kirbySecondQuantizedFermionicOperators2022,algabaLowdepthSimulationsFermionic2024} and experimentally demonstrated on highly reconfigurable quantum architectures \cite{nigmatullinExperimentalDemonstrationBreakeven2025,everedProbingKitaevHoneycomb2025,granetSuperconductingPairingCorrelations2025} require ancilla qubits and at least seven two-qubit entangling gates to implement even the simplest fermionic exchange operations, effectively increasing the error rate per fermionic operation seven-fold over a native implementation. 
This overhead is larger for spinful models such as the FHM and EFHM, or for complicated interaction terms \cite{granetSuperconductingPairingCorrelations2025}. 
The topological encodings also require preparing long-range-entangled stabilizer initial states, requiring either non-local unitary circuits or mid-circuit measurements \cite{guaita_locality_2024,chienSimulatingFermionsDigital2026,tantivasadakarnLongRangeEntanglementMeasuring2024}.
Additionally, the encoded fermion exchange statistics and particle number could be corrupted by gate noise, whereas physical fermions have naturally protected exchange statistics and lose particles slowly. 
We note that fault tolerance can be achieved in fermionic platforms via recently proposed methods~\cite{schuckertFermionqubitFaulttolerantQuantum2024a,ottErrorcorrectedFermionicQuantum2024}. While this is  also achievable in fermion-to-qubit mapping in conventional architectures, the fermionic error correction may have an advantage by avoiding this two-layer procedure~\cite{chenErrorcorrectingCodesFermionic2024}. 

\section*{Acknowledgments}

We thank Vaibhav Sharma, Visal So, Sohail Dasgupta, Zewen Zhang, Han Pu, Diego Fallas Padilla, Fang Xie, Kevin Slagle, Joseph Desroches, Ao Chen, Nan Cheng, Feng Chen,  Ziwei Su, Run Hou, Yi Xu,  Arielle Sanford, Zoe Yan, and Jacob Covey for discussions. Computing resources were supported in part by Rice University's Center for Research Computing (CRC) and the Google Cloud Research Credits Program.  We acknowledge support in part from the National Science Foundation (PHY-1848304), the Department of Energy (DE-SC0024301),   the Office of Naval Research (N00014-20-1-2695), and the W. M. Keck Foundation (Grant No.
995764).

\appendix

\counterwithin{figure}{section}
\renewcommand{\thefigure}{\Alph{section}\arabic{figure}}

\section{Preparing initial states}
\label{app:initState}

\begin{figure}
    \centering
    \includegraphics[width=0.6\linewidth]{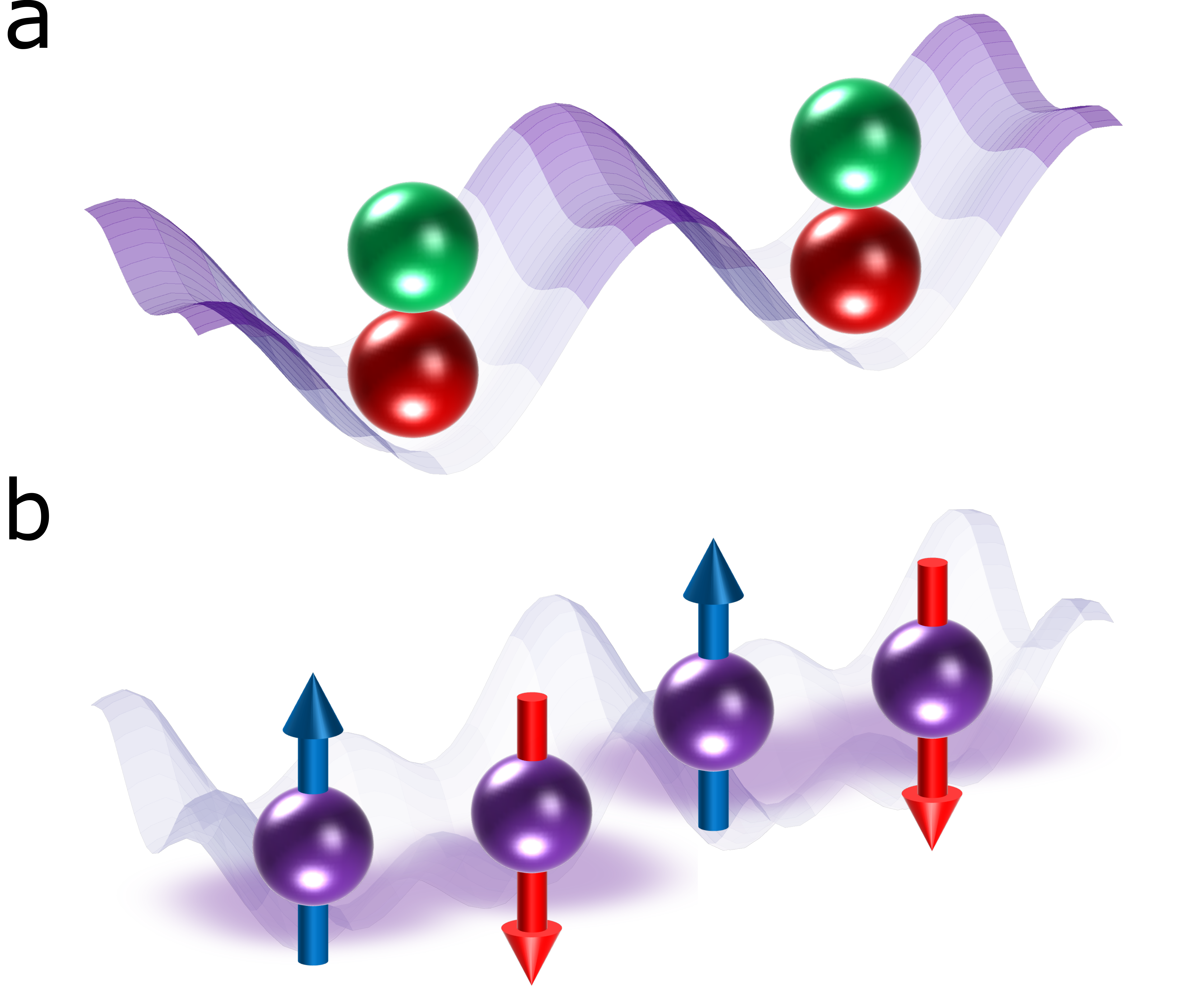}
    \caption{A preparation scheme for the one-particle-per-site SU(2) spin singlet product state. In the experiment setup, there are two interfering laser beams with a $2:1$ wavelength ratio, which we call the long wavelength lattice and the short wavelength lattice. 
    (a) By forming a band insulator, paired atoms with balanced spins are loaded into each trap of the long wavelength optical lattice. (b) Slowly ramping up the short wavelength optical lattice with balanced trap depth, each doublon in the long wavelength lattice trap is equally split into a spin singlet in the double-well superlattice.}
    \label{fig:statePrep}
\end{figure}

In this section, we present our proposal to prepare the initial states used in the main text. Our considerations focus on the experimental accessibility and fidelity of the states created. For the FHM and the EFHM, we choose the one-particle-per-site SU(2) singlet product state as our initial state
\begin{equation}
    \ket{\psi_0}=\ket{\text{singlet}}\ket{\text{singlet}}\dots\ket{\text{singlet}},
\end{equation}
where
\begin{equation}
    \ket{\text{singlet}}=\frac{1}{\sqrt{2}}\left( c^\dagger_{L\uparrow}c^\dagger_{R\downarrow} + c^\dagger_{R\uparrow}c^\dagger_{L\downarrow}\right)\ket{0}.
\end{equation}

This is a state that can be experimentally realized with high fidelity by various methods. 
In our proposed experiment setup, there are two interfering laser beams with a $2:1$ wavelength ratio. One straightforward method,  demonstrated in Ref.~\cite{chalopinOpticalSuperlatticeEngineering2025a}, is first turning on only the long wavelength lattice, and loading two atoms into each site of the lattice (\fig{statePrep}a). Then, one adiabatically ramps up the short-wavelength lattice such that the doublons in the original long-wavelength lattice  split into a singlet across the two wells of the double-well superlattice (\fig{statePrep}b). 
Alternatively, as  Ref.~\cite{bojović2025highfidelitycollisionalquantumgates} recently demonstrated, if one ramps up the short wavelength lattice with a strong magnetic gradient instead, this prepares an $\ket{\uparrow,\downarrow}$ state, and then one can perform a $\sqrt{\mathrm{SWAP}}$ entangling gate followed by a quarter-cycle singlet-triplet oscillation using another magnetic field gradient, with a fidelity for the entangling gate of $\sim 99.75\%$. 

\begin{figure}
    \centering
    \includegraphics[width=\linewidth]{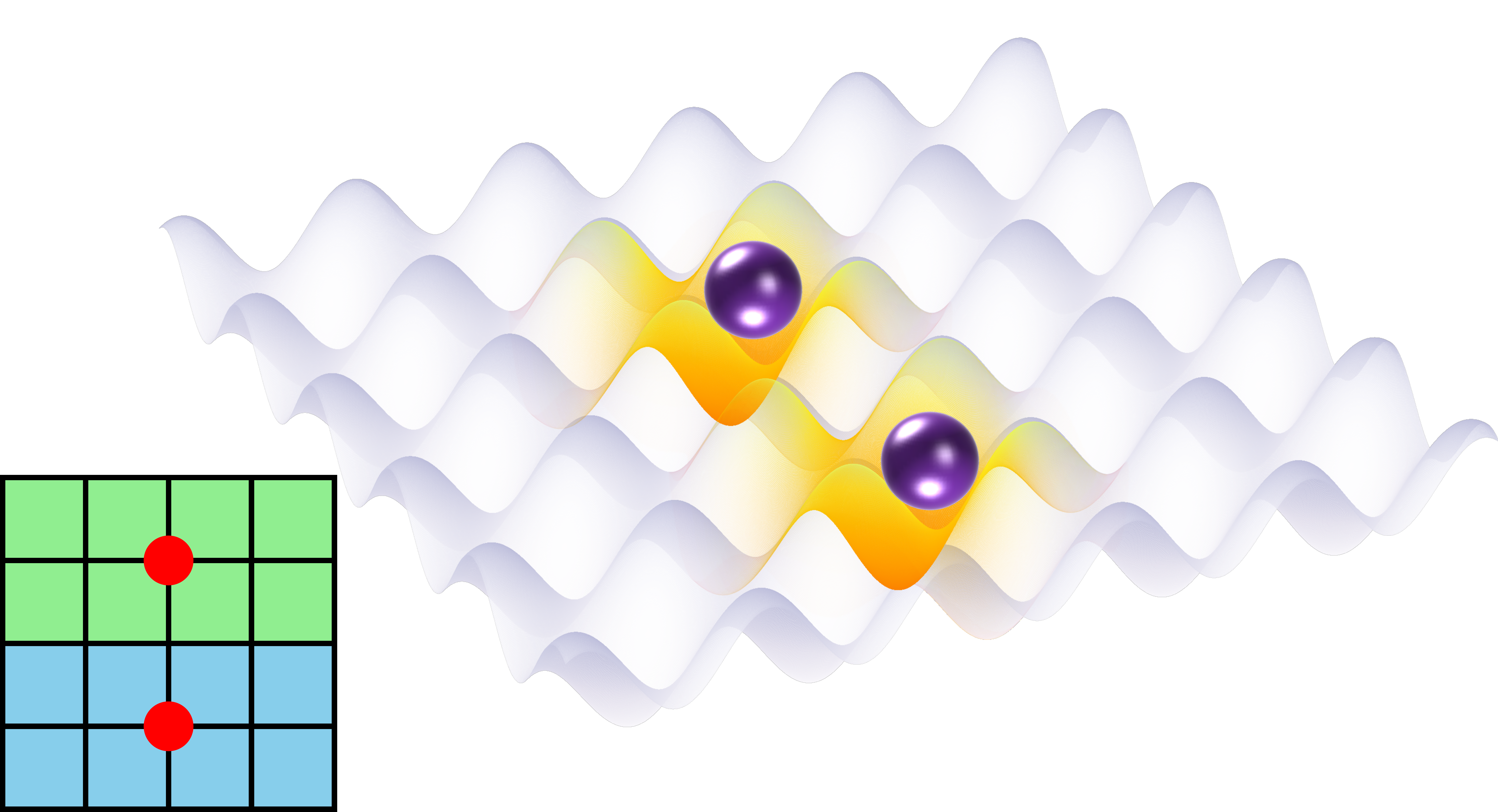}
    \caption{The product state of one particle (the red dot in the inset) localized at a single site in each $4\times 2$ cell (colored blocks in the inset).}
    \label{fig:HHInitState}
\end{figure}

Our initial state choice is slightly different for the hardcore bosonic HHM. 
To get the right density for the filling $\nu=1/m$ Laughlin state, we need $N_p=\frac{\phi N_\Box}{2\pi m}$ particles, where $N_\Box$ is the number of plaquettes (i.e., the area). For $\phi=2\pi/4$, this implies  $N_p=N_{\Box}/8$. Therefore, we consider systems with tiled 8-site unit cells, giving $N_s=L_x\times L_y = (1+4n_x)\times (1+2n_y)$ site lattices ($n_x, n_y \in \mathbb Z$), resulting in $N_\Box=(L_x-1)\times(L_y-1)=8n_xn_y$ plaquettes. Accordingly, we choose an initial state that is a product state with one particle localized at a single site in each $4\times 2$ cell, as illustrated in \fig{HHInitState}. This can be prepared by selectively tuning on-site potentials at specific sites during atom loading.

Our initial state choice is different from the  strategy used in many qubit-based VQE methods. In VQE methods such as the unitary  coupled-cluster (UCCSD) ansatzes~\cite{grimsleyAdaptiveVariationalAlgorithm2019}, the initial state in the ansatz to find the Hamiltonian ground state is usually the Hartree-Fock product state, which shortens the ``distance'' to the target state.
The combination of the Hartree-Fock or other initial state preparation 
is an interesting direction for future study.

\section{Energy error scalings}
\label{app:epsScaling}

\begin{table*}[hbt]
    \centering
    \begin{tabular}{c|c|c|c}
    \hline
     Systems & VQE & ED & MPS ($D=1$) \\
    \hline
     Critical & $T^{-\frac{p}{D(q+1)+z}}$ & $\frac{1}{\log^{p/D} T}$ & $T^{-\frac{p}{1+3p/(2\kappa)}}$ \\
     Gapped & $\exp\left[-c_VT^\frac{1}{Dq+2D+1}\right]$ & $\exp\left[-c_E\log^{1/D} T\right]$ & $\exp\left[-c_M\log^{4/3} T\right]$ \\
    \hline
    \end{tabular}
    \caption{Scaling of the algorithm error $\epsilon$ with computation time $T$ for local observables in $D$-dimensional systems. The constants $c_{V,E,M}$, $p$, $q$, $z$, and $\kappa$ are model-dependent parameters defined in \sect{efficiency} and in this Appendix.}
    \label{tab:scaling}
\end{table*}

\sect{efficiency} estimates the leading asymptotic scaling of the VQE error with computational time $T$ for critical systems and compares it with the same scaling for exact diagonalization (ED). In this Appendix, we extend that discussion in three directions. First, we refine the VQE derivation to include the first subleading contribution from the logarithmic simulation overhead in Eq.~\eqref{eq:Ngate}, which produces an algebraic prefactor for $\epsilon$ as a function of $T$. Second, we derive  the corresponding leading-order scalings for gapped systems. Third, we compare the leading-order results with the scaling for matrix product state (MPS) methods for both gapped and critical systems in 1D. The resulting asymptotic scalings are summarized in Table~\ref{tab:scaling}.

\subsection{Error scaling of critical systems for VQE}

\subsubsection{Refined asymptotic derivation of the VQE error scaling}
\label{app:vqePrefactor}

The main text gives the leading-order dependence of the quantum simulation error $\epsilon_Q$ on the total evolution time $T$ and the system size $L$. Here we derive the first subleading correction, an algebraic prefactor in $T(\epsilon_Q)$, arising from  the HHKL's logarithmic simulation overhead in Eq.~\eqref{eq:Ngate}. 

Starting from the adiabatic bound at a system size $N_s$ and a large $\tau$
\begin{equation}
    \epsilon_A \le C_A e^{-c_1(\Delta\tau)^{1/D}},
\end{equation}
we obtain
\begin{equation}
    \tau \sim \frac{1}{\Delta}\left[\log(C_A/\epsilon_A)\right]^D.
\end{equation}
Balancing this with the quantum simulation error arising from HHKL, $\epsilon_A=\epsilon_S$, gives a total error $\epsilon_Q =2 \epsilon_A$, so rewriting $\log(C_A/\epsilon_A)=\log(2C_A/\epsilon_Q)$ and defining
\begin{equation}
    x:=\log(1/\epsilon_Q),\quad w:=x+\log(2C_A)
\end{equation}
gives
\begin{equation}
    \tau \sim \frac{w^D}{\Delta}\sim \frac{1}{\Delta}\left[x+\log(2C_A)\right]^D. \label{eq:tau-D-x}
\end{equation}

Substituting   $\tau$ into the  estimate Eq.~\eqref{eq:Nl-depth-sim-plus-ovhd},
 $   N_l \sim N_s^{q+1}\tau\log(N_s\tau/\epsilon_Q)$,
 yields
\begin{align}
    N_l \sim\blank \frac{N_s^{q+1}}{\Delta}\left[x + \log (2C_A)\right]^D
    \log\!\left(\frac{N_s}{\Delta\,\epsilon_Q}\left[x + \log (2C_A)\right]^D\right).
\end{align}
Since  $T\propto N_l$ and $N_s\sim L^D$, $\Delta\sim L^{-z}$ so that  $\frac{N_s^{q+1}}{\Delta}\sim L^b$ with $b=D(q+1)+z$, we write
\begin{align}\label{eq:Tdef}
    T \sim\blank L^b\left[x + \log (2C_A)\right]^D \nonumber\\
    \blank\times 
    \left\{x+(D+z)\log\!L+D\log\!\left[x + \log (2C_A)\right]\right\}
\end{align}

\eq{Tdef} gives the relation between $\epsilon_Q$ and $T$. To understand it, we expand it in  the regimes of interest. For large-$x$, i.e. small-$\epsilon_Q$,  the binomial expansion of Eq.~\eqref{eq:tau-D-x} gives
\begin{equation}
\tau \sim \frac{1}{\Delta}\left[x^D + Dx^{D-1} \log (2C_A)+\dots\right],
\end{equation}
so \eq{Tdef} simplifies to
\begin{equation}
    \frac{T}{L^b}\sim x^D
    \left[x+D\log x+(D+z)\log\!L+B+\cdots\right],
\end{equation}
where $B$ is an $L$-independent constant.
We note that in the later theory derivation, we balance $\epsilon_Q\sim\epsilon_L$ which   implies $x\sim\!p\log\!L$. Defining
\begin{equation}
    y:=\left(\frac{T}{L^{b}}\right)^\frac{1}{D+1},
\end{equation}
\eq{Tdef} then becomes
\begin{equation}\label{eq:xyImplicit}
    y^{D+1}=c_Dx^{D+1}+D x^D\log x+B x^D+\cdots, 
\end{equation}
with $c_D$ a $D$-dependent constant.

To invert Eq.~\eqref{eq:xyImplicit} at the large-$x$ limit, we use the asymptotic ansatz with the unknowns $u$ and $v$:
\begin{equation}
    x=c_2(y-u\log y-v+\cdots),
\end{equation}
and we will see that  constant $u$ and $v$ suffice to asymptotically satisfy Eq.~\eqref{eq:xyImplicit}.
Substituting this ansatz into Eq.~\eqref{eq:xyImplicit}, the equation is satisfied to leading order ($y^{D+1}$ terms);  solving for $u$ and $v$ to obtain the next two orders ($y^D\log y$ and $y^D$ terms)  
gives
\begin{equation}
    x=c_2\left[y-\frac{D}{(D+1)c_D}\log y-\frac{B}{(D+1)c_D}+\cdots\right].
\end{equation}
Substituting the definitions of $x$ and $y$   leads to the final result
\begin{equation}\label{eq:vqeScalingPref}
    \epsilon_Q(T,L)\sim C
    \left(\frac{T}{L^{b}}\right)^{\gamma}
    \exp\left[-c_2\left(\frac{T}{L^{b}}\right)^\frac{1}{D+1}\right],
\end{equation}
where $\gamma$ and $C$ are independent of $T$ and $L$. This formula is the leading order result for  $\epsilon_Q \ll 1$. 

In addition to the $x\sim\log\!L$ limit, one can consider the fixed-$L$ limit with $L\ll x$, as visible at  long times limit in
\fig{vqe}, and a similar asymptotic analysis to the above leads to an extra $L$-dependence in the prefactor
\begin{equation}\label{eq:vqeScalingFixL}
    \epsilon_Q(T,L)\sim CL^\beta
    \left(\frac{T}{L^{b}}\right)^{\gamma}
    \exp\left[-c_2\left(\frac{T}{L^{b}}\right)^\frac{1}{D+1}\right],
\end{equation}
where $\beta$ is also independent of $T$ and $L$.

\subsubsection{Numerically evaluating the fixed-$T$ finite-size scaling of the VQE energy error}
\label{app:fixT}

\begin{figure*}[htb]
    \centering
    \includegraphics[width=0.8\linewidth]{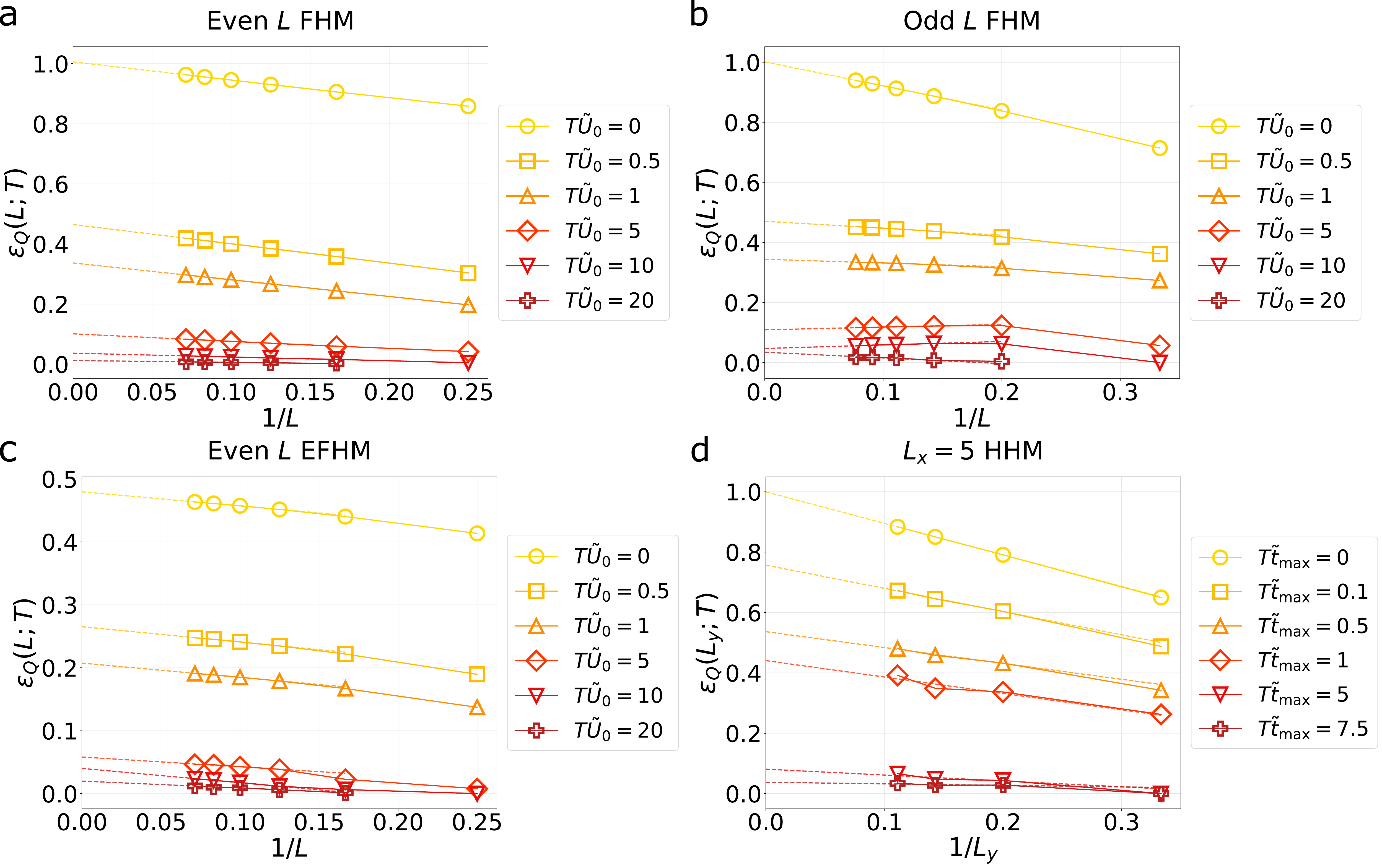}
    \caption{VQE energy error $\epsilon_Q(L;T)$ as a function of inverse linear system size $1/L$ at various fixed quantum evolution times $T$ for (a) even-$L$ FHM, (b) odd-$L$ FHM, (c) even $L$ EFHM, and (d) $L_x=5$ square-lattice bosonic HHM. The evolution time $T$ is defined in \app{qtime}. The dashed lines are fixed-$T$ linear fits to $\epsilon_Q(L;T)=\epsilon_Q(\infty;T)+A(T)/L$, performed on the  $4$ ($3$ for HHM) largest-$L$ data points for each $T$. The intercepts at $1/L=0$ estimate the thermodynamic-limit errors $\epsilon_Q(\infty;T)$, which are finite and decrease as $T$ is increased.}
    \label{fig:vqeFixT}
\end{figure*}

In the last subsection, we carefully derived the asymptotic VQE energy-error scaling in the $T\gg L^b$ regime, summarized by \eq{vqeScalingPref}. In particular, at fixed $L$ and large $T$, this scaling reduces to $\epsilon_Q \sim \exp\left[- f(L) T^{1/(D+1)}\right]$, which is consistent with the fixed-$L$ numerics shown in \fig{vqe} of the main text.

However, this asymptotic expression does not apply to the VQE error $\epsilon_Q(L;T)$ for fixed $T$ when taking the thermodynamic limit. This is because the assumption $T\gg L^b$ is violated when $L\to\infty$ for any fixed $T$. We note that this limit is not necessary for obtaining the scaling advantage discussed in the main text, but is nevertheless interesting to consider.

To understand  the behavior of $\epsilon_Q(L\to\infty;T)$ with $T$, we obtain the $L\to\infty$ values by numerically extrapolating the finite-$L$  $\epsilon_Q$ at fixed $T$ to $L=\infty$.
\fig{vqeFixT} plots the $\epsilon_Q(L;T)$ data versus $1/L$. 
The data collapse into straight lines at the largest $L$'s, which suggest that they follow a finite-size scaling
\begin{equation}\label{eq:fixTgeneral}
    \epsilon_Q(L;T) = \epsilon_Q(\infty;T)+\frac{A(T)}{L}.
\end{equation}
The parameters $\epsilon_Q(\infty;T)$ and $A(T)$ can then be extracted by the linear fits of data points of the largest $L$'s.
At each fixed $T$, the VQE error approaches a finite intercept $\epsilon_Q(\infty;T)$ as $1/L\to0$, which  decreases as $T$ increases -- the error remains finite in the $L\to \infty$ limit and increasing $T$   reduces the VQE error, even in the thermodynamic limit.

\subsection{Error scaling of gapped systems for ED and VQE}

We have shown the error scaling of ED and VQE for critical systems in \sect{efficiency}. The same analysis from Eq.~\eqref{eq:EDerror} to Eq.~\eqref{eq:VQEerror} can be readily extended to the gapped systems, with the replacement of the finite-size error $\epsilon_L \sim L^{-p}$ with $\sim e^{-L/L_0}$ \cite{zywang_bounding_2021}, where $L_0$ is the correlation length. Applying this change and that $L\sim\log^{1/D} T$ for ED,   the ED error is
\begin{equation}\label{eq:EDerrorGapped}
    \epsilon\sim e^{-L/L_0}\sim \exp\left[-\frac{\log^{1/D} T}{L_0}\right].
\end{equation}
For VQE, since the finite-size error is changed and the gap is present, the quantum simulation error $\epsilon_Q$ can be derived from \eq{Tdef} by assuming $x\sim L\gg 1$. This yields the same result, \eq{vqeScaling}, with $z=0$ and minor modifications to the prefactor. The optimal size $L(T^*)$ to satisfy \eq{TStar} becomes
\begin{equation}
    L(T^*)\sim (\Delta T^*L^{D+1}_0)^\frac{1}{Dq+2D+1},
\end{equation}
and the algorithm error for VQE when choosing the optimal $L$ for a time $T^*$ is
\begin{equation}
    \epsilon=\epsilon_Q+\epsilon_L\sim \exp\left[-\left(\frac{\Delta T^*}{L_0^{D(q+1)}}\right)^\frac{1}{Dq+2D+1}\right].
\end{equation}
Compared to the ED error scaling \eq{EDerrorGapped}, this is also an exponential improvement in the runtime required to reach fixed error, similar to the quantum improvement in the case of critical systems shown in the main text.

\subsection{Error scaling for 1D MPS}

In this subsection, we compare the fVQE error convergence with that of MPS algorithms by deriving how the finite-size and bond-dimension errors scale with computational time for the models considered in the main text. MPS is a variational ansatz that has errors from both its finite size $\epsilon_L$ and its finite bond dimensions $\epsilon_E$, i.e., the finite entanglement it captures by a truncated maximal bond dimension $\chi$. The finite-size error $\epsilon_L$ of MPS scales identically to that of ED and VQE. The finite-bond error $\epsilon_E$ has also been studied by past literature for 1D critical systems~\cite{pollmannTheoryFiniteEntanglementScaling2009,pirvuMatrixProductStates2012}, and a rigorous upper bound of $\epsilon_E$ has also been shown for 1D gapped systems~\cite{osborne_2007,dalzellLocallyAccurateMPS2019}. We translate these results into our framework for both critical and gapped systems.

\subsubsection{Critical system}

\begin{figure*}[htb]
    \centering
    \includegraphics[width=\linewidth]{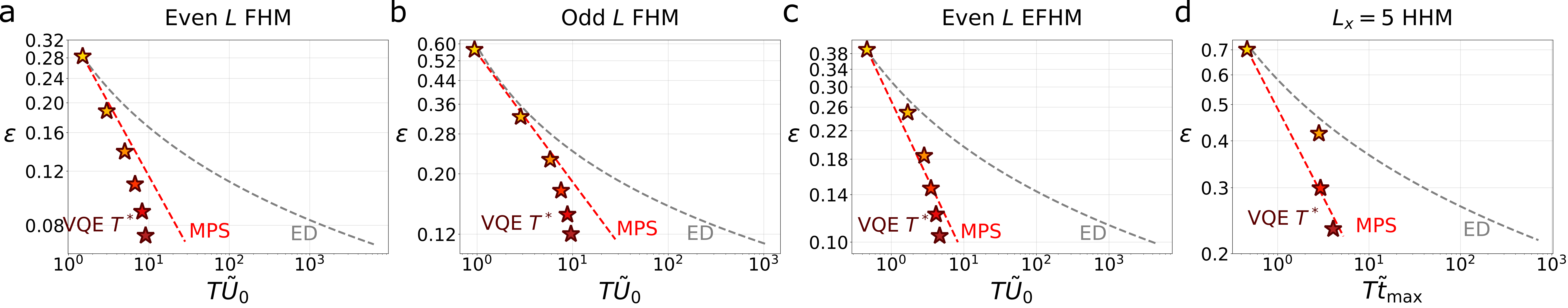}
    \caption{Variational energy relative error $\epsilon$ as a function of total evolution time $T$ for (a) even system size $L$ FHM, (b) odd $L$ FHM, (c) even $L$ EFHM, and (d) $L_x=5$ square lattice bosonic HHM, with all the notations the same as they are in \fig{vqe}. All plots suggest that the classical MPS computation time to reduce simulation error to a fixed target $\epsilon$ is comparable to, or longer than, the VQE optimal time $T^*$ on these systems. The red dashed line is the MPS algorithms' error scaling $\epsilon\sim T^{-\frac{1}{1+3/(2\kappa)}}$ derived in \app{epsScaling} using the estimate $p=1$ and $a=2$ appropriate for these models (the prefactor is arbitrary), with $T$ the classical computation time and $\kappa$ a constant depending on the model.}
    \label{fig:vqeMPS}
\end{figure*}

The FHM and EFHM in this work have gapless ground states in the thermodynamic limit, described by a conformal field theory (CFT) with  central charge  $c=1$. The HHM is gapped in the bulk, but hosts a gapless $c=1$ CFT on the boundary. For these models, $\epsilon_L\sim L^{-p}$, and we estimate $p=1$  from numerical extrapolation. 
On the other hand, for infinite system size, $L\to\infty$, the finite bond dimension $\chi$ for a theory with central charge $c$ gives an error for local observable~\cite{pollmannTheoryFiniteEntanglementScaling2009,pirvuMatrixProductStates2012} 
\begin{equation}\label{eq:FES}
    \epsilon_E\sim\frac{1}{\xi^a}\sim\frac{1}{\chi^{a\kappa}},
\end{equation}
where $a$ is an observable-dependent constant and $a=2$ for energy density, the large but finite correlation length $\xi\sim\chi^\kappa$ is the effective correlation length that the finite bond dimension MPS can capture, and the  exponent is
\begin{equation}
    \kappa=\frac{6}{c\left(\sqrt{12/c}+1\right)},
\end{equation}
To achieve a desired total error, both finite-size error and finite-entanglement error must be controlled, i.e. sufficiently large $L$ and $\chi$ must be taken simultaneously. 
To minimize the asymptotic total error $\epsilon=\epsilon_L+\epsilon_E$ for a given computational time, which scales as $T\sim L \chi^3$ for each MPS sweep, one should take the optimal $\chi^* \sim L^{p/(a\kappa)}$ to balance $\epsilon_L=\epsilon_E$. This choice also gives $\xi\propto L^{p/a}\ll L$, self-consistently ensuring that \eq{FES} is valid at the chosen $\chi^*$. Then the total error depends on the run-time as
\begin{equation}
    \epsilon=\epsilon_L+\epsilon_E\sim T^{-\frac{p}{1+3p/(a\kappa)}}.
\end{equation}
For $p=c=1$, $a=2$, the exponent is about $0.47$.

The power law scaling holds for space complexity as well, as the spatial resources  scale as $M\sim L\chi^2 \sim L^{1+2p/(a\kappa)}$, resulting in the MPS energy error decay with classical memory
\begin{equation}
    \epsilon\sim M^{-\frac{p}{1+p/\kappa}}.
\end{equation}

\fig{vqeMPS} displays the same data as \fig{vqe} of the main text, now with the MPS error scaling versus time. 
Only the slopes on the log-log plots are meaningful for comparison, as the prefactors are arbitrary and set for visual convenience.
Remarkably,  even in these one-dimensional cases where MPS is expected to be efficient, the VQE matches or exceeds the convergence rate of MPS. This strongly suggests that the VQE will have an even larger advantage in higher dimensions, where we expect the VQE scaling to be similar, while the MPS complexity will scale exponentially with the width of the system.  

\subsubsection{Gapped system}

Finally, we briefly discuss the MPS error scaling of the gapped systems. 
We note that the bound quoted here is a rigorous upper bound on local-approximation error as a function of bond dimension, and is generally considerably more pessimistic than what is observed in  DMRG calculations on simple gapped chains.

According to Refs.~\cite{AKLV2013,dalzellLocallyAccurateMPS2019}, the finite bond-dimension error $\epsilon_E$ of approximating the exact ground state of an $L$-site 1D gapped local Hamiltonian is upper bounded by the maximal bond dimension $\chi$ as
\begin{equation}
    \epsilon_E\leq C_ELe^{-c_\chi\log^{4/3}{\chi}},
\end{equation}
with $C_E$ and $c_\chi$ $L$-independent constants. 
As the finite-size error $\epsilon_L$ scales identically as in the ED case, to minimize the asymptotic total error $\epsilon_L+\epsilon_E$ when $L$ approaches the thermodynamic limit, we increase the bond dimension $\chi^*$ simultaneously with $L$ so that $\epsilon_E(\chi^*)=\epsilon_L(L)$. At large $L$, such an optimal increase of $\chi^*$ is asymptotically given by
\begin{equation}
    \chi^* \sim e^{c'_\chi L^{3/4}},
\end{equation}
with constant $c'_\chi$.
With the time complexity $T\sim L\chi^3$, we obtain the MPS error scaling for the gapped 1D chain
\begin{equation}
    \epsilon\sim \exp\left[-c_M\log^{4/3} T\right].
\end{equation}
This is a decay faster than any power law.

From the above discussions, we find that the MPS methods are comparably efficient with VQE for 1D and quasi-1D systems (a power law speedup for fVQE, if any). However, as we pointed out in \sect{efficiency}, if one uses a 1D MPS representation for a higher-dimensional lattice, the required bond dimension grows exponentially with the transverse width, resulting in an overall computational complexity that is significantly higher than that of the fVQE method.

\section{Optimization method}
\label{app:optim}

This section details  the optimization method
we use. We perform a standard SLSQP optimization algorithm 
where each parameter is constrained to an interval as described in \app{param-window},
and we iterate the algorithm until consecutive cost function values agree within $10^{-6}$.

The variational parameters used in the optimization algorithm are dimensionless. For the FHM and EFHM the parameters are $\m{t}_{1,2}/\Ut$, $\m{\mu}_{F,1}/\Ut$, $\m{\mu}_{F,2}/\Ut$, and $\m{T}\Ut$.  For the HHM, since $\Ut \to\infty$, we instead use the dimensionless parameters $\m{t_{x1}}\m{T},\,\m{t_{x2}}\m{T},\,\m{t_y}\m{T},\, \m{\mu_Q}\m{T}$ and $\m{\mu_S}\m{T}$ .

To efficiently seed the SLSQP initial values, we use what we call ``bootstrap" and ``transfer" techniques that optimize large $N_l$ and $L$ ansatzes using information obtained from smaller $N_l$ and $L$ solutions. To do this, for each data point in \fig{vqe}, we perform $101$ runs -- in three classes described momentarily -- and choose the best result. $50$ of the runs start from random initial parameter guesses sampled from a uniform distribution in the allowed windows that will be given in Table~\ref{tab:paramFHM} and~\ref{tab:paramHH} of App.~\ref{app:param-window}. Another $50$ runs start from initial guesses using the bootstrap technique described in App.~\ref{app:bootstrap}, which uses results from $N_l-1$ layers (if any exist) to seed the $N_l$-layer calculation.
A final run uses the transfer technique described in \app{transfer}, which uses results from $L-2$ (if they exist) to seed the $L$-site calculation. The calculation starts from $N_l=1$ and small $L$ (usually around $L=4$) and iteratively increases $N_l$ to the maximum value considered, then increments $L$ and repeats. For the largest $L$ ($L=13$ and $14$ for the FHM and $L=14$ for the EFHM), doing $101$ runs for each case became too expensive, so we used a single run based on the transfer technique, which was found to work well. 

Despite its simplicity, this optimization procedure is efficient enough to be useful, with the number of iterations $N_\text{iter}$ growing slowly with the system size $L$ and (roughly linearly) with the number of layers $N_l$. Although improving the optimization cost for the VQE is not a major focus of this work, these results suggest that the ansatz may be optimized efficiently even for much larger systems, in particular in the sense that the optimization costs do not eliminate the exponential scaling advantage of runtime.  There are many routes to improved optimization, which are left for future work.

\subsection{Parameter windows}
\label{app:param-window}

Our variational parameter space is restricted by experimental concerns and to enhance the efficiency and analysis of scaling of our optimization. 
Experimental concerns such as avoiding band mixing and rapid parameter changes constrain the parameters. Some such  constraints -- for instance, fixing the large Hubbard interaction at $\m{U}=\Ut$ -- have been reflected in our ansatz.
The experimental capabilities restrict the tunneling and on-site potential to be not much larger than $U$. Also, the signs of all tunnelings $t$'s need to be positive.
So we limit $0\leq t/\Ut\leq 1$ and $|\mu/\Ut|\leq 1$. To avoid the uncontrolled growth of ansatz evolution time, we limit the time parameter $0\leq\m{T}\Ut\leq\pi$.

Second, limiting parameter windows enhances the efficiency of the optimization algorithm. While a larger parameter window may contain better local minima on the variational energy landscape, searching this larger space is significantly slower and may contain many local minima and barren plateaus.  
To find the most efficient parameter windows, we started with the broadest ranges allowed by experimental restrictions and then systematically tested progressively narrower windows. For FHM and EFHM, we performed these tests on small systems with $L=6$ ($L=5$ for odd $L$). Specifically, we tested the windows of the drive Hamiltonian parameters $t/\Ut$'s and $|\mu/\Ut|$ progressively from $[0,1]$ to $[0,0.5]$ and $[0,0.25]$. 
The optimal windows are shown in Table~\ref{tab:paramFHM}.

\begin{table}[hb]
    \centering
    \begin{tabular}{c|c|c|c}
    \hline
     Parameters & Even $L$ FHM & Odd $L$ FHM & EFHM \\
    \hline
     $\m{t}_{1,2}/\Ut$ & $[0,0.25]$ & $[0,0.5]$ & $[0,0.5]$ \\
     $\m{\mu}_{F,r}/\Ut$ & - & - & $[-0.5,0.5]$ \\
     $\m{T}\Ut$ & $[0,\pi]$ & $[0,\pi]$ & $[0,\pi]$ \\
    \hline
    \end{tabular}
    \caption{The parameter intervals used for the optimization algorithm for FHM and EFHM.}
\label{tab:paramFHM}
\end{table}

For the HHM with parameters $\m{\mu_Q}\m{T}$ and $\m{\mu_S}\m{T}$, we compare between $[-\pi,\pi]$ and $[-0.5\pi,0.5\pi]$.
The optimal parameter window is shown in Table~\ref{tab:paramHH}.

\begin{table}[htb]
    \centering
    \begin{tabular}{c|c}
    \hline
     Parameters & HHM \\
    \hline
     $\m{t}_{x1,x2,y}\m{T}$ & $[0,0.5\pi]$ \\
     $\m{\mu}_{Q,S}\m{T}$ & $[-0.5\pi,0.5\pi]$ \\
    \hline
    \end{tabular}
    \caption{The parameter intervals used for the optimization algorithm for HHM.}
    \label{tab:paramHH}
\end{table}

\subsection{Bootstrap technique}
\label{app:bootstrap}
 
\def\ad#1{\textrm{ad}_#1}

\begin{figure*}
    \centering
    \includegraphics[width=\linewidth]{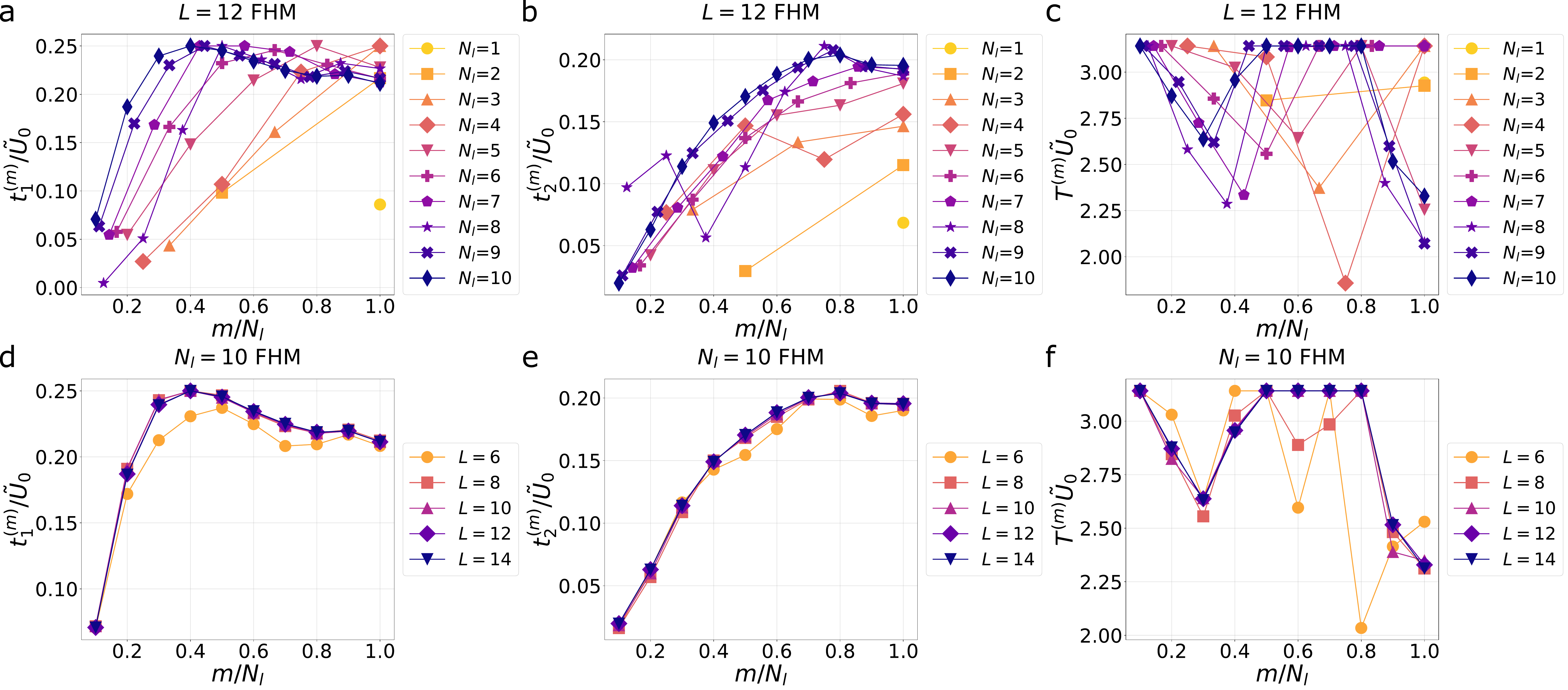}
    \caption{(a\textasciitilde c) The (a) $\m{t_1}/\Ut$, (b) $\m{t_2}/\Ut$ and (c) $\m{T}\Ut$ values of the $L=12$ FHM for ansatz depths from $N_l=1$ to $10$ as a function of $m/N_l$. The (d) $\m{t_1}/\Ut$, (e) $\m{t_2}/\Ut$ and (f) $\m{T}\Ut$ values of FHM solved by the $N_l=10$ ansatz of even $L$'s as a function of $m/N_l$.  In both cases, the $t^{(m)}$ converges to a single, smooth curve at large $N_l$.}
    \label{fig:trajectory}
\end{figure*}


To optimize large $N_l$ ansatzes efficiently, we use the bootstrap technique to generate good initial guesses given results from smaller $N_l$ ansatzes. For a number of layers $N_l$, we use an initial guess where the first $ N_l-1$ layers use the solution for the $N_l-1$-layer problem, and the last layer's parameters are chosen randomly.
As shown in Figs.~\ref{fig:trajectory}a-- c for the $L=12$ FHM, the ansatz parameters converge to a smooth function of the layer index $m$, as the total number of layers $N_l$ increases. This convergence suggests the applicability of the bootstrap scheme -- effectively, it implies that parameters in the earlier layers need only minor adjustments when a new layer is added.

\subsection{Transfer technique}
\label{app:transfer}

In addition to the fully random initial guesses and the bootstrap (using  random initial guesses for the last layer), we also use the transfer technique to optimize ansatzes of large $L$ lattices using smaller $L$ results without extra random guesses.
This is particularly straightforward since our ansatzes use $L$-independent parameterizations.
Additionally, we expect the parameters to depend weakly on $L$ for large $L$. (This is suggested by finite-size scaling bounding how the true result varies and Lieb-Robinson arguments applied to the \textit{VQE dynamics}.) 
This suggests the transfer procedure: we run one optimization of size-$L$ $N_l$-layer VQE, using the optimal $L-2$ 
$N_l$-layer VQE
result as the initial guess. 
The number of optimization steps following this guess grows more slowly with $L$ and $N_l$ than for fully random guesses. 
Figs.~\ref{fig:trajectory}d--f show that the parameters in the $N_l=10$ ansatz for the even-$L$ FHM converge to  smooth function of $m$ at large $L$, analogous to the convergence found for our bootstrap technique with large $N_l$. This shows the utility of the transfer technique.

\subsection{Iteration number scaling}

\begin{figure*}
    \centering
    \includegraphics[width=\linewidth]{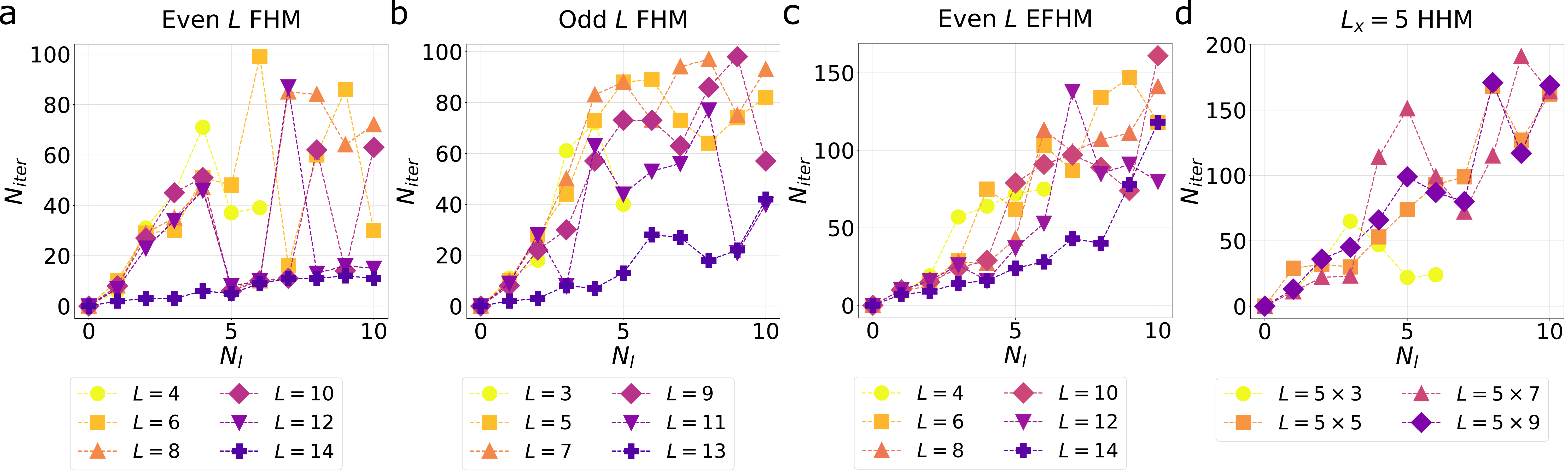}
    \caption{The number of optimization algorithm iterations as a function of ansatz layers $N_l$ for (a) even $L$ FHM, (b) odd $L$ FHM, (c) even $L$ EFHM, and (d) $L_x=5$ square lattice bosonic HHM. We observe a moderate linear-like growth of the number of iterations given a larger number of parameters.}
    \label{fig:Niter}
\end{figure*}

Finally,~\fig{Niter} shows the growth of the iteration numbers needed to reach convergence for each ansatz depth $N_l$ and size $L$ for all models in \fig{vqe}. The total number of iterations $N_\text{iter}$ does not grow significantly with $L$, perhaps even decreasing, and increases slowly (very roughly at most linearly) with $N_l$.
Notably, $N_\text{iter}$ for the largest $L$ runs using only the transfer technique is \textit{less} than the iterations used to optimize at smaller $L$.
As the total quantum algorithm run-time has to be multiplied by a factor $N_\text{iter}$ as $N_\text{iter} T$, the slow growth of $N_\text{iter}$ with both $L$ and $N_l$ suggests that the exponential quantum speedup we claim in the main text based on the scaling of $T$ can survive even after accounting for the many shots needed for optimization.

\section{Total quantum evolution time}
\label{app:qtime}

A key consideration in this work is the numerical scaling of the VQE algorithm error as a function of the total quantum evolution time $T$, the sum of the time parameters over all ansatz layers:
\begin{equation}\label{eq:T}
    T=\sum_m \m{T}.
\end{equation}
However, the time parameters $\m{T}$ have a physical unit of time, which makes their absolute values depend on the physical setup; in contrast, the optimization parameters in the VQE algorithm are dimensionless, e.g. $t_{x1}^{(m)} T^{(m)}$. To relate the total quantum evolution time $T$ to the dimensionless variational parameters, it is helpful to convert $T$ to a dimensionless parameter independent of the actual experimental time scale. This section addresses this for the FHM, EFHM, and HHM.

As mentioned in \sect{framework}, for the FHM and EFHM, we take $\m{U}=\Ut$ to be fixed at each layer for simplicity. (This is also the frequency unit, since we set $\hbar=1$). It is straightforward to make $T$ dimensionless by multiplying $\Ut$ on  both sides of \eq{T}:
\begin{equation}\label{eq:qtime}
    T\Ut=\sum_m \m{T}\Ut.
\end{equation}
This means the dimensionless total time $T\Ut$ is also the sum of dimensionless variational parameters $\m{T}\Ut$ used in the numerical optimization.

For the hardcore HHM,  $\m{U}\to\infty$ is no longer a useful reference energy. Thus we need to find another fixed frequency scale $\ftm$ to make $T$ dimensionless. As an aside, we note this is necessary only for characterizing the total optimization time, not for performing the  numerical optimization, which can work directly with the dimensionless variational parameters $\theta_{x1}=\m{t_{x1}}\m{T},\,\theta_{x2}=\m{t_{x2}}\m{T},\,\theta_{y}=\m{t_y}\m{T}$, $\theta_{Q}=\m{\mu_Q}\m{T}$ and $\theta_{S}=\m{\mu_S}\m{T}$.
To determine $T=\sum_m T^{(m)}$ from the dimensionless variational parameters, we define
\begin{equation}\label{eq:Tm}
    \m{T}=\max\left\{\frac{\theta_{x1}}{\maxttm},\,
    \frac{\theta_{x2}}{\maxttm},\,
    \frac{\theta_y}{\maxttm},\,
    \frac{|\theta_Q|}{Q\maxttm},\,
    \frac{|\theta_S|}{Q\maxttm}\right\}
\end{equation}
where ${\tilde t}_{\text{max}}$ is the maximum tunneling rate that can be applied experimentally, and $Q {\tilde t}_{\text{max}}$ with $Q>1$ is the maximum on-site potential that can be applied experimentally. This choice of $T^{(m)}$ characterizes the total time relative to $T\maxttm$ if one maximizes the tunnelings and on-site potential consistent with these constraints, thus giving the shortest time evolution consistent with the numerically chosen dimensionless optimization parameters. In this work, we choose $Q=3/2$.

Summing over all ansatz layers, we relate the dimensionless total evolution time $T$ to the dimensionless variational parameters by
\begin{equation}
    T\maxttm=\sum_m\max\left\{\theta_{x1},\,\theta_{x2},\,\theta_y,\,\frac{|\theta_Q|}{Q},\,\frac{|\theta_S|}{Q} \right\}.
\end{equation}

\section{Universal gate set}
\label{app:universality}

\def\gr{G^\mathrm{tun}}
\def\gi{G^\mathrm{int}}
\def\im{\mathcal{I}_M}
\def\dm{\Delta\mu}
\def\dn{\Delta n}
\def\nn#1{n_{#1\uparrow}n_{#1\downarrow}}
\def\doublon{\uparrow\downarrow}

In this section, we demonstrate that the native  Hamiltonians in our framework can implement a universal set of fermionic operations.
Universal fermionic quantum gate sets have been constructed using local fermionic (but number non-conserving) operations \cite{BK2002} and also from particle-number-conserving operations \cite{oszmaniec_universal_2017,huUniversalDynamicsGlobally2025}.
Specifically, from the theory of simple Lie algebra, Ref.~\cite{oszmaniec_universal_2017} proves that a fermionic quantum gate set generates the full $\mathfrak{su}(\mathcal{H})$ algebra in any $N_p$-particle subspace $\mathcal{H}$ of the $M$ fermion modes Fock space if it includes two ingredients: (1) arbitrary evolutions generated by two-mode  tunneling Hamiltonians: 
\begin{align}\label{eq:Grot}
    \gr_{ab}(\bm{\theta}) &=& \!\!\! \exp\Big\{\!-i \blank \Big[ \frac{\theta_1}{2}\left(e^{-i\theta_2} c_a^\dagger c_b + \mathrm{h.c.}\right) \nonumber\\
    &&\blank\hspace{0.0in} {}+ \frac{\theta_3}{2}\left(n_a-n_b\right)\Big]\Big\} \hspace{0.15in} \forall a, b\in \im,
\end{align}
and (2) any two-mode non-quadratic ``interaction'' gate 
\begin{equation}\label{eq:Gint}
    \gi_{cd} (\theta) = \exp\left\{-i\theta n_c n_d\right\}
\end{equation}
for some $c$ and $d \in \im$,
where $\im=\{(1,\uparrow),(1,\downarrow),\dots,(N_s,\downarrow)\}$ is the set of fermion mode indices.
Here and in the rest of the section, unless explicitly specified, we use $a,b,\dots$ to refer to lumped  site and spin indices $(i,\sigma)$ (with the spin index taking two values for the FHM and EFHM, and a single value for the HHM). 

Therefore, to show our fVQE framework's universality in any particle-number-conserving sector, we show 
that the Hamiltonian \eq{drive} augmented with local spin-dependent terms,
\begin{equation}\label{eq:FH+SSR}
    \m{H}_\mathrm{SR} = \m{H} + \sum_i \m{\bm{B}}_i\cdot \bm{S}_i,
\end{equation}
can be used to engineer arbitrary $\gr_{ab}$ and one $\gi_{cd}$. 
We note that the spin-dependent terms involving fermionic spin-$1/2$ operators $\bm{S}_i=(S^x_i,S^y_i,S^z_i)$ with tunable magnetic field parameters $\bm{B}_i=(B^x_i,B^y_i,B^z_i)$ are not used for the protocols in the main text, a point we discuss further at the end of this subsection. 
The single-spin rotations can be engineered as follows: local $S^z$ rotations can be implemented by a single-site resolved laser with a vector light shift~\cite{weitenbergSinglespinAddressingAtomic2011,trubkoPotassiumTuneoutwavelengthMeasurement2017} that couples the two (pseudo-)spin levels $\ket{\uparrow}$, $\ket{\downarrow}$, and local $S^x$, $S^y$ rotations can be realized by Raman coupling between the two spin levels with a focused laser. 
For simplicity in the rest of this section, we suppress the circuit layer index $(m)$, and we use $t_1$ and $t_2$ to denote the strong (intra-cell) and weak (inter-cell) tunnelings in the optical superlattice. We also use $\mu_{i,\sigma}$ to denote a mode-specific potential $\mu_{i,\sigma}=\mu_i+\sigma\frac{B^z_i}{2}$, using $\sigma = +1$ and $-1$ and $\uparrow$ and $\downarrow$ interchangeably. 

\subsection{Synthesizing the universal gate set}
\label{app:local-gate}

Now, we  illustrate how to synthesize the universal gate set in the particle number conserving sector. We show that given local control of  $\mu_{i,\sigma}$ and spin rotations, only global control of the interaction and tunneling, as in Eq.~\eqref{eq:FH+SSR}, suffice for universality.

The uniform Hubbard interaction  $U n_{i,\uparrow} n_{i,\downarrow}$ gives the required interaction gate $\gi_{cd}$~\cite{oszmaniec_universal_2017}. It then remains to show we can generate arbitrary two-mode tunnelings Eq.~\eqref{eq:Grot}.

We now construct a universal set of tunneling gates. It will be helpful to use, as intermediate constructions, two-mode ``rotations'' $R^{x,y,z}_{ab}(\theta)=e^{i\theta W_{ab}/2}$ where $W\in\{X,Y,Z\}$,  as  
\begin{align}\label{eq:modeRotation}
    X_{ab}\equiv\blank c^\dagger_{a}c^{\phantom\dagger}_{b} + c^\dagger_{b}c^{\phantom\dagger}_{a} \nonumber\\
    Y_{ab}\equiv\blank -ic^\dagger_{a}c^{\phantom\dagger}_{b} + ic^\dagger_{b}c^{\phantom\dagger}_{a} \\
    Z_{ab}\equiv\blank n_{a} - n_{b}. \nonumber
\end{align}
These provide rotations between the  two modes $a$ and $b$,
\begin{equation}
    \ket{0}_{ab}\equiv c^\dagger_a\ket{0},\,\quad \ket{1}_{ab}\equiv c^\dagger_{b}\ket{0}.
\end{equation} 
Now we can see how the native operations allow one to generate the arbitrary two-mode rotation gates $\gr_{ab}$. We begin by considering nearest-neighbor operations:
\begin{enumerate}
    \item On-site $\gr_{(i,\uparrow),(i,\downarrow)}$ between two spin flavors is synthesized by local single-spin $S^x_i$, $S^y_i$, and $S^z_i$ operators.
    \item The single-mode $\gr_{aa}(\theta,0,0)=e^{-i\theta n_a}$ is realized by the time evolution of the mode-specific potential term.
    \item The nearest-neighbor $X$-rotation is synthesized as shown immediately below, through \eq{localTunneling}.
    \item The $Z$-rotation between neighboring-site modes $(i,\sigma)\leftrightarrow(i+1,\sigma)$ is realized by the local potential gradient $\mu (n_{i,\sigma} - n_{i+1,\sigma})$.
   This requires evolution in the absence of a $U$ term, which may not be straightforward to rapidly control in the experiment. However,   the effect of non-zero $U$ terms can be canceled by a subsequent phase-reverse gate. A similar construction is  shown explicitly for the $X$-rotation gate below, and the  can be straightforwardly used to eliminate $U$ when generating the $Z$-rotation.
    \item The neighboring-site $Y$-rotation can be synthesized with $X$- and $Z$-rotations via $R^y(\theta)=R^z(\frac{\pi}{2})R^x(\theta)R^z(-\frac{\pi}{2})$; combining all $X$, $Y$ and $Z$ rotations we can realize the arbitrary neighboring-site $\gr_{(i,\sigma),(i+1,\sigma)}$.
\end{enumerate}

Once one has the nearest-neighbor operations, one can extend them to arbirary mode pairs as follows. We move  mode $b$ to the nearest neighbor of $a$ (call this $c$) and then use the operations above:
    \begin{align}\label{eq:swap-to-nn}
        \gr_{ab}=R^{x}_{bc}(\pi/2)\, \gr_{ac} \,R^{x}_{cb}(-\pi/2)
    \end{align}
    where $\gr_{ac}$ now acts on nearest neighbors, and $\uparrow$ and $\downarrow$ are considered ``nearest-neighbors" along an artificial extra dimension.
    This equation follows from $R^{x}_{bc}(\pi/2)\, c^\dagger_c \,R^{x}_{cb}(-\pi/2)=c^\dagger_b$, which can be calculated straightforwardly. The long-distance swap $R^{x}_{bc}(\pi/2)$ can be built from nearest-neighbors iteratively,  $R^{x}_{bc}(\pi/2)=R^{x}_{b b_1}(\pi/2)R^{x}_{b_1 b_2}(\pi/2)\cdots R^{x}_{b_\ell c}(\pi/2)$ for a path of nearest neighbor sites $(b,b_1,b_2,\cdots,b_\ell,c)$ connecting $b$ to $c$.

Now we turn to the only missing piece in the above construction, the neighboring-site tunneling gate $R^x_{ab}$. We show how to synthesize this gate, without loss of generality, on bond $(1,\uparrow)\leftrightarrow(2,\uparrow)$
using the drive Hamiltonian \eq{FH+SSR}. The idea is to suppress the tunneling on all bonds not involved in the tunneling by applying a strong staggered on-site potential $\mu_{i,\sigma}$ across them; on the target bond, we keep the coherent tunneling by tuning each pair of tunnel-coupled sites to resonance. We now present this idea in detail.


We start from \eq{FH+SSR}:
\begin{equation}\label{eq:initTunneling}
    U^\mathrm{tun}\equiv\exp\left [-iTH_\mathrm{SR}(t_1=\theta/T,t_2=0,\mu_{i,\sigma})\right ]
\end{equation}
with a vanishing inter-cell tunneling $t_2=0$ to isolate each double-well from the others. 

After this tunneling operation, we switch to applying a ``phase-reverse'' gate with a tunneling-free Hamiltonian: 
\begin{align}\label{eq:phaseReverse}
    U^\mathrm{PR}
    \equiv\blank \exp\left [-iT'H_\mathrm{SR}(t_1=t_2=0,\mu'_{i,\sigma})\right ]\nonumber\\
    =\blank\exp\left [-iT'\left(U \sum_{i} n_{i,\uparrow} n_{i,\downarrow} - \sum_{i,\sigma} \mu'_{i,\sigma} n_{i,\sigma}\right)\right],
\end{align}
where $T'=2n\pi/U-T$ with the integer $n$ such that $T'>0$, and $\mu'_{i,\sigma}=-\mu_{i,\sigma}T/T'$. 

The combination of \eq{initTunneling} and \eq{phaseReverse}, with the appropriate parameters $\mu_{i,\sigma}$, suppresses all unwanted tunneling except the target one with up spin on the $(1,2)$ pair. We first analyze how the strong staggered potential suppresses tunneling, and then analyze how the target tunneling is energetically allowed. These will show that $U^{\text{PR}} U^{\text{tun}}$ repeated with three sets of offsets on the target sites realizes the desired $R^x_{(1\uparrow)(2\uparrow)}(\theta)$ gate. 

The derivation is based on standard time-dependent perturbation theory. 
If $\ket{i}$ is a Fock basis state, $H_\mathrm{SR}$ can be separated into its diagonal term $H_d$ and off-diagonal tunneling term $H_t$
\begin{equation}
    \bra{i}H_d\ket{j} = d_i\delta_{ij},\quad \bra{i}H_t\ket{j} = V_{ij}(1-\delta_{ij}).
\end{equation}
In the interaction picture:
\begin{equation}\label{eq:Hi}
    \bra{i}(H_t)_I(\tau)\ket{j} = \bra{i}e^{iH_d\tau}H_te^{-iH_d\tau}\ket{j}
    =V_{ij} e^{i\Delta_{ij}\tau},
\end{equation}
where $\Delta_{ij}=d_i-d_j$.
Importantly, with our carefully chosen $\mu_{i,\sigma}'$ and $T'$,
\begin{equation}
    U^\mathrm{PR}=e^{-iT'H'_\mathrm{SR}}=e^{iTH_d},
\end{equation}
effectively transforms the time evolution $U^\mathrm{tun}$ into the interaction picture:
\begin{equation}\label{eq:intTimeEvo}
    U^\mathrm{PR}U^\mathrm{tun}
    =\left(U^\mathrm{tun}\right)_I
    =\exp\left[-i\int^T_0d\tau(H_t)_I(\tau)\right].
\end{equation}
For strong staggered potential,  $|\Delta_{ij}|\gg V_{ij}$, the fast-oscillating $(H_t)_I$ averages out at the time scale $T\sim 1/V_{ij}$. A detailed expansion, such as the Dyson series, finds the error of dropping $(H_t)_I$ is  $O(V_{ij}/\Delta_{ij})$.

We can analyze each double well individually, since the Hamiltonians that generate the time evolution commute between different double wells. We begin by considering a double-well $(a,b)$, and the general tunneling term in the interaction picture is  
\begin{equation}
    (H_t)_I(\tau) = -t_1\sum_{\sigma} e^{i(U\Delta n_{ab,{\bar\sigma}}-\Delta\mu_{ab,\sigma})\tau}c^\dagger_{a,\sigma}c^{\phantom\dagger}_{b,\sigma} + \text{h.c.},
\end{equation}
where $\bar \sigma$ is the  complement of $\sigma$, i.e. $\bar\sigma=\uparrow$ if $\sigma=\downarrow$ and $\bar \sigma=\downarrow$ if $\sigma=\uparrow$; $\dn_{ab,\sigma}=n_{a,\sigma}-n_{b,\sigma}$; and $\dm_{ab,\sigma}=\mu_{a,\sigma} - \mu_{b,\sigma}$.


If $(a,b)$ is the unwanted cell with $a,b>2$, effective tunneling terms are suppressed by a fast-oscillating phase for all $\dn_{ab,\sigma'}=0,\pm 1$ cases with both $\sigma=\uparrow$ and $\downarrow$. This is achieved  by setting $|\dm_{ab}|\gg t_1$, $|U+\dm_{ab}|\gg t_1$ and $|U-\dm_{ab}|\gg t_1$. 

If $(a,b)=(1,2)$ is the target cell, we want to suppress all tunneling terms but the $c^\dagger_{1,\uparrow}c^{\phantom\dagger}_{2,\uparrow}+\rm h.c.$ one. To suppress spin-down tunneling, we keep $\dm_{12,\downarrow}\gg t_1$, as in the unwanted cells. $\dm_{12,\downarrow}$ can take a wide range of values as long as $|U\dn_{12,\uparrow}-\dm_{12,\downarrow}|\gg t_1$; for concreteness, we set $\dm_{12,\downarrow}=U/2$. 

Finally, we consider the desired resonant tunneling terms. The value  of $\dn_{12,\uparrow}$ to achieve resonant tunneling depends on the densities via
\begin{equation}
    \dm_{12,\uparrow}=U\dn_{12,\downarrow},
\end{equation}
where $\dm_{12,\downarrow}$ can take values $0$ and $\pm1$. 
To complete the full target tunneling gate, we propose a three-step scheme, each step of which executes the pure tunneling process for one value of the $\dm_{12,\downarrow}$ while leaving the other sectors unaffected:
\begin{align}\label{eq:localTunneling}
    R^{x}_{(1,\uparrow)(2,\uparrow)} 
    =\blank U^\mathrm{PR}U^\mathrm{tun}|_{\dm_{12,\uparrow}=-U} \nonumber\\
    \blank\hspace{-0.14in} \times U^\mathrm{PR}U^\mathrm{tun}|_{\dm_{12,\uparrow}=U} \nonumber\\
    \blank\hspace{-0.14in} \times U^\mathrm{PR}U^\mathrm{tun}|_{\dm_{12,\uparrow}=0}.
\end{align}
Each step can be analyzed by  \eq{intTimeEvo}. 
This provides the last operation necessary to realize the arbitrary two-mode tunneling gate $\gr_{ab}$.

\subsection{Discussions}

We note that the derivation in this section differs from the previous literature, as we never assume we have local controllability over $U$, nor do we need to tune to exactly $U=0$. While our framework demonstrates universality with the assistance of local on-site potential control, an alternative is provided by the recent finding on global controllability~\cite{huUniversalDynamicsGlobally2025} which eliminates the need for such local control but  increases the cost to $O(N^2_s)$ global controls, where $N_s$ is the number of sites.

The construction shown here requires an additional ingredient compared to  the drive Hamiltonian  in the main text, the local spin operators. These are necessary  to obtain universality. 
While this control is experimentally realizable as described earlier, we exclude it from our $SU(2)$-symmetric fVQE ansatzes, because the dynamics and target Hamiltonians in the examples demonstrated in the main text are all $SU(2)$-symmetric, and empirically the calculated fVQE energy errors converge to zero within machine precision without such terms.   The rigorous proof of the universality of $SU(2)$-symmetric gates into any $SU(2)$-symmetric state space is beyond the scope of this paper (it will be given in future work).

\section{Correlation measurement}
\label{app:measurement}

In this section, we discuss the experimentally accessible range of observables in our framework, which is important for two reasons. First, this tells us which properties of the ground state wavefunction we will have access to. Second, as discussed in \sect{framework}, to simulate a Hamiltonian $H_T$ by the fVQE method one must obtain $\braket{H_T}$, which can be done by measuring subsets of Hamiltonian terms and adding them together as necessary. In other words, the Hamiltonians that can be simulated are those that are sums of terms that can be experimentally measured. We now briefly discuss several important types of correlations -- both for their interest in characterizing the ground states in the main text and as potential terms in the target Hamiltonian -- and how they can be measured using the current experiment techniques.

\subsection{Densities \& two-point correlations}

Site- and spin-resolved densities $\braket{n_{i,\sigma}}$ and their correlations $\braket{n_{i,\sigma}n_{j,\tau}}$ can be directly measured by quantum gas microscopy, as can general density-related moments $\braket{n_{i,\sigma}n_{j,\tau}\dots}$. 
This spin-dependent density measurement is the fundamental ingredient for all measurements we consider. 

\def\I{(i,\sigma)(j,\sigma)}
\def\K{(k,\tau)(l,\tau)}

A crucial observable beyond local densities $\braket{n_{i,\sigma}}=\braket{c^\dagger_{i,\sigma}c_{i,\sigma}^{\phantom \dagger}}$ is off-diagonal coherence
$\braket{c^\dagger_{i,\sigma}c^{\phantom \dagger}_{j,\sigma}}$,
such as the tunneling terms in the kinetic energy, possibly including  a gauge field. These
can be measured by isolating the two wells $i$ and $j$, performing the double-well rotation $R^a_{\I}(\pi/2)$ -- one set of measurements with $a=x$ and another with $a=y$ -- introduced in the prior subsection (this takes just a few time evolutions under the system Hamiltonian with appropriate double-well offsets), and then measuring the populations of each site. We are primarily interested in nearest-neighbor $i$ and $j$, but longer-range coherences also can be measured by bringing $i$ and $j$  to neighbors using the swap techniques of \app{local-gate}, or re-arrangement if one is using optical tweezers.


Specifically, using the operator notation $X_{\I}$, $Y_{\I}$, and $Z_{\I}$ defined in \eq{modeRotation}, we rewrite
\begin{equation}\label{eq:Jij-decompose}
    c^\dagger_{i,\sigma} c^{{\phantom \dagger}}_{j,\sigma}
    =\frac{X_{\I} + i Y_{\I}}{2}. 
\end{equation}
Then
\begin{equation}\label{eq:Yij}
    Y_{\I} =- R^x_{\I}(\pi/2)Z_{\I}R^x_{\I}(-\pi/2),
\end{equation}and
\begin{align}\label{eq:Xij}
X_{\I} = R^y_{\I}(\pi/2)Z_{\I}R^y_{\I}(-\pi/2).
\end{align}
So the average populations after rotation, $(1/2)(1+\braket{Z_{\I}})$, allow one to determine $Z_{\I}$ and therefore  the expectation values needed to obtain the coherence $\braket{c_{i,\sigma}^{\dagger} c_{j,\sigma}^{\phantom \dagger}}$ from \eq{Jij-decompose}.



In addition to the spin-invariant measurement introduced above, spin-flip two-point correlations can be measured via spin swaps. 

\subsection{Pairing \& four-point correlations}

In addition to the two-point measurements,  off-diagonal two-body four-point correlations, such as the singlet pairing correlations
\begin{align}\label{eq:pair-coherence}
    \braket{\Delta^\dagger_{S,i}\Delta_{S,i+r}^{\phantom \dagger}}
    =\blank\frac{1}{2}\Big[\braket{c^\dagger_{i,\uparrow}c^{\phantom\dagger}_{i+r,\uparrow}c^\dagger_{i+1,\downarrow}c^{\phantom\dagger}_{i+r+1,\downarrow}} \nonumber\\
    \blank\hspace{0.3in}+ \braket{c^\dagger_{i,\uparrow}c^{\phantom\dagger}_{i+r+1,\uparrow}c^\dagger_{i+1,\downarrow}c^{\phantom\dagger}_{i+r,\downarrow}} \nonumber\\
    \blank\hspace{0.3in}+ \braket{c^\dagger_{i+1,\uparrow}c^{\phantom\dagger}_{i+r,\uparrow}c^\dagger_{i,\downarrow}c^{\phantom\dagger}_{i+r+1,\downarrow}} \nonumber\\
    \blank\hspace{0.3in}+ \braket{c^\dagger_{i+1,\uparrow}c^{\phantom\dagger}_{i+r+1,\uparrow}c^\dagger_{i,\downarrow}c^{\phantom\dagger}_{i+r,\downarrow}}\Big],
\end{align}
are also measurable in experiments via a combination of experimental techniques, each of which has been demonstrated. 

To measure $\braket{c^\dagger_{i,\sigma}c^{\phantom\dagger}_{j,\sigma}c^\dagger_{k,\tau}c^{\phantom\dagger}_{l,\tau}}$ we can use a method similar to the last  subsection, simultaneously rotating on the pair of double wells $(i,j)$ and $(k,l)$. Using the identity \eq{Jij-decompose},
\begin{widetext}
\begin{align}\label{eq:cccc}\braket{c^\dagger_{i,\sigma}c^{\phantom\dagger}_{j,\sigma}c^\dagger_{k,\tau}c^{\phantom\dagger}_{l,\tau}} =\blank \frac{1}{4} \big[\braket{X_{\I}X_{\K}} - \braket{Y_{\I}Y_{\K}} \nonumber\\
    \blank + i\left(\braket{X_{\I}Y_{\K}} + \braket{Y_{\I}X_{\K}}\right)\big].
\end{align}
From Eqs.~\eqref{eq:Yij} and~\eqref{eq:Xij}, all terms in \eq{cccc} can be rotated into diagonal correlators. For example, the first term is
\begin{align}\label{eq:XX-corre}
    \braket{X_{\I}X_{\K}}
    =\blank \braket{R^y_{\I}(\pi/2)R^y_{\K}(\pi/2) Z_{\I} Z_{\K} R^y_{\I}(-\pi/2)R^y_{\K}(-\pi/2)}. 
\end{align}
\end{widetext}
The two two-site rotations $R^y_{\I}(\pi/2)R^y_{\K}(\pi/2)$ can be performed simultaneously. With site-resolved, spin-selective readout via the quantum gas microscope, we can measure \eq{XX-corre} by summing four density-density correlations.  

The general four-point correlations $\braket{c^\dagger_{i,\sigma}c^\dagger_{j,\tau}c_{k,\eta}^{\phantom \dagger}c_{l,\xi}^{\phantom \dagger}}$ can also be measured using the above technique, by first swapping sites using techniques introduced above or additional tweezers. Alternatively, previous studies have also used information scrambling techniques to capture the general four-point correlations~\cite{naldesi_fermionic_2023,tran_measuring_2023}.

\subsection{Girvin-MacDonald correlation}

Another type of correlation -- the GMD correlation $C^{\text{GMD}}_{ij}$, which can be written as sums of  $\braket{a^\dagger_i n_k n_l \dots a^{\phantom\dagger}_j}$ by expanding the exponential in Eq.~\eqref{eq:GMD-def} and using $n_k^2=n_k$ -- can also be measured using the double-well rotation on the pair of sites $i$ and $j$ and the site-resolved density readout over the lattice. After re-arranging and isolating the double wells $(i,j)$ with methods similar to previous cases, the $\pi/2$ $X$ and $Y$ rotations transfer the correlation into sums of populations:
\begin{align}
    \braket{a^\dagger_i n_k \dots a^{\phantom\dagger}_j} 
    =\blank \frac{1}{2}\braket{X_{ij} n_k \dots} + \frac{i}{2}\braket{Y_{ij} n_k \dots} \nonumber\\
    =\blank \frac{1}{2}\braket{R^y_{ij}(-\pi/2) Z_{ij} R^y_{ij}(\pi/2) n_k \dots} \nonumber\\
    \blank - \frac{i}{2}\braket{R^x_{ij}(-\pi/2) Z_{ij} R^x_{ij}(\pi/2) n_k \dots},
\end{align}
where $X_{ij}$, $Y_{ij}$ and $Z_{ij}$ are defined analogously to  the fermion bilinears in \eq{modeRotation}.
The site-resolved density readout in a quantum gas microscope  then provides the expectation values $\braket{(n_i-n_j)n_{k}n_{l}\dots}$. 

In addition to GMD correlations, the above technique allows us to measure  density- or spin-conditioned tunneling $c^\dagger_in^{\phantom\dagger}_kc^{\phantom\dagger}_j$, extending the range of Hamiltonians that can be simulated.

\section{Experiment run-time estimate}
\label{app:run-time}

This section discusses the number of shots and the total time required for an experiment on quantum hardware to run iterations to reach a $1\%$ VQE algorithm error in our framework. For these  proof-of-principle computations, we use the same simple optimization methods described in this text, and we also use only the most naive techniques to estimate the variational energy and its gradient.   We expect the optimization cost can be dramatically decreased from these estimates.

We start by considering the experimental time required to measure the energy gradient required by the optimizer, done via naive finite-differences: 
\begin{equation}
    \frac{\partial\braket{H_T}}{\partial\theta_i} \approx \frac{\braket{H_T(\theta_i + \Delta\theta_i)} - \braket{H_T(\theta_i)}}{\Delta\theta_i} + O(\Delta \theta_i)
\end{equation}
for small $\Delta\theta_i$.
To reach an error in the gradient $\delta$, one needs $\sim 1/\delta^2$ measurements. 
We consider a gradient error target $\Delta\theta=10\%$, requiring $\sim 10^2$ measurements, which leads to an energy error after optimization of $\sim 1\%$. We note that these measurements need not be done serially -- in a large system, different regions contribute effectively independent measurements when they are sufficiently separated.  A 2D $L_m=100$  optical lattice allows one to effectively measure of $L^2_m/L^2_c$ independent systems of size $L_c$ per experiment cycle, where $L_c$ is a scale large enough that there is little correlation in the measurements.

To obtain the gradient for each optimization iteration, an $N_l$-layer, $N_\text{param}$-parameter-per-layer ansatz requires measurements of $N_\text{param}N_l+1$ energy evaluations. For the ansatzes used in this work, this is about $5N_l$ evaluations.
 
The next step is to consider the number of gradients that must be evaluated for the optimizer. We will consider $N_l=10$, which typically gives less than $1\%$ variational error. We start by optimizing the $N_l=10$ parameters for  classically tractable systems, say with $L=4$ in 2D.



Next, we use the transfer technique by increasing the system size $L$ by $2$. At a target $1\%$ error level, this would usually need $10$ iterations for each $L$, as suggested by   \fig{Niter}. (Note that while \fig{Niter} used around $10 N_l$ iterations, this converge the error four orders of magnitude more accurately than required here.) For the finite-size error to be $1\%$, we need to reach roughly $L=20$. For each $L$, the transfer process needs $5N_lN_{\rm iter}\approx 500$ energy values. With  $10^2$ measurements needed for each energy value as discussed above,  the number of experimental shots for each energy value would be
\begin{equation}
    \sum^{20}_{L=4\text{ step by 2}} \frac{10^2}{L^2_m/L^2} \approx 15.
\end{equation}

Altogether,  the total fVQE calculation for variational error, optimization error, and  finite-size error each about 1$\%$ requires about $15 \times  500 \sim  8 \times 10^3$ shots. Such shot numbers are routinely employed in papers with quantum gas microscopes \cite{bourgundFormationIndividualStripes2025a,hirtheMagneticallyMediatedHole2023}. 
For current fermion experiments, the cycle time is about $\sim 10\text{s}$  \cite{yangSiteResolvedImagingUltracold2021}, so the full calculation takes roughly $22$ hours. We note that the slowest part is the evaporative atom cooling, which might be optimized further or -- using tweezer arrays -- improved by an order of magnitude or more. A calculation of this accuracy would surpass that achievable for many Hubbard-like problems on classical hardware \cite{wuVariationalBenchmarksQuantum2024}.

The optimization approach used for these estimates leaves substantial room for improvement, possibly by orders of magnitude. More efficient transfer and bootstrapping, more hardware parallelism, faster platforms, and better optimization methods can further reduce the run-time. For example, as alternatives to gradient-based algorithms, we note that gradient-free optimization approaches, such as simplex methods and surrogate model methods, are fully compatible with our framework. The recently proposed fermion-gate-based ansatzes \cite{Preiss2024,QNP2021} estimate the gradient via the parameter shift method. This reduces the need for high-accuracy quantum measurements, but does not eliminate the essential $O(N_\text{param}N_l)$ gradient measurement overhead. Meanwhile, the variational quantum imaginary-time evolution (VarQITE) method \cite{tabaresProgrammingOpticallatticeFermiHubbard2025} circumvents finite-difference gradient measurements by measuring the quantum geometric tensor $\braket{H^{(m)}H^{(m')}}$ in each iteration, incurring a severe $O(N^2_\text{param}N^2_l)$ measurement overhead. In general, benchmarking and developing efficient and experiment-friendly optimization methods is an important subject for future study. Nevertheless, the estimates here show that straightforward applications of existing methods plausibly can achieve interesting, difficult to classically simulate, results on existing quantum platforms.

Finally, similar to the discussion of the optimization overhead in \app{optim}, we briefly discuss how measurement repetitions enter the scaling analysis. As discussed in this Appendix, independent measurements used to estimate an observable with statistical error $\epsilon_M$ require $O(1/\epsilon_M^2)$ repetitions. Thus the accumulated evolution time for the readout scales as
\begin{equation}
    T_{\rm tot}\propto\frac{T}{\epsilon_M^2}.
\end{equation}
However, like the optimization overhead discussed in \app{optim}, we find that they do not affect the major conclusion of our scaling analysis, as taking $\epsilon_M\sim\epsilon$ only changes the polynomial exponent in $1/\epsilon$, and therefore $T_{\rm tot}$ remains $O(\mathrm{poly}(1/\epsilon))$.

\section{Exact ground state of Hofstadter-Hubbard model}
\label{app:fqh}

\begin{figure}
    \centering
    \includegraphics[width=0.9\linewidth]{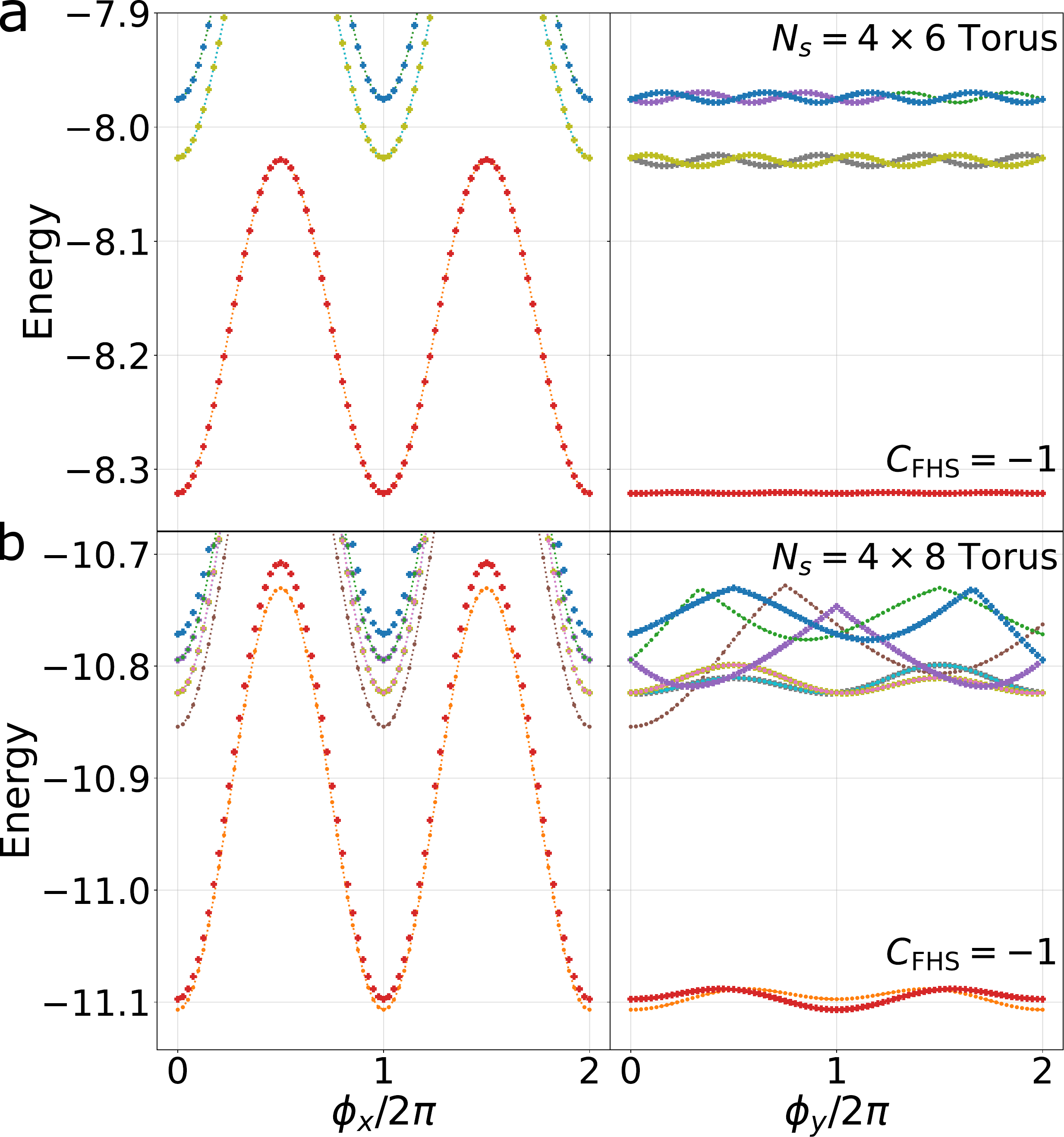}
    \caption{Lowest $10$ energy levels of the HHM, with an adiabatically inserted external flux threading the $x$- (left) and $y$-cycle (right) of (a) $L_x\times L_y=4\times6$ and (b) $L_x\times L_y=4\times8$ tori. The gapped ground-state sector has an $m=2$-fold quasi-degeneracy with a bundle Chern number $C_\text{FHS}=-1$, and more explicitly for $Lx\times L_y=4\times 8$ both degenerate states evolve $2m\pi$-periodically.}
    \label{fig:flux}
\end{figure}

This section shows that the ground state of the HHM  is a fractional Quantum Hall (FQH) state, with evidence from ED. While extensive studies \cite{leonard_realization_2023,FQH2005,rosson_bosonic_2019,pauw_detecting_2024,repellin_fractional_2020,wangMeasurableSignaturesBosonic2022a} on the ground state of the fractionally filled 2D HHM have confirmed that its ground state is an FQH state, reproducing the FQH physics for  model parameters and system sizes similar to our fVQE results is useful. We consider  two aspects: the periodic energy spectrum of the topologically quasi-degenerate ground states when an external flux is inserted adiabatically into $4\times6$ and $4\times8$ tori, and the existence of edge currents and the vanishing bulk density-density correlations on the open-boundary plane $L_x\times L_y=5\times7$ grid.

\begin{figure}[tb]
    \centering
    \includegraphics[width=\linewidth]{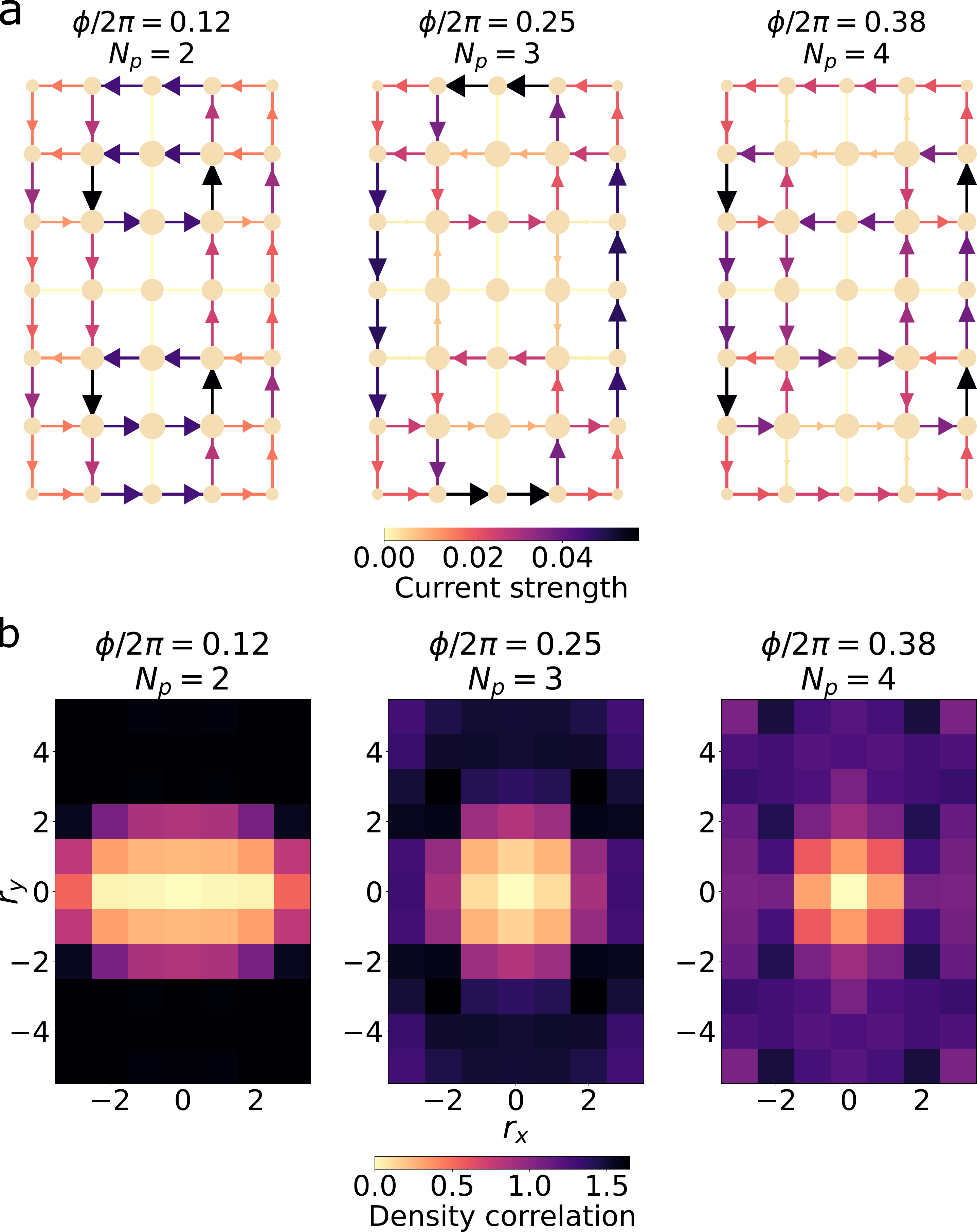}
    \caption{(a) Currents on each bond and (b) the bulk density-density correlations at $\phi/2\pi=1/8$, $1/4$ and $3/8$ in the ground state of an HHM on a $5\times7$ lattice with the open boundary condition and particle numbers appropriate to an $m=2$ FQH state (see text). The plots follow the same conventions as \fig{current}a\textasciitilde f.}
    \label{fig:EDlocalFQHObservables}
\end{figure}

We first show that the ground state of the $L_x=5$ tori with $\phi=2\pi/4$ at an appropriate filling is the fractionalized $m=2$ FQH state. For an FQH system on a torus, the topologically ordered ground-state manifold is expected to exhibit an $m$-fold quasi-degeneracy and a nontrivial (integer) Chern number when the entire manifold is treated as a bundle over the two-dimensional space of twisted boundary conditions $(\theta_x,\theta_y)$. 
Using the discretized, gauge-invariant Chern-number algorithm by Fukui, Hatsugai and Suzuki~\cite{fukuiChernNumbersDiscretized2005}, we find in both \fig{flux}a and \fig{flux}b that the $m$-fold quasi-degenerate ground-state band carries a total bundle Chern number $C_{\rm FHS}=-1$, providing direct evidence for fractionalized topological degeneracy. 
Moreover, due to the topological nature of the ground state, inserting a flux $2\pi$ can transfer the system from one ground state to another. The ED computation confirms this behavior for threading a flux with $\theta_y$ at a system size comparable to that of our fVQE results in the main text. In \fig{flux}b, we observe the expected behavior: a $2m\pi$-periodic pattern in the $\theta_y$ dependence of the $m$-fold quasi-degenerate energy spectrum, while ground states are interchanged after a $2\pi$ flux insertion, further providing signatures of an FQH state with filling factor $m$. A similar flux insertion along $x$ reveals a more conventional behavior, with a ground state returning to itself after $2\pi$ flux insertion, presumably due to the shorter length of the system in this direction.

In addition to the torus results, the ED ground state on a $5\times7$ lattice with open boundary conditions also exhibits signatures of the FQH state.  We calculate currents and density-density correlations for lattices with various fluxes per plaquette $\phi/(2\pi)=1/8$, $1/4$, and $3/8$, with particle number $N_p=2$, $3$, and $4$ appropriate for the $m=2$ FQH state by the formula $N_p=\phi N_\Box/(2\pi m)$ where $N_\Box=(L_x-1)\times(L_y-1)$ is the number of plaquettes. \fig{EDlocalFQHObservables} shows that the bulk cyclotron currents are suppressed as $N_p$ is increased, while chiral edge currents develop.  
The bulk density-density correlations (as defined in Eq.~\eqref{eq:FQH-dd-corre}) near the lattice center $\bm{r}=\bm{0}$ vanish at $\phi/(2\pi)=1/4$ filling, indicating the anti-correlation typical of a Laughlin state.
These results further show that the ground state of the HHM at a flux per plaquette $\phi/(2\pi)=1/4$ with appropriate particle filling indeed exhibits features characteristic of the FQH state. 


\bibliography{refs.bib}

\begin{thebibliography}{126}%
\makeatletter
\providecommand \@ifxundefined [1]{%
 \@ifx{#1\undefined}
}%
\providecommand \@ifnum [1]{%
 \ifnum #1\expandafter \@firstoftwo
 \else \expandafter \@secondoftwo
 \fi
}%
\providecommand \@ifx [1]{%
 \ifx #1\expandafter \@firstoftwo
 \else \expandafter \@secondoftwo
 \fi
}%
\providecommand \natexlab [1]{#1}%
\providecommand \enquote  [1]{``#1''}%
\providecommand \bibnamefont  [1]{#1}%
\providecommand \bibfnamefont [1]{#1}%
\providecommand \citenamefont [1]{#1}%
\providecommand \href@noop [0]{\@secondoftwo}%
\providecommand \href [0]{\begingroup \@sanitize@url \@href}%
\providecommand \@href[1]{\@@startlink{#1}\@@href}%
\providecommand \@@href[1]{\endgroup#1\@@endlink}%
\providecommand \@sanitize@url [0]{\catcode `\\12\catcode `\$12\catcode `\&12\catcode `\#12\catcode `\^12\catcode `\_12\catcode `\%12\relax}%
\providecommand \@@startlink[1]{}%
\providecommand \@@endlink[0]{}%
\providecommand \url  [0]{\begingroup\@sanitize@url \@url }%
\providecommand \@url [1]{\endgroup\@href {#1}{\urlprefix }}%
\providecommand \urlprefix  [0]{URL }%
\providecommand \Eprint [0]{\href }%
\providecommand \doibase [0]{https://doi.org/}%
\providecommand \selectlanguage [0]{\@gobble}%
\providecommand \bibinfo  [0]{\@secondoftwo}%
\providecommand \bibfield  [0]{\@secondoftwo}%
\providecommand \translation [1]{[#1]}%
\providecommand \BibitemOpen [0]{}%
\providecommand \bibitemStop [0]{}%
\providecommand \bibitemNoStop [0]{.\EOS\space}%
\providecommand \EOS [0]{\spacefactor3000\relax}%
\providecommand \BibitemShut  [1]{\csname bibitem#1\endcsname}%
\let\auto@bib@innerbib\@empty
\bibitem [{\citenamefont {Georgescu}\ \emph {et~al.}(2014)\citenamefont {Georgescu}, \citenamefont {Ashhab},\ and\ \citenamefont {Nori}}]{QuantSim}%
  \BibitemOpen
  \bibfield  {author} {\bibinfo {author} {\bibfnamefont {I.~M.}\ \bibnamefont {Georgescu}}, \bibinfo {author} {\bibfnamefont {S.}~\bibnamefont {Ashhab}},\ and\ \bibinfo {author} {\bibfnamefont {F.}~\bibnamefont {Nori}},\ }\bibfield  {title} {\bibinfo {title} {Quantum simulation},\ }\href {https://doi.org/10.1103/RevModPhys.86.153} {\bibfield  {journal} {\bibinfo  {journal} {Review of Modern Physics}\ }\textbf {\bibinfo {volume} {86}},\ \bibinfo {pages} {153} (\bibinfo {year} {2014})}\BibitemShut {NoStop}%
\bibitem [{\citenamefont {Feynman}(1982)}]{Feynman1982}%
  \BibitemOpen
  \bibfield  {author} {\bibinfo {author} {\bibfnamefont {R.~P.}\ \bibnamefont {Feynman}},\ }\bibfield  {title} {\bibinfo {title} {Simulating physics with computers},\ }\href {https://doi.org/10.1007/BF02650179} {\bibfield  {journal} {\bibinfo  {journal} {International Journal of Theoretical Physics}\ }\textbf {\bibinfo {volume} {21}},\ \bibinfo {pages} {467} (\bibinfo {year} {1982})}\BibitemShut {NoStop}%
\bibitem [{\citenamefont {Esslinger}(2010)}]{Esslingerreview}%
  \BibitemOpen
  \bibfield  {author} {\bibinfo {author} {\bibfnamefont {T.}~\bibnamefont {Esslinger}},\ }\bibfield  {title} {\bibinfo {title} {Fermi-{H}ubbard physics with atoms in an optical lattice},\ }\href {https://doi.org/10.1146/annurev-conmatphys-070909-104059} {\bibfield  {journal} {\bibinfo  {journal} {Annual Review of Condensed Matter Physics}\ }\textbf {\bibinfo {volume} {1}},\ \bibinfo {pages} {129} (\bibinfo {year} {2010})}\BibitemShut {NoStop}%
\bibitem [{\citenamefont {Tarruell}\ and\ \citenamefont {Sanchez-Palencia}(2018)}]{Tarruell_review}%
  \BibitemOpen
  \bibfield  {author} {\bibinfo {author} {\bibfnamefont {L.}~\bibnamefont {Tarruell}}\ and\ \bibinfo {author} {\bibfnamefont {L.}~\bibnamefont {Sanchez-Palencia}},\ }\bibfield  {title} {\bibinfo {title} {Quantum simulation of the {H}ubbard model with ultracold fermions in optical lattices},\ }\href {https://doi.org/https://doi.org/10.1016/j.crhy.2018.10.013} {\bibfield  {journal} {\bibinfo  {journal} {Comptes Rendus Physique}\ }\textbf {\bibinfo {volume} {19}},\ \bibinfo {pages} {365} (\bibinfo {year} {2018})}\BibitemShut {NoStop}%
\bibitem [{\citenamefont {Bloch}\ \emph {et~al.}(2012)\citenamefont {Bloch}, \citenamefont {Dalibard},\ and\ \citenamefont {Nascimb{\`e}ne}}]{Bloch2012}%
  \BibitemOpen
  \bibfield  {author} {\bibinfo {author} {\bibfnamefont {I.}~\bibnamefont {Bloch}}, \bibinfo {author} {\bibfnamefont {J.}~\bibnamefont {Dalibard}},\ and\ \bibinfo {author} {\bibfnamefont {S.}~\bibnamefont {Nascimb{\`e}ne}},\ }\bibfield  {title} {\bibinfo {title} {Quantum simulations with ultracold quantum gases},\ }\href {https://doi.org/10.1038/nphys2259} {\bibfield  {journal} {\bibinfo  {journal} {Nature Physics}\ }\textbf {\bibinfo {volume} {8}},\ \bibinfo {pages} {267} (\bibinfo {year} {2012})}\BibitemShut {NoStop}%
\bibitem [{\citenamefont {Gross}\ and\ \citenamefont {Bloch}(2017)}]{Gross2017}%
  \BibitemOpen
  \bibfield  {author} {\bibinfo {author} {\bibfnamefont {C.}~\bibnamefont {Gross}}\ and\ \bibinfo {author} {\bibfnamefont {I.}~\bibnamefont {Bloch}},\ }\bibfield  {title} {\bibinfo {title} {Quantum simulations with ultracold atoms in optical lattices},\ }\href {https://doi.org/10.1126/science.aal3837} {\bibfield  {journal} {\bibinfo  {journal} {Science}\ }\textbf {\bibinfo {volume} {357}},\ \bibinfo {pages} {995} (\bibinfo {year} {2017})}\BibitemShut {NoStop}%
\bibitem [{\citenamefont {Altman}\ \emph {et~al.}(2021)\citenamefont {Altman}, \citenamefont {Brown}, \citenamefont {Carleo}, \citenamefont {Carr}, \citenamefont {Demler}, \citenamefont {Chin}, \citenamefont {DeMarco}, \citenamefont {Economou}, \citenamefont {Eriksson}, \citenamefont {Fu}, \citenamefont {Greiner}, \citenamefont {Hazzard}, \citenamefont {Hulet}, \citenamefont {Kollár}, \citenamefont {Lev}, \citenamefont {Lukin}, \citenamefont {Ma}, \citenamefont {Mi}, \citenamefont {Misra}, \citenamefont {Monroe}, \citenamefont {Murch}, \citenamefont {Nazario}, \citenamefont {Ni}, \citenamefont {Potter}, \citenamefont {Roushan}, \citenamefont {Saffman}, \citenamefont {Schleier-Smith}, \citenamefont {Siddiqi}, \citenamefont {Simmonds}, \citenamefont {Singh}, \citenamefont {Spielman}, \citenamefont {Temme}, \citenamefont {Weiss}, \citenamefont {Vučković}, \citenamefont {Vuletić}, \citenamefont {Ye},\ and\ \citenamefont {Zwierlein}}]{altman_quantum_2021}%
  \BibitemOpen
  \bibfield  {author} {\bibinfo {author} {\bibfnamefont {E.}~\bibnamefont {Altman}}, \bibinfo {author} {\bibfnamefont {K.~R.}\ \bibnamefont {Brown}}, \bibinfo {author} {\bibfnamefont {G.}~\bibnamefont {Carleo}}, \bibinfo {author} {\bibfnamefont {L.~D.}\ \bibnamefont {Carr}}, \bibinfo {author} {\bibfnamefont {E.}~\bibnamefont {Demler}}, \bibinfo {author} {\bibfnamefont {C.}~\bibnamefont {Chin}}, \bibinfo {author} {\bibfnamefont {B.}~\bibnamefont {DeMarco}}, \bibinfo {author} {\bibfnamefont {S.~E.}\ \bibnamefont {Economou}}, \bibinfo {author} {\bibfnamefont {M.~A.}\ \bibnamefont {Eriksson}}, \bibinfo {author} {\bibfnamefont {K.-M.~C.}\ \bibnamefont {Fu}}, \bibinfo {author} {\bibfnamefont {M.}~\bibnamefont {Greiner}}, \bibinfo {author} {\bibfnamefont {K.~R.}\ \bibnamefont {Hazzard}}, \bibinfo {author} {\bibfnamefont {R.~G.}\ \bibnamefont {Hulet}}, \bibinfo {author} {\bibfnamefont {A.~J.}\ \bibnamefont {Kollár}}, \bibinfo {author} {\bibfnamefont {B.~L.}\ \bibnamefont {Lev}}, \bibinfo {author} {\bibfnamefont
  {M.~D.}\ \bibnamefont {Lukin}}, \bibinfo {author} {\bibfnamefont {R.}~\bibnamefont {Ma}}, \bibinfo {author} {\bibfnamefont {X.}~\bibnamefont {Mi}}, \bibinfo {author} {\bibfnamefont {S.}~\bibnamefont {Misra}}, \bibinfo {author} {\bibfnamefont {C.}~\bibnamefont {Monroe}}, \bibinfo {author} {\bibfnamefont {K.}~\bibnamefont {Murch}}, \bibinfo {author} {\bibfnamefont {Z.}~\bibnamefont {Nazario}}, \bibinfo {author} {\bibfnamefont {K.-K.}\ \bibnamefont {Ni}}, \bibinfo {author} {\bibfnamefont {A.~C.}\ \bibnamefont {Potter}}, \bibinfo {author} {\bibfnamefont {P.}~\bibnamefont {Roushan}}, \bibinfo {author} {\bibfnamefont {M.}~\bibnamefont {Saffman}}, \bibinfo {author} {\bibfnamefont {M.}~\bibnamefont {Schleier-Smith}}, \bibinfo {author} {\bibfnamefont {I.}~\bibnamefont {Siddiqi}}, \bibinfo {author} {\bibfnamefont {R.}~\bibnamefont {Simmonds}}, \bibinfo {author} {\bibfnamefont {M.}~\bibnamefont {Singh}}, \bibinfo {author} {\bibfnamefont {I.}~\bibnamefont {Spielman}}, \bibinfo {author} {\bibfnamefont {K.}~\bibnamefont
  {Temme}}, \bibinfo {author} {\bibfnamefont {D.~S.}\ \bibnamefont {Weiss}}, \bibinfo {author} {\bibfnamefont {J.}~\bibnamefont {Vučković}}, \bibinfo {author} {\bibfnamefont {V.}~\bibnamefont {Vuletić}}, \bibinfo {author} {\bibfnamefont {J.}~\bibnamefont {Ye}},\ and\ \bibinfo {author} {\bibfnamefont {M.}~\bibnamefont {Zwierlein}},\ }\bibfield  {title} {\bibinfo {title} {Quantum {Simulators}: {Architectures} and {Opportunities}},\ }\href {https://doi.org/10.1103/PRXQuantum.2.017003} {\bibfield  {journal} {\bibinfo  {journal} {PRX Quantum}\ }\textbf {\bibinfo {volume} {2}},\ \bibinfo {pages} {017003} (\bibinfo {year} {2021})}\BibitemShut {NoStop}%
\bibitem [{\citenamefont {K{\"o}hl}\ \emph {et~al.}(2005)\citenamefont {K{\"o}hl}, \citenamefont {Moritz}, \citenamefont {St{\"o}ferle}, \citenamefont {G{\"u}nter},\ and\ \citenamefont {Esslinger}}]{kohlFermionicAtomsThree2005}%
  \BibitemOpen
  \bibfield  {author} {\bibinfo {author} {\bibfnamefont {M.}~\bibnamefont {K{\"o}hl}}, \bibinfo {author} {\bibfnamefont {H.}~\bibnamefont {Moritz}}, \bibinfo {author} {\bibfnamefont {T.}~\bibnamefont {St{\"o}ferle}}, \bibinfo {author} {\bibfnamefont {K.}~\bibnamefont {G{\"u}nter}},\ and\ \bibinfo {author} {\bibfnamefont {T.}~\bibnamefont {Esslinger}},\ }\bibfield  {title} {\bibinfo {title} {Fermionic {{Atoms}} in a {{Three Dimensional Optical Lattice}}: {{Observing Fermi Surfaces}}, {{Dynamics}}, and {{Interactions}}},\ }\href {https://doi.org/10.1103/PhysRevLett.94.080403} {\bibfield  {journal} {\bibinfo  {journal} {Physical Review Letters}\ }\textbf {\bibinfo {volume} {94}},\ \bibinfo {pages} {080403} (\bibinfo {year} {2005})}\BibitemShut {NoStop}%
\bibitem [{\citenamefont {Hart}\ \emph {et~al.}(2015)\citenamefont {Hart}, \citenamefont {Duarte}, \citenamefont {Yang}, \citenamefont {Liu}, \citenamefont {Paiva}, \citenamefont {Khatami}, \citenamefont {Scalettar}, \citenamefont {Trivedi}, \citenamefont {Huse},\ and\ \citenamefont {Hulet}}]{hartObservationAntiferromagneticCorrelations2015a}%
  \BibitemOpen
  \bibfield  {author} {\bibinfo {author} {\bibfnamefont {R.~A.}\ \bibnamefont {Hart}}, \bibinfo {author} {\bibfnamefont {P.~M.}\ \bibnamefont {Duarte}}, \bibinfo {author} {\bibfnamefont {T.-L.}\ \bibnamefont {Yang}}, \bibinfo {author} {\bibfnamefont {X.}~\bibnamefont {Liu}}, \bibinfo {author} {\bibfnamefont {T.}~\bibnamefont {Paiva}}, \bibinfo {author} {\bibfnamefont {E.}~\bibnamefont {Khatami}}, \bibinfo {author} {\bibfnamefont {R.~T.}\ \bibnamefont {Scalettar}}, \bibinfo {author} {\bibfnamefont {N.}~\bibnamefont {Trivedi}}, \bibinfo {author} {\bibfnamefont {D.~A.}\ \bibnamefont {Huse}},\ and\ \bibinfo {author} {\bibfnamefont {R.~G.}\ \bibnamefont {Hulet}},\ }\bibfield  {title} {\bibinfo {title} {Observation of antiferromagnetic correlations in the {{Hubbard}} model with ultracold atoms},\ }\href {https://doi.org/10.1038/nature14223} {\bibfield  {journal} {\bibinfo  {journal} {Nature}\ }\textbf {\bibinfo {volume} {519}},\ \bibinfo {pages} {211} (\bibinfo {year} {2015})}\BibitemShut {NoStop}%
\bibitem [{\citenamefont {Shao}\ \emph {et~al.}(2024)\citenamefont {Shao}, \citenamefont {Wang}, \citenamefont {Zhu}, \citenamefont {Zhu}, \citenamefont {Sun}, \citenamefont {Chen}, \citenamefont {Zhang}, \citenamefont {Fan}, \citenamefont {Deng}, \citenamefont {Yao}, \citenamefont {Chen},\ and\ \citenamefont {Pan}}]{shaoAntiferromagneticPhaseTransition2024}%
  \BibitemOpen
  \bibfield  {author} {\bibinfo {author} {\bibfnamefont {H.-J.}\ \bibnamefont {Shao}}, \bibinfo {author} {\bibfnamefont {Y.-X.}\ \bibnamefont {Wang}}, \bibinfo {author} {\bibfnamefont {D.-Z.}\ \bibnamefont {Zhu}}, \bibinfo {author} {\bibfnamefont {Y.-S.}\ \bibnamefont {Zhu}}, \bibinfo {author} {\bibfnamefont {H.-N.}\ \bibnamefont {Sun}}, \bibinfo {author} {\bibfnamefont {S.-Y.}\ \bibnamefont {Chen}}, \bibinfo {author} {\bibfnamefont {C.}~\bibnamefont {Zhang}}, \bibinfo {author} {\bibfnamefont {Z.-J.}\ \bibnamefont {Fan}}, \bibinfo {author} {\bibfnamefont {Y.}~\bibnamefont {Deng}}, \bibinfo {author} {\bibfnamefont {X.-C.}\ \bibnamefont {Yao}}, \bibinfo {author} {\bibfnamefont {Y.-A.}\ \bibnamefont {Chen}},\ and\ \bibinfo {author} {\bibfnamefont {J.-W.}\ \bibnamefont {Pan}},\ }\bibfield  {title} {\bibinfo {title} {Antiferromagnetic phase transition in a {{3D}} fermionic {{Hubbard}} model},\ }\href {https://doi.org/10.1038/s41586-024-07689-2} {\bibfield  {journal} {\bibinfo  {journal} {Nature}\ }\textbf
  {\bibinfo {volume} {632}},\ \bibinfo {pages} {267} (\bibinfo {year} {2024})}\BibitemShut {NoStop}%
\bibitem [{\citenamefont {Taie}\ \emph {et~al.}(2022)\citenamefont {Taie}, \citenamefont {Ibarra-Garc{\'i}a-Padilla}, \citenamefont {Nishizawa}, \citenamefont {Takasu}, \citenamefont {Kuno}, \citenamefont {Wei}, \citenamefont {Scalettar}, \citenamefont {Hazzard},\ and\ \citenamefont {Takahashi}}]{Taie2022}%
  \BibitemOpen
  \bibfield  {author} {\bibinfo {author} {\bibfnamefont {S.}~\bibnamefont {Taie}}, \bibinfo {author} {\bibfnamefont {E.}~\bibnamefont {Ibarra-Garc{\'i}a-Padilla}}, \bibinfo {author} {\bibfnamefont {N.}~\bibnamefont {Nishizawa}}, \bibinfo {author} {\bibfnamefont {Y.}~\bibnamefont {Takasu}}, \bibinfo {author} {\bibfnamefont {Y.}~\bibnamefont {Kuno}}, \bibinfo {author} {\bibfnamefont {H.-T.}\ \bibnamefont {Wei}}, \bibinfo {author} {\bibfnamefont {R.~T.}\ \bibnamefont {Scalettar}}, \bibinfo {author} {\bibfnamefont {K.~R.~A.}\ \bibnamefont {Hazzard}},\ and\ \bibinfo {author} {\bibfnamefont {Y.}~\bibnamefont {Takahashi}},\ }\bibfield  {title} {\bibinfo {title} {{Observation of antiferromagnetic correlations in an ultracold SU(N) Hubbard model}},\ }\href {https://doi.org/10.1038/s41567-022-01725-6} {\bibfield  {journal} {\bibinfo  {journal} {Nature Physics}\ }\textbf {\bibinfo {volume} {18}},\ \bibinfo {pages} {1356} (\bibinfo {year} {2022})}\BibitemShut {NoStop}%
\bibitem [{\citenamefont {Xu}\ \emph {et~al.}(2025)\citenamefont {Xu}, \citenamefont {Kendrick}, \citenamefont {Kale}, \citenamefont {Gang}, \citenamefont {Feng}, \citenamefont {Zhang}, \citenamefont {Young}, \citenamefont {Lebrat},\ and\ \citenamefont {Greiner}}]{xuNeutralatomHubbardQuantum2025}%
  \BibitemOpen
  \bibfield  {author} {\bibinfo {author} {\bibfnamefont {M.}~\bibnamefont {Xu}}, \bibinfo {author} {\bibfnamefont {L.~H.}\ \bibnamefont {Kendrick}}, \bibinfo {author} {\bibfnamefont {A.}~\bibnamefont {Kale}}, \bibinfo {author} {\bibfnamefont {Y.}~\bibnamefont {Gang}}, \bibinfo {author} {\bibfnamefont {C.}~\bibnamefont {Feng}}, \bibinfo {author} {\bibfnamefont {S.}~\bibnamefont {Zhang}}, \bibinfo {author} {\bibfnamefont {A.~W.}\ \bibnamefont {Young}}, \bibinfo {author} {\bibfnamefont {M.}~\bibnamefont {Lebrat}},\ and\ \bibinfo {author} {\bibfnamefont {M.}~\bibnamefont {Greiner}},\ }\bibfield  {title} {\bibinfo {title} {A neutral-atom {{Hubbard}} quantum simulator in the cryogenic regime},\ }\href {https://doi.org/10.1038/s41586-025-09112-w} {\bibfield  {journal} {\bibinfo  {journal} {Nature}\ }\textbf {\bibinfo {volume} {642}},\ \bibinfo {pages} {909} (\bibinfo {year} {2025})}\BibitemShut {NoStop}%
\bibitem [{\citenamefont {Gross}\ and\ \citenamefont {Bakr}(2021)}]{GrossBakr_review}%
  \BibitemOpen
  \bibfield  {author} {\bibinfo {author} {\bibfnamefont {C.}~\bibnamefont {Gross}}\ and\ \bibinfo {author} {\bibfnamefont {W.~S.}\ \bibnamefont {Bakr}},\ }\bibfield  {title} {\bibinfo {title} {Quantum gas microscopy for single atom and spin detection},\ }\href {https://doi.org/https://doi.org/10.1038/s41567-021-01370-5} {\bibfield  {journal} {\bibinfo  {journal} {Nat. Phys.}\ }\textbf {\bibinfo {volume} {17}},\ \bibinfo {pages} {1316} (\bibinfo {year} {2021})}\BibitemShut {NoStop}%
\bibitem [{\citenamefont {Bohrdt}\ \emph {et~al.}(2018)\citenamefont {Bohrdt}, \citenamefont {Greif}, \citenamefont {Demler}, \citenamefont {Knap},\ and\ \citenamefont {Grusdt}}]{bohrdtAngleresolvedPhotoemissionSpectroscopy2018}%
  \BibitemOpen
  \bibfield  {author} {\bibinfo {author} {\bibfnamefont {A.}~\bibnamefont {Bohrdt}}, \bibinfo {author} {\bibfnamefont {D.}~\bibnamefont {Greif}}, \bibinfo {author} {\bibfnamefont {E.}~\bibnamefont {Demler}}, \bibinfo {author} {\bibfnamefont {M.}~\bibnamefont {Knap}},\ and\ \bibinfo {author} {\bibfnamefont {F.}~\bibnamefont {Grusdt}},\ }\bibfield  {title} {\bibinfo {title} {Angle-resolved photoemission spectroscopy with quantum gas microscopes},\ }\href {https://doi.org/10.1103/PhysRevB.97.125117} {\bibfield  {journal} {\bibinfo  {journal} {Physical Review B}\ }\textbf {\bibinfo {volume} {97}},\ \bibinfo {pages} {125117} (\bibinfo {year} {2018})}\BibitemShut {NoStop}%
\bibitem [{\citenamefont {Kaufman}\ \emph {et~al.}(2014)\citenamefont {Kaufman}, \citenamefont {Lester}, \citenamefont {Reynolds}, \citenamefont {Wall}, \citenamefont {Foss-Feig}, \citenamefont {Hazzard}, \citenamefont {Rey},\ and\ \citenamefont {Regal}}]{BosonDoubleWell}%
  \BibitemOpen
  \bibfield  {author} {\bibinfo {author} {\bibfnamefont {A.~M.}\ \bibnamefont {Kaufman}}, \bibinfo {author} {\bibfnamefont {B.~J.}\ \bibnamefont {Lester}}, \bibinfo {author} {\bibfnamefont {C.~M.}\ \bibnamefont {Reynolds}}, \bibinfo {author} {\bibfnamefont {M.~L.}\ \bibnamefont {Wall}}, \bibinfo {author} {\bibfnamefont {M.}~\bibnamefont {Foss-Feig}}, \bibinfo {author} {\bibfnamefont {K.~R.~A.}\ \bibnamefont {Hazzard}}, \bibinfo {author} {\bibfnamefont {A.~M.}\ \bibnamefont {Rey}},\ and\ \bibinfo {author} {\bibfnamefont {C.~A.}\ \bibnamefont {Regal}},\ }\bibfield  {title} {\bibinfo {title} {Two-particle quantum interference in tunnel-coupled optical tweezers},\ }\href {https://doi.org/10.1126/science.1250057} {\bibfield  {journal} {\bibinfo  {journal} {Science}\ }\textbf {\bibinfo {volume} {345}},\ \bibinfo {pages} {306} (\bibinfo {year} {2014})}\BibitemShut {NoStop}%
\bibitem [{\citenamefont {Murmann}\ \emph {et~al.}(2015)\citenamefont {Murmann}, \citenamefont {Bergschneider}, \citenamefont {Klinkhamer}, \citenamefont {Z\"urn}, \citenamefont {Lompe},\ and\ \citenamefont {Jochim}}]{FermionDoubleWell}%
  \BibitemOpen
  \bibfield  {author} {\bibinfo {author} {\bibfnamefont {S.}~\bibnamefont {Murmann}}, \bibinfo {author} {\bibfnamefont {A.}~\bibnamefont {Bergschneider}}, \bibinfo {author} {\bibfnamefont {V.~M.}\ \bibnamefont {Klinkhamer}}, \bibinfo {author} {\bibfnamefont {G.}~\bibnamefont {Z\"urn}}, \bibinfo {author} {\bibfnamefont {T.}~\bibnamefont {Lompe}},\ and\ \bibinfo {author} {\bibfnamefont {S.}~\bibnamefont {Jochim}},\ }\bibfield  {title} {\bibinfo {title} {Two fermions in a double well: Exploring a fundamental building block of the {Hubbard} model},\ }\href {https://doi.org/10.1103/PhysRevLett.114.080402} {\bibfield  {journal} {\bibinfo  {journal} {Physical Review Letters}\ }\textbf {\bibinfo {volume} {114}},\ \bibinfo {pages} {080402} (\bibinfo {year} {2015})}\BibitemShut {NoStop}%
\bibitem [{\citenamefont {Bergschneider}\ \emph {et~al.}(2019)\citenamefont {Bergschneider}, \citenamefont {Klinkhamer}, \citenamefont {Becher}, \citenamefont {Klemt}, \citenamefont {Palm}, \citenamefont {Z{\"u}rn}, \citenamefont {Jochim},\ and\ \citenamefont {Preiss}}]{Bergschneider2019}%
  \BibitemOpen
  \bibfield  {author} {\bibinfo {author} {\bibfnamefont {A.}~\bibnamefont {Bergschneider}}, \bibinfo {author} {\bibfnamefont {V.~M.}\ \bibnamefont {Klinkhamer}}, \bibinfo {author} {\bibfnamefont {J.~H.}\ \bibnamefont {Becher}}, \bibinfo {author} {\bibfnamefont {R.}~\bibnamefont {Klemt}}, \bibinfo {author} {\bibfnamefont {L.}~\bibnamefont {Palm}}, \bibinfo {author} {\bibfnamefont {G.}~\bibnamefont {Z{\"u}rn}}, \bibinfo {author} {\bibfnamefont {S.}~\bibnamefont {Jochim}},\ and\ \bibinfo {author} {\bibfnamefont {P.~M.}\ \bibnamefont {Preiss}},\ }\bibfield  {title} {\bibinfo {title} {Experimental characterization of two-particle entanglement through position and momentum correlations},\ }\href {https://doi.org/10.1038/s41567-019-0508-6} {\bibfield  {journal} {\bibinfo  {journal} {Nature Physics}\ }\textbf {\bibinfo {volume} {15}},\ \bibinfo {pages} {640} (\bibinfo {year} {2019})}\BibitemShut {NoStop}%
\bibitem [{\citenamefont {Becher}\ \emph {et~al.}(2020)\citenamefont {Becher}, \citenamefont {Sindici}, \citenamefont {Klemt}, \citenamefont {Jochim}, \citenamefont {Daley},\ and\ \citenamefont {Preiss}}]{Becher2020}%
  \BibitemOpen
  \bibfield  {author} {\bibinfo {author} {\bibfnamefont {J.~H.}\ \bibnamefont {Becher}}, \bibinfo {author} {\bibfnamefont {E.}~\bibnamefont {Sindici}}, \bibinfo {author} {\bibfnamefont {R.}~\bibnamefont {Klemt}}, \bibinfo {author} {\bibfnamefont {S.}~\bibnamefont {Jochim}}, \bibinfo {author} {\bibfnamefont {A.~J.}\ \bibnamefont {Daley}},\ and\ \bibinfo {author} {\bibfnamefont {P.~M.}\ \bibnamefont {Preiss}},\ }\bibfield  {title} {\bibinfo {title} {Measurement of identical particle entanglement and the influence of antisymmetrization},\ }\href {https://doi.org/10.1103/PhysRevLett.125.180402} {\bibfield  {journal} {\bibinfo  {journal} {Physical Review Letters}\ }\textbf {\bibinfo {volume} {125}},\ \bibinfo {pages} {180402} (\bibinfo {year} {2020})}\BibitemShut {NoStop}%
\bibitem [{\citenamefont {Spar}\ \emph {et~al.}(2022)\citenamefont {Spar}, \citenamefont {Guardado-Sanchez}, \citenamefont {Chi}, \citenamefont {Yan},\ and\ \citenamefont {Bakr}}]{Spar2022}%
  \BibitemOpen
  \bibfield  {author} {\bibinfo {author} {\bibfnamefont {B.~M.}\ \bibnamefont {Spar}}, \bibinfo {author} {\bibfnamefont {E.}~\bibnamefont {Guardado-Sanchez}}, \bibinfo {author} {\bibfnamefont {S.}~\bibnamefont {Chi}}, \bibinfo {author} {\bibfnamefont {Z.~Z.}\ \bibnamefont {Yan}},\ and\ \bibinfo {author} {\bibfnamefont {W.~S.}\ \bibnamefont {Bakr}},\ }\bibfield  {title} {\bibinfo {title} {Realization of a fermi-hubbard optical tweezer array},\ }\href {https://doi.org/10.1103/PhysRevLett.128.223202} {\bibfield  {journal} {\bibinfo  {journal} {Physical Review Letters}\ }\textbf {\bibinfo {volume} {128}},\ \bibinfo {pages} {223202} (\bibinfo {year} {2022})}\BibitemShut {NoStop}%
\bibitem [{\citenamefont {Florshaim}\ \emph {et~al.}(2024)\citenamefont {Florshaim}, \citenamefont {Zohar}, \citenamefont {Koplovich}, \citenamefont {Meltzer}, \citenamefont {Weill}, \citenamefont {Nemirovsky}, \citenamefont {Stern},\ and\ \citenamefont {Sagi}}]{Florshaim2023}%
  \BibitemOpen
  \bibfield  {author} {\bibinfo {author} {\bibfnamefont {Y.}~\bibnamefont {Florshaim}}, \bibinfo {author} {\bibfnamefont {E.}~\bibnamefont {Zohar}}, \bibinfo {author} {\bibfnamefont {D.~Z.}\ \bibnamefont {Koplovich}}, \bibinfo {author} {\bibfnamefont {I.}~\bibnamefont {Meltzer}}, \bibinfo {author} {\bibfnamefont {R.}~\bibnamefont {Weill}}, \bibinfo {author} {\bibfnamefont {J.}~\bibnamefont {Nemirovsky}}, \bibinfo {author} {\bibfnamefont {A.}~\bibnamefont {Stern}},\ and\ \bibinfo {author} {\bibfnamefont {Y.}~\bibnamefont {Sagi}},\ }\bibfield  {title} {\bibinfo {title} {Spatial adiabatic passage of ultracold atoms in optical tweezers},\ }\href {https://doi.org/10.1126/sciadv.adl1220} {\bibfield  {journal} {\bibinfo  {journal} {Science Advances}\ }\textbf {\bibinfo {volume} {10}},\ \bibinfo {pages} {eadl1220} (\bibinfo {year} {2024})}\BibitemShut {NoStop}%
\bibitem [{\citenamefont {Wei}\ \emph {et~al.}(2024)\citenamefont {Wei}, \citenamefont {Ibarra-García-Padilla}, \citenamefont {Wall},\ and\ \citenamefont {Hazzard}}]{wei_hubbard_2024}%
  \BibitemOpen
  \bibfield  {author} {\bibinfo {author} {\bibfnamefont {H.-T.}\ \bibnamefont {Wei}}, \bibinfo {author} {\bibfnamefont {E.}~\bibnamefont {Ibarra-García-Padilla}}, \bibinfo {author} {\bibfnamefont {M.~L.}\ \bibnamefont {Wall}},\ and\ \bibinfo {author} {\bibfnamefont {K.~R.~A.}\ \bibnamefont {Hazzard}},\ }\bibfield  {title} {\bibinfo {title} {Hubbard parameters for programmable tweezer arrays},\ }\href {https://doi.org/10.1103/PhysRevA.109.013318} {\bibfield  {journal} {\bibinfo  {journal} {Physical Review A}\ }\textbf {\bibinfo {volume} {109}},\ \bibinfo {pages} {013318} (\bibinfo {year} {2024})}\BibitemShut {NoStop}%
\bibitem [{\citenamefont {Wall}\ \emph {et~al.}(2015)\citenamefont {Wall}, \citenamefont {Hazzard},\ and\ \citenamefont {Rey}}]{wallEffectiveManybodyParameters2015}%
  \BibitemOpen
  \bibfield  {author} {\bibinfo {author} {\bibfnamefont {M.~L.}\ \bibnamefont {Wall}}, \bibinfo {author} {\bibfnamefont {K.~R.~A.}\ \bibnamefont {Hazzard}},\ and\ \bibinfo {author} {\bibfnamefont {A.~M.}\ \bibnamefont {Rey}},\ }\bibfield  {title} {\bibinfo {title} {Effective many-body parameters for atoms in nonseparable {Gaussian} optical potentials},\ }\href {https://doi.org/10.1103/PhysRevA.92.013610} {\bibfield  {journal} {\bibinfo  {journal} {Physical Review A}\ }\textbf {\bibinfo {volume} {92}},\ \bibinfo {pages} {013610} (\bibinfo {year} {2015})}\BibitemShut {NoStop}%
\bibitem [{\citenamefont {Tao}\ \emph {et~al.}(2024)\citenamefont {Tao}, \citenamefont {Ammenwerth}, \citenamefont {Gyger}, \citenamefont {Bloch},\ and\ \citenamefont {Zeiher}}]{taoHighFidelityDetectionLargeScale2024}%
  \BibitemOpen
  \bibfield  {author} {\bibinfo {author} {\bibfnamefont {R.}~\bibnamefont {Tao}}, \bibinfo {author} {\bibfnamefont {M.}~\bibnamefont {Ammenwerth}}, \bibinfo {author} {\bibfnamefont {F.}~\bibnamefont {Gyger}}, \bibinfo {author} {\bibfnamefont {I.}~\bibnamefont {Bloch}},\ and\ \bibinfo {author} {\bibfnamefont {J.}~\bibnamefont {Zeiher}},\ }\bibfield  {title} {\bibinfo {title} {High-{{Fidelity Detection}} of {{Large-Scale Atom Arrays}} in an {{Optical Lattice}}},\ }\href {https://doi.org/10.1103/PhysRevLett.133.013401} {\bibfield  {journal} {\bibinfo  {journal} {Physical Review Letters}\ }\textbf {\bibinfo {volume} {133}},\ \bibinfo {pages} {013401} (\bibinfo {year} {2024})}\BibitemShut {NoStop}%
\bibitem [{\citenamefont {Young}\ \emph {et~al.}(2024)\citenamefont {Young}, \citenamefont {Geller}, \citenamefont {Eckner}, \citenamefont {Schine}, \citenamefont {Glancy}, \citenamefont {Knill},\ and\ \citenamefont {Kaufman}}]{youngAtomicBosonSampler2024}%
  \BibitemOpen
  \bibfield  {author} {\bibinfo {author} {\bibfnamefont {A.~W.}\ \bibnamefont {Young}}, \bibinfo {author} {\bibfnamefont {S.}~\bibnamefont {Geller}}, \bibinfo {author} {\bibfnamefont {W.~J.}\ \bibnamefont {Eckner}}, \bibinfo {author} {\bibfnamefont {N.}~\bibnamefont {Schine}}, \bibinfo {author} {\bibfnamefont {S.}~\bibnamefont {Glancy}}, \bibinfo {author} {\bibfnamefont {E.}~\bibnamefont {Knill}},\ and\ \bibinfo {author} {\bibfnamefont {A.~M.}\ \bibnamefont {Kaufman}},\ }\bibfield  {title} {\bibinfo {title} {An atomic boson sampler},\ }\href {https://doi.org/10.1038/s41586-024-07304-4} {\bibfield  {journal} {\bibinfo  {journal} {Nature}\ }\textbf {\bibinfo {volume} {629}},\ \bibinfo {pages} {311} (\bibinfo {year} {2024})}\BibitemShut {NoStop}%
\bibitem [{\citenamefont {Young}\ \emph {et~al.}(2022)\citenamefont {Young}, \citenamefont {Eckner}, \citenamefont {Schine}, \citenamefont {Childs},\ and\ \citenamefont {Kaufman}}]{youngTweezerprogrammable2DQuantum2022}%
  \BibitemOpen
  \bibfield  {author} {\bibinfo {author} {\bibfnamefont {A.~W.}\ \bibnamefont {Young}}, \bibinfo {author} {\bibfnamefont {W.~J.}\ \bibnamefont {Eckner}}, \bibinfo {author} {\bibfnamefont {N.}~\bibnamefont {Schine}}, \bibinfo {author} {\bibfnamefont {A.~M.}\ \bibnamefont {Childs}},\ and\ \bibinfo {author} {\bibfnamefont {A.~M.}\ \bibnamefont {Kaufman}},\ }\bibfield  {title} {\bibinfo {title} {Tweezer-programmable {{2D}} quantum walks in a {{Hubbard-regime}} lattice},\ }\href {https://doi.org/10.1126/science.abo0608} {\bibfield  {journal} {\bibinfo  {journal} {Science}\ }\textbf {\bibinfo {volume} {377}},\ \bibinfo {pages} {885} (\bibinfo {year} {2022})}\BibitemShut {NoStop}%
\bibitem [{\citenamefont {Parker}\ \emph {et~al.}(2013)\citenamefont {Parker}, \citenamefont {Ha},\ and\ \citenamefont {Chin}}]{parkerDirectObservationEffective2013}%
  \BibitemOpen
  \bibfield  {author} {\bibinfo {author} {\bibfnamefont {C.~V.}\ \bibnamefont {Parker}}, \bibinfo {author} {\bibfnamefont {L.-C.}\ \bibnamefont {Ha}},\ and\ \bibinfo {author} {\bibfnamefont {C.}~\bibnamefont {Chin}},\ }\bibfield  {title} {\bibinfo {title} {Direct observation of effective ferromagnetic domains of cold atoms in a shaken optical lattice},\ }\href {https://doi.org/10.1038/nphys2789} {\bibfield  {journal} {\bibinfo  {journal} {Nature Physics}\ }\textbf {\bibinfo {volume} {9}},\ \bibinfo {pages} {769} (\bibinfo {year} {2013})}\BibitemShut {NoStop}%
\bibitem [{\citenamefont {McArdle}\ \emph {et~al.}(2020{\natexlab{a}})\citenamefont {McArdle}, \citenamefont {Endo}, \citenamefont {Aspuru-Guzik}, \citenamefont {Benjamin},\ and\ \citenamefont {Yuan}}]{QuantCompChem}%
  \BibitemOpen
  \bibfield  {author} {\bibinfo {author} {\bibfnamefont {S.}~\bibnamefont {McArdle}}, \bibinfo {author} {\bibfnamefont {S.}~\bibnamefont {Endo}}, \bibinfo {author} {\bibfnamefont {A.}~\bibnamefont {Aspuru-Guzik}}, \bibinfo {author} {\bibfnamefont {S.~C.}\ \bibnamefont {Benjamin}},\ and\ \bibinfo {author} {\bibfnamefont {X.}~\bibnamefont {Yuan}},\ }\bibfield  {title} {\bibinfo {title} {Quantum computational chemistry},\ }\href {https://doi.org/10.1103/RevModPhys.92.015003} {\bibfield  {journal} {\bibinfo  {journal} {Reviews of Modern Physics}\ }\textbf {\bibinfo {volume} {92}},\ \bibinfo {pages} {015003} (\bibinfo {year} {2020}{\natexlab{a}})}\BibitemShut {NoStop}%
\bibitem [{\citenamefont {Bohrdt}\ \emph {et~al.}(2021)\citenamefont {Bohrdt}, \citenamefont {Homeier}, \citenamefont {Reinmoser}, \citenamefont {Demler},\ and\ \citenamefont {Grusdt}}]{Bohrdt2021}%
  \BibitemOpen
  \bibfield  {author} {\bibinfo {author} {\bibfnamefont {A.}~\bibnamefont {Bohrdt}}, \bibinfo {author} {\bibfnamefont {L.}~\bibnamefont {Homeier}}, \bibinfo {author} {\bibfnamefont {C.}~\bibnamefont {Reinmoser}}, \bibinfo {author} {\bibfnamefont {E.}~\bibnamefont {Demler}},\ and\ \bibinfo {author} {\bibfnamefont {F.}~\bibnamefont {Grusdt}},\ }\bibfield  {title} {\bibinfo {title} {Exploration of doped quantum magnets with ultracold atoms},\ }\href {https://doi.org/https://doi.org/10.1016/j.aop.2021.168651} {\bibfield  {journal} {\bibinfo  {journal} {Annals of Physics}\ }\textbf {\bibinfo {volume} {435}},\ \bibinfo {pages} {168651} (\bibinfo {year} {2021})}\BibitemShut {NoStop}%
\bibitem [{\citenamefont {Imada}\ \emph {et~al.}(1998)\citenamefont {Imada}, \citenamefont {Fujimori},\ and\ \citenamefont {Tokura}}]{imada1998}%
  \BibitemOpen
  \bibfield  {author} {\bibinfo {author} {\bibfnamefont {M.}~\bibnamefont {Imada}}, \bibinfo {author} {\bibfnamefont {A.}~\bibnamefont {Fujimori}},\ and\ \bibinfo {author} {\bibfnamefont {Y.}~\bibnamefont {Tokura}},\ }\bibfield  {title} {\bibinfo {title} {Metal-insulator transitions},\ }\href {https://doi.org/10.1103/RevModPhys.70.1039} {\bibfield  {journal} {\bibinfo  {journal} {Review of Modern Physics}\ }\textbf {\bibinfo {volume} {70}},\ \bibinfo {pages} {1039} (\bibinfo {year} {1998})}\BibitemShut {NoStop}%
\bibitem [{\citenamefont {Montorsi}(1992)}]{montorsi1992}%
  \BibitemOpen
  \bibinfo {editor} {\bibfnamefont {A.}~\bibnamefont {Montorsi}},\ ed.,\ \href@noop {} {\emph {\bibinfo {title} {The Hubbard Model: A reprint volume}}}\ (\bibinfo  {publisher} {World Scientific},\ \bibinfo {year} {1992})\BibitemShut {NoStop}%
\bibitem [{\citenamefont {Tasaki}(1998)}]{Tasaki1998}%
  \BibitemOpen
  \bibfield  {author} {\bibinfo {author} {\bibfnamefont {H.}~\bibnamefont {Tasaki}},\ }\bibfield  {title} {\bibinfo {title} {The {H}ubbard model - an introduction and selected rigorous results},\ }\href {https://doi.org/10.1088/0953-8984/10/20/004} {\bibfield  {journal} {\bibinfo  {journal} {Journal of Physics: Condensed Matter}\ }\textbf {\bibinfo {volume} {10}},\ \bibinfo {pages} {4353} (\bibinfo {year} {1998})}\BibitemShut {NoStop}%
\bibitem [{\citenamefont {Arovas}\ \emph {et~al.}(2022)\citenamefont {Arovas}, \citenamefont {Berg}, \citenamefont {Kivelson},\ and\ \citenamefont {Raghu}}]{Arovas2022}%
  \BibitemOpen
  \bibfield  {author} {\bibinfo {author} {\bibfnamefont {D.~P.}\ \bibnamefont {Arovas}}, \bibinfo {author} {\bibfnamefont {E.}~\bibnamefont {Berg}}, \bibinfo {author} {\bibfnamefont {S.~A.}\ \bibnamefont {Kivelson}},\ and\ \bibinfo {author} {\bibfnamefont {S.}~\bibnamefont {Raghu}},\ }\bibfield  {title} {\bibinfo {title} {The {H}ubbard model},\ }\href {https://doi.org/10.1146/annurev-conmatphys-031620-102024} {\bibfield  {journal} {\bibinfo  {journal} {Annual Review of Condensed Matter Physics}\ }\textbf {\bibinfo {volume} {13}},\ \bibinfo {pages} {239} (\bibinfo {year} {2022})}\BibitemShut {NoStop}%
\bibitem [{\citenamefont {Qin}\ \emph {et~al.}(2022)\citenamefont {Qin}, \citenamefont {Sch\"{a}fer}, \citenamefont {Andergassen}, \citenamefont {Corboz},\ and\ \citenamefont {Gull}}]{Qin2022}%
  \BibitemOpen
  \bibfield  {author} {\bibinfo {author} {\bibfnamefont {M.}~\bibnamefont {Qin}}, \bibinfo {author} {\bibfnamefont {T.}~\bibnamefont {Sch\"{a}fer}}, \bibinfo {author} {\bibfnamefont {S.}~\bibnamefont {Andergassen}}, \bibinfo {author} {\bibfnamefont {P.}~\bibnamefont {Corboz}},\ and\ \bibinfo {author} {\bibfnamefont {E.}~\bibnamefont {Gull}},\ }\bibfield  {title} {\bibinfo {title} {The {H}ubbard model: A computational perspective},\ }\href {https://doi.org/10.1146/annurev-conmatphys-090921-033948} {\bibfield  {journal} {\bibinfo  {journal} {Annual Review of Condensed Matter Physics}\ }\textbf {\bibinfo {volume} {13}},\ \bibinfo {pages} {275} (\bibinfo {year} {2022})}\BibitemShut {NoStop}%
\bibitem [{Hub(2013)}]{Hubbard_model}%
  \BibitemOpen
  \bibfield  {title} {\bibinfo {title} {The {Hubbard} model at half a century},\ }\href {https://doi.org/10.1038/nphys2759} {\bibfield  {journal} {\bibinfo  {journal} {Nat. Phys.}\ }\textbf {\bibinfo {volume} {9}},\ \bibinfo {pages} {523} (\bibinfo {year} {2013})}\BibitemShut {NoStop}%
\bibitem [{\citenamefont {Li}\ \emph {et~al.}(2026)\citenamefont {Li}, \citenamefont {De}, \citenamefont {Sivakumar}, \citenamefont {Huie}, \citenamefont {Wei}, \citenamefont {Piel}, \citenamefont {Greene}, \citenamefont {Hazzard}, \citenamefont {Yan},\ and\ \citenamefont {Covey}}]{liQuantumScienceArrays2026}%
  \BibitemOpen
  \bibfield  {author} {\bibinfo {author} {\bibfnamefont {Z.}~\bibnamefont {Li}}, \bibinfo {author} {\bibfnamefont {R.}~\bibnamefont {De}}, \bibinfo {author} {\bibfnamefont {R.}~\bibnamefont {Sivakumar}}, \bibinfo {author} {\bibfnamefont {W.}~\bibnamefont {Huie}}, \bibinfo {author} {\bibfnamefont {H.-T.}\ \bibnamefont {Wei}}, \bibinfo {author} {\bibfnamefont {J.~D.}\ \bibnamefont {Piel}}, \bibinfo {author} {\bibfnamefont {C.~H.}\ \bibnamefont {Greene}}, \bibinfo {author} {\bibfnamefont {K.~R.~A.}\ \bibnamefont {Hazzard}}, \bibinfo {author} {\bibfnamefont {Z.~Z.}\ \bibnamefont {Yan}},\ and\ \bibinfo {author} {\bibfnamefont {J.~P.}\ \bibnamefont {Covey}},\ }\href@noop {} {\bibinfo {title} {Quantum science with arrays of metastable helium-3 atoms}} (\bibinfo {year} {2026}),\ \Eprint {https://arxiv.org/abs/2601.06763} {arXiv:2601.06763 [quant-ph]} \BibitemShut {NoStop}%
\bibitem [{\citenamefont {Defenu}\ \emph {et~al.}(2023)\citenamefont {Defenu}, \citenamefont {Donner}, \citenamefont {Macr\`{\i}}, \citenamefont {Pagano}, \citenamefont {Ruffo},\ and\ \citenamefont {Trombettoni}}]{DefenuLRI2021}%
  \BibitemOpen
  \bibfield  {author} {\bibinfo {author} {\bibfnamefont {N.}~\bibnamefont {Defenu}}, \bibinfo {author} {\bibfnamefont {T.}~\bibnamefont {Donner}}, \bibinfo {author} {\bibfnamefont {T.}~\bibnamefont {Macr\`{\i}}}, \bibinfo {author} {\bibfnamefont {G.}~\bibnamefont {Pagano}}, \bibinfo {author} {\bibfnamefont {S.}~\bibnamefont {Ruffo}},\ and\ \bibinfo {author} {\bibfnamefont {A.}~\bibnamefont {Trombettoni}},\ }\bibfield  {title} {\bibinfo {title} {Long-range interacting quantum systems},\ }\href {https://doi.org/10.1103/RevModPhys.95.035002} {\bibfield  {journal} {\bibinfo  {journal} {Review of Modern Physics}\ }\textbf {\bibinfo {volume} {95}},\ \bibinfo {pages} {035002} (\bibinfo {year} {2023})}\BibitemShut {NoStop}%
\bibitem [{\citenamefont {Johnson}\ and\ \citenamefont {Rolston}(2010)}]{JohnsonRydbergDress2010}%
  \BibitemOpen
  \bibfield  {author} {\bibinfo {author} {\bibfnamefont {J.~E.}\ \bibnamefont {Johnson}}\ and\ \bibinfo {author} {\bibfnamefont {S.~L.}\ \bibnamefont {Rolston}},\ }\bibfield  {title} {\bibinfo {title} {Interactions between {R}ydberg-dressed atoms},\ }\href {https://doi.org/10.1103/PhysRevA.82.033412} {\bibfield  {journal} {\bibinfo  {journal} {Phys. Rev. A}\ }\textbf {\bibinfo {volume} {82}},\ \bibinfo {pages} {033412} (\bibinfo {year} {2010})}\BibitemShut {NoStop}%
\bibitem [{\citenamefont {Li}\ and\ \citenamefont {Sarma}(2015)}]{RydbergHubbard2015}%
  \BibitemOpen
  \bibfield  {author} {\bibinfo {author} {\bibfnamefont {X.}~\bibnamefont {Li}}\ and\ \bibinfo {author} {\bibfnamefont {S.~D.}\ \bibnamefont {Sarma}},\ }\bibfield  {title} {\bibinfo {title} {Exotic topological density waves in cold atomic {R}ydberg-dressed fermions},\ }\href {https://doi.org/10.1038/ncomms8137} {\bibfield  {journal} {\bibinfo  {journal} {Nat. Comm.}\ }\textbf {\bibinfo {volume} {6}},\ \bibinfo {pages} {7137} (\bibinfo {year} {2015})}\BibitemShut {NoStop}%
\bibitem [{\citenamefont {Jaksch}\ and\ \citenamefont {Zoller}(2003)}]{jaksch_zoller_2003}%
  \BibitemOpen
  \bibfield  {author} {\bibinfo {author} {\bibfnamefont {D.}~\bibnamefont {Jaksch}}\ and\ \bibinfo {author} {\bibfnamefont {P.}~\bibnamefont {Zoller}},\ }\bibfield  {title} {\bibinfo {title} {Creation of effective magnetic fields in optical lattices: the {Hofstadter} butterfly for cold neutral atoms},\ }\href {https://doi.org/10.1088/1367-2630/5/1/356} {\bibfield  {journal} {\bibinfo  {journal} {New Journal of Physics}\ }\textbf {\bibinfo {volume} {5}},\ \bibinfo {pages} {56} (\bibinfo {year} {2003})}\BibitemShut {NoStop}%
\bibitem [{\citenamefont {Aidelsburger}\ \emph {et~al.}(2013)\citenamefont {Aidelsburger}, \citenamefont {Atala}, \citenamefont {Lohse}, \citenamefont {Barreiro}, \citenamefont {Paredes},\ and\ \citenamefont {Bloch}}]{aidelsburger_realization_2013}%
  \BibitemOpen
  \bibfield  {author} {\bibinfo {author} {\bibfnamefont {M.}~\bibnamefont {Aidelsburger}}, \bibinfo {author} {\bibfnamefont {M.}~\bibnamefont {Atala}}, \bibinfo {author} {\bibfnamefont {M.}~\bibnamefont {Lohse}}, \bibinfo {author} {\bibfnamefont {J.~T.}\ \bibnamefont {Barreiro}}, \bibinfo {author} {\bibfnamefont {B.}~\bibnamefont {Paredes}},\ and\ \bibinfo {author} {\bibfnamefont {I.}~\bibnamefont {Bloch}},\ }\bibfield  {title} {\bibinfo {title} {Realization of the {Hofstadter} {Hamiltonian} with {Ultracold} {Atoms} in {Optical} {Lattices}},\ }\href {https://doi.org/10.1103/PhysRevLett.111.185301} {\bibfield  {journal} {\bibinfo  {journal} {Physical Review Letters}\ }\textbf {\bibinfo {volume} {111}},\ \bibinfo {pages} {185301} (\bibinfo {year} {2013})}\BibitemShut {NoStop}%
\bibitem [{\citenamefont {Jotzu}\ \emph {et~al.}(2014)\citenamefont {Jotzu}, \citenamefont {Messer}, \citenamefont {Desbuquois}, \citenamefont {Lebrat}, \citenamefont {Uehlinger}, \citenamefont {Greif},\ and\ \citenamefont {Esslinger}}]{jotzuExperimentalRealizationTopological2014}%
  \BibitemOpen
  \bibfield  {author} {\bibinfo {author} {\bibfnamefont {G.}~\bibnamefont {Jotzu}}, \bibinfo {author} {\bibfnamefont {M.}~\bibnamefont {Messer}}, \bibinfo {author} {\bibfnamefont {R.}~\bibnamefont {Desbuquois}}, \bibinfo {author} {\bibfnamefont {M.}~\bibnamefont {Lebrat}}, \bibinfo {author} {\bibfnamefont {T.}~\bibnamefont {Uehlinger}}, \bibinfo {author} {\bibfnamefont {D.}~\bibnamefont {Greif}},\ and\ \bibinfo {author} {\bibfnamefont {T.}~\bibnamefont {Esslinger}},\ }\bibfield  {title} {\bibinfo {title} {Experimental realization of the topological {{Haldane}} model with ultracold fermions},\ }\href {https://doi.org/10.1038/nature13915} {\bibfield  {journal} {\bibinfo  {journal} {Nature}\ }\textbf {\bibinfo {volume} {515}},\ \bibinfo {pages} {237} (\bibinfo {year} {2014})}\BibitemShut {NoStop}%
\bibitem [{\citenamefont {Kennedy}\ \emph {et~al.}(2015)\citenamefont {Kennedy}, \citenamefont {Burton}, \citenamefont {Chung},\ and\ \citenamefont {Ketterle}}]{kennedyObservationBoseEinstein2015}%
  \BibitemOpen
  \bibfield  {author} {\bibinfo {author} {\bibfnamefont {C.~J.}\ \bibnamefont {Kennedy}}, \bibinfo {author} {\bibfnamefont {W.~C.}\ \bibnamefont {Burton}}, \bibinfo {author} {\bibfnamefont {W.~C.}\ \bibnamefont {Chung}},\ and\ \bibinfo {author} {\bibfnamefont {W.}~\bibnamefont {Ketterle}},\ }\bibfield  {title} {\bibinfo {title} {Observation of {{Bose}}--{{Einstein}} condensation in a strong synthetic magnetic field},\ }\href {https://doi.org/10.1038/nphys3421} {\bibfield  {journal} {\bibinfo  {journal} {Nature Physics}\ }\textbf {\bibinfo {volume} {11}},\ \bibinfo {pages} {859} (\bibinfo {year} {2015})}\BibitemShut {NoStop}%
\bibitem [{\citenamefont {Léonard}\ \emph {et~al.}(2023)\citenamefont {Léonard}, \citenamefont {Kim}, \citenamefont {Kwan}, \citenamefont {Segura}, \citenamefont {Grusdt}, \citenamefont {Repellin}, \citenamefont {Goldman},\ and\ \citenamefont {Greiner}}]{leonard_realization_2023}%
  \BibitemOpen
  \bibfield  {author} {\bibinfo {author} {\bibfnamefont {J.}~\bibnamefont {Léonard}}, \bibinfo {author} {\bibfnamefont {S.}~\bibnamefont {Kim}}, \bibinfo {author} {\bibfnamefont {J.}~\bibnamefont {Kwan}}, \bibinfo {author} {\bibfnamefont {P.}~\bibnamefont {Segura}}, \bibinfo {author} {\bibfnamefont {F.}~\bibnamefont {Grusdt}}, \bibinfo {author} {\bibfnamefont {C.}~\bibnamefont {Repellin}}, \bibinfo {author} {\bibfnamefont {N.}~\bibnamefont {Goldman}},\ and\ \bibinfo {author} {\bibfnamefont {M.}~\bibnamefont {Greiner}},\ }\bibfield  {title} {\bibinfo {title} {Realization of a fractional quantum {Hall} state with ultracold atoms},\ }\href {https://doi.org/10.1038/s41586-023-06122-4} {\bibfield  {journal} {\bibinfo  {journal} {Nature}\ }\textbf {\bibinfo {volume} {619}},\ \bibinfo {pages} {495} (\bibinfo {year} {2023})}\BibitemShut {NoStop}%
\bibitem [{\citenamefont {González-Cuadra}\ \emph {et~al.}(2023)\citenamefont {González-Cuadra}, \citenamefont {Bluvstein}, \citenamefont {Kalinowski}, \citenamefont {Kaubruegger}, \citenamefont {Maskara}, \citenamefont {Naldesi}, \citenamefont {Zache}, \citenamefont {Kaufman}, \citenamefont {Lukin}, \citenamefont {Pichler}, \citenamefont {Vermersch}, \citenamefont {Ye},\ and\ \citenamefont {Zoller}}]{ZollerLukin2023}%
  \BibitemOpen
  \bibfield  {author} {\bibinfo {author} {\bibfnamefont {D.}~\bibnamefont {González-Cuadra}}, \bibinfo {author} {\bibfnamefont {D.}~\bibnamefont {Bluvstein}}, \bibinfo {author} {\bibfnamefont {M.}~\bibnamefont {Kalinowski}}, \bibinfo {author} {\bibfnamefont {R.}~\bibnamefont {Kaubruegger}}, \bibinfo {author} {\bibfnamefont {N.}~\bibnamefont {Maskara}}, \bibinfo {author} {\bibfnamefont {P.}~\bibnamefont {Naldesi}}, \bibinfo {author} {\bibfnamefont {T.~V.}\ \bibnamefont {Zache}}, \bibinfo {author} {\bibfnamefont {A.~M.}\ \bibnamefont {Kaufman}}, \bibinfo {author} {\bibfnamefont {M.~D.}\ \bibnamefont {Lukin}}, \bibinfo {author} {\bibfnamefont {H.}~\bibnamefont {Pichler}}, \bibinfo {author} {\bibfnamefont {B.}~\bibnamefont {Vermersch}}, \bibinfo {author} {\bibfnamefont {J.}~\bibnamefont {Ye}},\ and\ \bibinfo {author} {\bibfnamefont {P.}~\bibnamefont {Zoller}},\ }\bibfield  {title} {\bibinfo {title} {Fermionic quantum processing with programmable neutral atom arrays},\ }\href
  {https://doi.org/10.1073/pnas.2304294120} {\bibfield  {journal} {\bibinfo  {journal} {Proceedings of the National Academy of Sciences}\ }\textbf {\bibinfo {volume} {120}},\ \bibinfo {pages} {e2304294120} (\bibinfo {year} {2023})}\BibitemShut {NoStop}%
\bibitem [{\citenamefont {Bravyi}\ and\ \citenamefont {Kitaev}(2002)}]{BK2002}%
  \BibitemOpen
  \bibfield  {author} {\bibinfo {author} {\bibfnamefont {S.~B.}\ \bibnamefont {Bravyi}}\ and\ \bibinfo {author} {\bibfnamefont {A.~Y.}\ \bibnamefont {Kitaev}},\ }\bibfield  {title} {\bibinfo {title} {Fermionic quantum computation},\ }\href {https://doi.org/https://doi.org/10.1006/aphy.2002.6254} {\bibfield  {journal} {\bibinfo  {journal} {Annals of Physics}\ }\textbf {\bibinfo {volume} {298}},\ \bibinfo {pages} {210} (\bibinfo {year} {2002})}\BibitemShut {NoStop}%
\bibitem [{\citenamefont {Bojovi{\'c}}\ \emph {et~al.}(2026)\citenamefont {Bojovi{\'c}}, \citenamefont {Hilker}, \citenamefont {Wang}, \citenamefont {Obermeyer}, \citenamefont {Barendregt}, \citenamefont {Tell}, \citenamefont {Chalopin}, \citenamefont {Preiss}, \citenamefont {Bloch},\ and\ \citenamefont {Franz}}]{bojović2025highfidelitycollisionalquantumgates}%
  \BibitemOpen
  \bibfield  {author} {\bibinfo {author} {\bibfnamefont {P.}~\bibnamefont {Bojovi{\'c}}}, \bibinfo {author} {\bibfnamefont {T.}~\bibnamefont {Hilker}}, \bibinfo {author} {\bibfnamefont {S.}~\bibnamefont {Wang}}, \bibinfo {author} {\bibfnamefont {J.}~\bibnamefont {Obermeyer}}, \bibinfo {author} {\bibfnamefont {M.}~\bibnamefont {Barendregt}}, \bibinfo {author} {\bibfnamefont {D.}~\bibnamefont {Tell}}, \bibinfo {author} {\bibfnamefont {T.}~\bibnamefont {Chalopin}}, \bibinfo {author} {\bibfnamefont {P.~M.}\ \bibnamefont {Preiss}}, \bibinfo {author} {\bibfnamefont {I.}~\bibnamefont {Bloch}},\ and\ \bibinfo {author} {\bibfnamefont {T.}~\bibnamefont {Franz}},\ }\href {https://doi.org/10.1038/s41586-026-10356-3} {\bibinfo {title} {High-fidelity collisional quantum gates with fermionic atoms}} (\bibinfo {year} {2026})\BibitemShut {NoStop}%
\bibitem [{\citenamefont {Peruzzo}\ \emph {et~al.}(2014)\citenamefont {Peruzzo}, \citenamefont {McClean}, \citenamefont {Shadbolt}, \citenamefont {Yung}, \citenamefont {Zhou}, \citenamefont {Love}, \citenamefont {Aspuru-Guzik},\ and\ \citenamefont {O’Brien}}]{peruzzo_variational_2014}%
  \BibitemOpen
  \bibfield  {author} {\bibinfo {author} {\bibfnamefont {A.}~\bibnamefont {Peruzzo}}, \bibinfo {author} {\bibfnamefont {J.}~\bibnamefont {McClean}}, \bibinfo {author} {\bibfnamefont {P.}~\bibnamefont {Shadbolt}}, \bibinfo {author} {\bibfnamefont {M.-H.}\ \bibnamefont {Yung}}, \bibinfo {author} {\bibfnamefont {X.-Q.}\ \bibnamefont {Zhou}}, \bibinfo {author} {\bibfnamefont {P.~J.}\ \bibnamefont {Love}}, \bibinfo {author} {\bibfnamefont {A.}~\bibnamefont {Aspuru-Guzik}},\ and\ \bibinfo {author} {\bibfnamefont {J.~L.}\ \bibnamefont {O’Brien}},\ }\bibfield  {title} {\bibinfo {title} {A variational eigenvalue solver on a photonic quantum processor},\ }\href {https://doi.org/10.1038/ncomms5213} {\bibfield  {journal} {\bibinfo  {journal} {Nature Communications}\ }\textbf {\bibinfo {volume} {5}},\ \bibinfo {pages} {4213} (\bibinfo {year} {2014})}\BibitemShut {NoStop}%
\bibitem [{\citenamefont {Grimsley}\ \emph {et~al.}(2019)\citenamefont {Grimsley}, \citenamefont {Economou}, \citenamefont {Barnes},\ and\ \citenamefont {Mayhall}}]{grimsleyAdaptiveVariationalAlgorithm2019}%
  \BibitemOpen
  \bibfield  {author} {\bibinfo {author} {\bibfnamefont {H.~R.}\ \bibnamefont {Grimsley}}, \bibinfo {author} {\bibfnamefont {S.~E.}\ \bibnamefont {Economou}}, \bibinfo {author} {\bibfnamefont {E.}~\bibnamefont {Barnes}},\ and\ \bibinfo {author} {\bibfnamefont {N.~J.}\ \bibnamefont {Mayhall}},\ }\bibfield  {title} {\bibinfo {title} {An adaptive variational algorithm for exact molecular simulations on a quantum computer},\ }\href {https://doi.org/10.1038/s41467-019-10988-2} {\bibfield  {journal} {\bibinfo  {journal} {Nature Communications}\ }\textbf {\bibinfo {volume} {10}},\ \bibinfo {pages} {3007} (\bibinfo {year} {2019})}\BibitemShut {NoStop}%
\bibitem [{\citenamefont {Kokail}\ \emph {et~al.}(2019)\citenamefont {Kokail}, \citenamefont {Maier}, \citenamefont {Van~Bijnen}, \citenamefont {Brydges}, \citenamefont {Joshi}, \citenamefont {Jurcevic}, \citenamefont {Muschik}, \citenamefont {Silvi}, \citenamefont {Blatt}, \citenamefont {Roos},\ and\ \citenamefont {Zoller}}]{kokail_self-verifying_2019}%
  \BibitemOpen
  \bibfield  {author} {\bibinfo {author} {\bibfnamefont {C.}~\bibnamefont {Kokail}}, \bibinfo {author} {\bibfnamefont {C.}~\bibnamefont {Maier}}, \bibinfo {author} {\bibfnamefont {R.}~\bibnamefont {Van~Bijnen}}, \bibinfo {author} {\bibfnamefont {T.}~\bibnamefont {Brydges}}, \bibinfo {author} {\bibfnamefont {M.~K.}\ \bibnamefont {Joshi}}, \bibinfo {author} {\bibfnamefont {P.}~\bibnamefont {Jurcevic}}, \bibinfo {author} {\bibfnamefont {C.~A.}\ \bibnamefont {Muschik}}, \bibinfo {author} {\bibfnamefont {P.}~\bibnamefont {Silvi}}, \bibinfo {author} {\bibfnamefont {R.}~\bibnamefont {Blatt}}, \bibinfo {author} {\bibfnamefont {C.~F.}\ \bibnamefont {Roos}},\ and\ \bibinfo {author} {\bibfnamefont {P.}~\bibnamefont {Zoller}},\ }\bibfield  {title} {\bibinfo {title} {Self-verifying variational quantum simulation of lattice models},\ }\href {https://doi.org/10.1038/s41586-019-1177-4} {\bibfield  {journal} {\bibinfo  {journal} {Nature}\ }\textbf {\bibinfo {volume} {569}},\ \bibinfo {pages} {355} (\bibinfo {year}
  {2019})}\BibitemShut {NoStop}%
\bibitem [{\citenamefont {Yuan}\ \emph {et~al.}(2019)\citenamefont {Yuan}, \citenamefont {Endo}, \citenamefont {Zhao}, \citenamefont {Li},\ and\ \citenamefont {Benjamin}}]{yuan_theory_2019}%
  \BibitemOpen
  \bibfield  {author} {\bibinfo {author} {\bibfnamefont {X.}~\bibnamefont {Yuan}}, \bibinfo {author} {\bibfnamefont {S.}~\bibnamefont {Endo}}, \bibinfo {author} {\bibfnamefont {Q.}~\bibnamefont {Zhao}}, \bibinfo {author} {\bibfnamefont {Y.}~\bibnamefont {Li}},\ and\ \bibinfo {author} {\bibfnamefont {S.}~\bibnamefont {Benjamin}},\ }\bibfield  {title} {\bibinfo {title} {Theory of variational quantum simulation},\ }\href {https://doi.org/10.22331/q-2019-10-07-191} {\bibfield  {journal} {\bibinfo  {journal} {Quantum}\ }\textbf {\bibinfo {volume} {3}},\ \bibinfo {pages} {191} (\bibinfo {year} {2019})}\BibitemShut {NoStop}%
\bibitem [{\citenamefont {Ho}\ and\ \citenamefont {Hsieh}(2019)}]{ho_efficient_2019}%
  \BibitemOpen
  \bibfield  {author} {\bibinfo {author} {\bibfnamefont {W.~W.}\ \bibnamefont {Ho}}\ and\ \bibinfo {author} {\bibfnamefont {T.~H.}\ \bibnamefont {Hsieh}},\ }\bibfield  {title} {\bibinfo {title} {Efficient variational simulation of non-trivial quantum states},\ }\href {https://doi.org/10.21468/SciPostPhys.6.3.029} {\bibfield  {journal} {\bibinfo  {journal} {SciPost Physics}\ }\textbf {\bibinfo {volume} {6}},\ \bibinfo {pages} {029} (\bibinfo {year} {2019})}\BibitemShut {NoStop}%
\bibitem [{\citenamefont {Cerezo}\ \emph {et~al.}(2021)\citenamefont {Cerezo}, \citenamefont {Arrasmith}, \citenamefont {Babbush}, \citenamefont {Benjamin}, \citenamefont {Endo}, \citenamefont {Fujii}, \citenamefont {McClean}, \citenamefont {Mitarai}, \citenamefont {Yuan}, \citenamefont {Cincio},\ and\ \citenamefont {Coles}}]{cerezo_variational_2021}%
  \BibitemOpen
  \bibfield  {author} {\bibinfo {author} {\bibfnamefont {M.}~\bibnamefont {Cerezo}}, \bibinfo {author} {\bibfnamefont {A.}~\bibnamefont {Arrasmith}}, \bibinfo {author} {\bibfnamefont {R.}~\bibnamefont {Babbush}}, \bibinfo {author} {\bibfnamefont {S.~C.}\ \bibnamefont {Benjamin}}, \bibinfo {author} {\bibfnamefont {S.}~\bibnamefont {Endo}}, \bibinfo {author} {\bibfnamefont {K.}~\bibnamefont {Fujii}}, \bibinfo {author} {\bibfnamefont {J.~R.}\ \bibnamefont {McClean}}, \bibinfo {author} {\bibfnamefont {K.}~\bibnamefont {Mitarai}}, \bibinfo {author} {\bibfnamefont {X.}~\bibnamefont {Yuan}}, \bibinfo {author} {\bibfnamefont {L.}~\bibnamefont {Cincio}},\ and\ \bibinfo {author} {\bibfnamefont {P.~J.}\ \bibnamefont {Coles}},\ }\bibfield  {title} {\bibinfo {title} {Variational quantum algorithms},\ }\href {https://doi.org/10.1038/s42254-021-00348-9} {\bibfield  {journal} {\bibinfo  {journal} {Nature Reviews Physics}\ }\textbf {\bibinfo {volume} {3}},\ \bibinfo {pages} {625} (\bibinfo {year} {2021})}\BibitemShut
  {NoStop}%
\bibitem [{\citenamefont {Nam}\ \emph {et~al.}(2020)\citenamefont {Nam}, \citenamefont {Chen}, \citenamefont {Pisenti}, \citenamefont {Wright}, \citenamefont {Delaney}, \citenamefont {Maslov}, \citenamefont {Brown}, \citenamefont {Allen}, \citenamefont {Amini}, \citenamefont {Apisdorf}, \citenamefont {Beck}, \citenamefont {Blinov}, \citenamefont {Chaplin}, \citenamefont {Chmielewski}, \citenamefont {Collins}, \citenamefont {Debnath}, \citenamefont {Hudek}, \citenamefont {Ducore}, \citenamefont {Keesan}, \citenamefont {Kreikemeier}, \citenamefont {Mizrahi}, \citenamefont {Solomon}, \citenamefont {Williams}, \citenamefont {Wong-Campos}, \citenamefont {Moehring}, \citenamefont {Monroe},\ and\ \citenamefont {Kim}}]{nam_gs_2020}%
  \BibitemOpen
  \bibfield  {author} {\bibinfo {author} {\bibfnamefont {Y.}~\bibnamefont {Nam}}, \bibinfo {author} {\bibfnamefont {J.-S.}\ \bibnamefont {Chen}}, \bibinfo {author} {\bibfnamefont {N.~C.}\ \bibnamefont {Pisenti}}, \bibinfo {author} {\bibfnamefont {K.}~\bibnamefont {Wright}}, \bibinfo {author} {\bibfnamefont {C.}~\bibnamefont {Delaney}}, \bibinfo {author} {\bibfnamefont {D.}~\bibnamefont {Maslov}}, \bibinfo {author} {\bibfnamefont {K.~R.}\ \bibnamefont {Brown}}, \bibinfo {author} {\bibfnamefont {S.}~\bibnamefont {Allen}}, \bibinfo {author} {\bibfnamefont {J.~M.}\ \bibnamefont {Amini}}, \bibinfo {author} {\bibfnamefont {J.}~\bibnamefont {Apisdorf}}, \bibinfo {author} {\bibfnamefont {K.~M.}\ \bibnamefont {Beck}}, \bibinfo {author} {\bibfnamefont {A.}~\bibnamefont {Blinov}}, \bibinfo {author} {\bibfnamefont {V.}~\bibnamefont {Chaplin}}, \bibinfo {author} {\bibfnamefont {M.}~\bibnamefont {Chmielewski}}, \bibinfo {author} {\bibfnamefont {C.}~\bibnamefont {Collins}}, \bibinfo {author} {\bibfnamefont {S.}~\bibnamefont
  {Debnath}}, \bibinfo {author} {\bibfnamefont {K.~M.}\ \bibnamefont {Hudek}}, \bibinfo {author} {\bibfnamefont {A.~M.}\ \bibnamefont {Ducore}}, \bibinfo {author} {\bibfnamefont {M.}~\bibnamefont {Keesan}}, \bibinfo {author} {\bibfnamefont {S.~M.}\ \bibnamefont {Kreikemeier}}, \bibinfo {author} {\bibfnamefont {J.}~\bibnamefont {Mizrahi}}, \bibinfo {author} {\bibfnamefont {P.}~\bibnamefont {Solomon}}, \bibinfo {author} {\bibfnamefont {M.}~\bibnamefont {Williams}}, \bibinfo {author} {\bibfnamefont {J.~D.}\ \bibnamefont {Wong-Campos}}, \bibinfo {author} {\bibfnamefont {D.}~\bibnamefont {Moehring}}, \bibinfo {author} {\bibfnamefont {C.}~\bibnamefont {Monroe}},\ and\ \bibinfo {author} {\bibfnamefont {J.}~\bibnamefont {Kim}},\ }\bibfield  {title} {\bibinfo {title} {Ground-state energy estimation of the water molecule on a trapped-ion quantum computer},\ }\href {https://doi.org/10.1038/s41534-020-0259-3} {\bibfield  {journal} {\bibinfo  {journal} {npj Quantum Information}\ }\textbf {\bibinfo {volume} {6}},\ \bibinfo
  {pages} {33} (\bibinfo {year} {2020})}\BibitemShut {NoStop}%
\bibitem [{\citenamefont {Ebadi}\ \emph {et~al.}(2022)\citenamefont {Ebadi}, \citenamefont {Keesling}, \citenamefont {Cain}, \citenamefont {Wang}, \citenamefont {Levine}, \citenamefont {Bluvstein}, \citenamefont {Semeghini}, \citenamefont {Omran}, \citenamefont {Liu}, \citenamefont {Samajdar}, \citenamefont {Luo}, \citenamefont {Nash}, \citenamefont {Gao}, \citenamefont {Barak}, \citenamefont {Farhi}, \citenamefont {Sachdev}, \citenamefont {Gemelke}, \citenamefont {Zhou}, \citenamefont {Choi}, \citenamefont {Pichler}, \citenamefont {Wang}, \citenamefont {Greiner}, \citenamefont {Vuletić},\ and\ \citenamefont {Lukin}}]{ebadiQuantumOptimizationMaximum2022}%
  \BibitemOpen
  \bibfield  {author} {\bibinfo {author} {\bibfnamefont {S.}~\bibnamefont {Ebadi}}, \bibinfo {author} {\bibfnamefont {A.}~\bibnamefont {Keesling}}, \bibinfo {author} {\bibfnamefont {M.}~\bibnamefont {Cain}}, \bibinfo {author} {\bibfnamefont {T.~T.}\ \bibnamefont {Wang}}, \bibinfo {author} {\bibfnamefont {H.}~\bibnamefont {Levine}}, \bibinfo {author} {\bibfnamefont {D.}~\bibnamefont {Bluvstein}}, \bibinfo {author} {\bibfnamefont {G.}~\bibnamefont {Semeghini}}, \bibinfo {author} {\bibfnamefont {A.}~\bibnamefont {Omran}}, \bibinfo {author} {\bibfnamefont {J.-G.}\ \bibnamefont {Liu}}, \bibinfo {author} {\bibfnamefont {R.}~\bibnamefont {Samajdar}}, \bibinfo {author} {\bibfnamefont {X.-Z.}\ \bibnamefont {Luo}}, \bibinfo {author} {\bibfnamefont {B.}~\bibnamefont {Nash}}, \bibinfo {author} {\bibfnamefont {X.}~\bibnamefont {Gao}}, \bibinfo {author} {\bibfnamefont {B.}~\bibnamefont {Barak}}, \bibinfo {author} {\bibfnamefont {E.}~\bibnamefont {Farhi}}, \bibinfo {author} {\bibfnamefont {S.}~\bibnamefont {Sachdev}},
  \bibinfo {author} {\bibfnamefont {N.}~\bibnamefont {Gemelke}}, \bibinfo {author} {\bibfnamefont {L.}~\bibnamefont {Zhou}}, \bibinfo {author} {\bibfnamefont {S.}~\bibnamefont {Choi}}, \bibinfo {author} {\bibfnamefont {H.}~\bibnamefont {Pichler}}, \bibinfo {author} {\bibfnamefont {S.-T.}\ \bibnamefont {Wang}}, \bibinfo {author} {\bibfnamefont {M.}~\bibnamefont {Greiner}}, \bibinfo {author} {\bibfnamefont {V.}~\bibnamefont {Vuletić}},\ and\ \bibinfo {author} {\bibfnamefont {M.~D.}\ \bibnamefont {Lukin}},\ }\bibfield  {title} {\bibinfo {title} {Quantum optimization of maximum independent set using {Rydberg} atom arrays},\ }\href {https://doi.org/10.1126/science.abo6587} {\bibfield  {journal} {\bibinfo  {journal} {Science}\ }\textbf {\bibinfo {volume} {376}},\ \bibinfo {pages} {1209} (\bibinfo {year} {2022})}\BibitemShut {NoStop}%
\bibitem [{\citenamefont {{Foss-Feig}}\ \emph {et~al.}(2023)\citenamefont {{Foss-Feig}}, \citenamefont {Tikku}, \citenamefont {Lu}, \citenamefont {Mayer}, \citenamefont {Iqbal}, \citenamefont {Gatterman}, \citenamefont {Gerber}, \citenamefont {Gilmore}, \citenamefont {Gresh}, \citenamefont {Hankin}, \citenamefont {Hewitt}, \citenamefont {Horst}, \citenamefont {Matheny}, \citenamefont {Mengle}, \citenamefont {Neyenhuis}, \citenamefont {Dreyer}, \citenamefont {Hayes}, \citenamefont {Hsieh},\ and\ \citenamefont {Kim}}]{foss-feig_experimental_2023}%
  \BibitemOpen
  \bibfield  {author} {\bibinfo {author} {\bibfnamefont {M.}~\bibnamefont {{Foss-Feig}}}, \bibinfo {author} {\bibfnamefont {A.}~\bibnamefont {Tikku}}, \bibinfo {author} {\bibfnamefont {T.-C.}\ \bibnamefont {Lu}}, \bibinfo {author} {\bibfnamefont {K.}~\bibnamefont {Mayer}}, \bibinfo {author} {\bibfnamefont {M.}~\bibnamefont {Iqbal}}, \bibinfo {author} {\bibfnamefont {T.~M.}\ \bibnamefont {Gatterman}}, \bibinfo {author} {\bibfnamefont {J.~A.}\ \bibnamefont {Gerber}}, \bibinfo {author} {\bibfnamefont {K.}~\bibnamefont {Gilmore}}, \bibinfo {author} {\bibfnamefont {D.}~\bibnamefont {Gresh}}, \bibinfo {author} {\bibfnamefont {A.}~\bibnamefont {Hankin}}, \bibinfo {author} {\bibfnamefont {N.}~\bibnamefont {Hewitt}}, \bibinfo {author} {\bibfnamefont {C.~V.}\ \bibnamefont {Horst}}, \bibinfo {author} {\bibfnamefont {M.}~\bibnamefont {Matheny}}, \bibinfo {author} {\bibfnamefont {T.}~\bibnamefont {Mengle}}, \bibinfo {author} {\bibfnamefont {B.}~\bibnamefont {Neyenhuis}}, \bibinfo {author} {\bibfnamefont {H.}~\bibnamefont
  {Dreyer}}, \bibinfo {author} {\bibfnamefont {D.}~\bibnamefont {Hayes}}, \bibinfo {author} {\bibfnamefont {T.~H.}\ \bibnamefont {Hsieh}},\ and\ \bibinfo {author} {\bibfnamefont {I.~H.}\ \bibnamefont {Kim}},\ }\href@noop {} {\bibinfo {title} {Experimental demonstration of the advantage of adaptive quantum circuits}} (\bibinfo {year} {2023}),\ \Eprint {https://arxiv.org/abs/2302.03029} {arXiv:2302.03029 [cond-mat, physics:quant-ph]} \BibitemShut {NoStop}%
\bibitem [{\citenamefont {Li}\ \emph {et~al.}(2023)\citenamefont {Li}, \citenamefont {Mukhopadhyay},\ and\ \citenamefont {Bayat}}]{li_fermionic_2023}%
  \BibitemOpen
  \bibfield  {author} {\bibinfo {author} {\bibfnamefont {Q.}~\bibnamefont {Li}}, \bibinfo {author} {\bibfnamefont {C.}~\bibnamefont {Mukhopadhyay}},\ and\ \bibinfo {author} {\bibfnamefont {A.}~\bibnamefont {Bayat}},\ }\bibfield  {title} {\bibinfo {title} {Fermionic simulators for enhanced scalability of variational quantum simulation},\ }\href {https://doi.org/10.1103/PhysRevResearch.5.043175} {\bibfield  {journal} {\bibinfo  {journal} {Physical Review Research}\ }\textbf {\bibinfo {volume} {5}},\ \bibinfo {pages} {043175} (\bibinfo {year} {2023})}\BibitemShut {NoStop}%
\bibitem [{\citenamefont {Gkritsis}\ \emph {et~al.}(2025)\citenamefont {Gkritsis}, \citenamefont {Dux}, \citenamefont {Zhang}, \citenamefont {Jain}, \citenamefont {Gogolin},\ and\ \citenamefont {Preiss}}]{Preiss2024}%
  \BibitemOpen
  \bibfield  {author} {\bibinfo {author} {\bibfnamefont {F.}~\bibnamefont {Gkritsis}}, \bibinfo {author} {\bibfnamefont {D.}~\bibnamefont {Dux}}, \bibinfo {author} {\bibfnamefont {J.}~\bibnamefont {Zhang}}, \bibinfo {author} {\bibfnamefont {N.}~\bibnamefont {Jain}}, \bibinfo {author} {\bibfnamefont {C.}~\bibnamefont {Gogolin}},\ and\ \bibinfo {author} {\bibfnamefont {P.~M.}\ \bibnamefont {Preiss}},\ }\bibfield  {title} {\bibinfo {title} {Simulating {{Chemistry}} with {{Fermionic Optical Superlattices}}},\ }\href {https://doi.org/10.1103/PRXQuantum.6.010318} {\bibfield  {journal} {\bibinfo  {journal} {PRX Quantum}\ }\textbf {\bibinfo {volume} {6}},\ \bibinfo {pages} {010318} (\bibinfo {year} {2025})}\BibitemShut {NoStop}%
\bibitem [{\citenamefont {Tabares}\ \emph {et~al.}(2025)\citenamefont {Tabares}, \citenamefont {Kokail}, \citenamefont {Zoller}, \citenamefont {{Gonz{\'a}lez-Cuadra}},\ and\ \citenamefont {{Gonz{\'a}lez-Tudela}}}]{tabaresProgrammingOpticallatticeFermiHubbard2025}%
  \BibitemOpen
  \bibfield  {author} {\bibinfo {author} {\bibfnamefont {C.}~\bibnamefont {Tabares}}, \bibinfo {author} {\bibfnamefont {C.}~\bibnamefont {Kokail}}, \bibinfo {author} {\bibfnamefont {P.}~\bibnamefont {Zoller}}, \bibinfo {author} {\bibfnamefont {D.}~\bibnamefont {{Gonz{\'a}lez-Cuadra}}},\ and\ \bibinfo {author} {\bibfnamefont {A.}~\bibnamefont {{Gonz{\'a}lez-Tudela}}},\ }\bibfield  {title} {\bibinfo {title} {Programming {{Optical-Lattice Fermi-Hubbard Quantum Simulators}}},\ }\href {https://doi.org/10.1103/3nx4-bnyy} {\bibfield  {journal} {\bibinfo  {journal} {PRX Quantum}\ }\textbf {\bibinfo {volume} {6}},\ \bibinfo {pages} {030356} (\bibinfo {year} {2025})}\BibitemShut {NoStop}%
\bibitem [{\citenamefont {Yan}\ \emph {et~al.}(2022)\citenamefont {Yan}, \citenamefont {Spar}, \citenamefont {Prichard}, \citenamefont {Chi}, \citenamefont {Wei}, \citenamefont {Ibarra-Garc\'{\i}a-Padilla}, \citenamefont {Hazzard},\ and\ \citenamefont {Bakr}}]{Yan2022}%
  \BibitemOpen
  \bibfield  {author} {\bibinfo {author} {\bibfnamefont {Z.~Z.}\ \bibnamefont {Yan}}, \bibinfo {author} {\bibfnamefont {B.~M.}\ \bibnamefont {Spar}}, \bibinfo {author} {\bibfnamefont {M.~L.}\ \bibnamefont {Prichard}}, \bibinfo {author} {\bibfnamefont {S.}~\bibnamefont {Chi}}, \bibinfo {author} {\bibfnamefont {H.-T.}\ \bibnamefont {Wei}}, \bibinfo {author} {\bibfnamefont {E.}~\bibnamefont {Ibarra-Garc\'{\i}a-Padilla}}, \bibinfo {author} {\bibfnamefont {K.~R.~A.}\ \bibnamefont {Hazzard}},\ and\ \bibinfo {author} {\bibfnamefont {W.~S.}\ \bibnamefont {Bakr}},\ }\bibfield  {title} {\bibinfo {title} {Two-dimensional programmable tweezer arrays of fermions},\ }\href {https://doi.org/10.1103/PhysRevLett.129.123201} {\bibfield  {journal} {\bibinfo  {journal} {Physical Review Letters}\ }\textbf {\bibinfo {volume} {129}},\ \bibinfo {pages} {123201} (\bibinfo {year} {2022})}\BibitemShut {NoStop}%
\bibitem [{Note1()}]{Note1}%
  \BibitemOpen
  \bibinfo {note} {As a side remark, we note that the method is robust to the experimental calibration error of the Hamiltonian parameters during the time evolution since the time evolution is used only to generate variational quantum states.}\BibitemShut {Stop}%
\bibitem [{\citenamefont {Nocedal}\ and\ \citenamefont {Wright}(2006)}]{nocedal2006numerical}%
  \BibitemOpen
  \bibfield  {author} {\bibinfo {author} {\bibfnamefont {J.}~\bibnamefont {Nocedal}}\ and\ \bibinfo {author} {\bibfnamefont {S.~J.}\ \bibnamefont {Wright}},\ }\href@noop {} {\emph {\bibinfo {title} {Numerical optimization}}}\ (\bibinfo  {publisher} {Springer},\ \bibinfo {year} {2006})\BibitemShut {NoStop}%
\bibitem [{\citenamefont {Essler}\ \emph {et~al.}(2005)\citenamefont {Essler}, \citenamefont {Frahm}, \citenamefont {G{\"o}hmann}, \citenamefont {Kl{\"u}mper},\ and\ \citenamefont {Korepin}}]{1dHubbard}%
  \BibitemOpen
  \bibfield  {author} {\bibinfo {author} {\bibfnamefont {F.~H.}\ \bibnamefont {Essler}}, \bibinfo {author} {\bibfnamefont {H.}~\bibnamefont {Frahm}}, \bibinfo {author} {\bibfnamefont {F.}~\bibnamefont {G{\"o}hmann}}, \bibinfo {author} {\bibfnamefont {A.}~\bibnamefont {Kl{\"u}mper}},\ and\ \bibinfo {author} {\bibfnamefont {V.~E.}\ \bibnamefont {Korepin}},\ }\href@noop {} {\emph {\bibinfo {title} {The one-dimensional Hubbard model}}}\ (\bibinfo  {publisher} {Cambridge University Press},\ \bibinfo {year} {2005})\BibitemShut {NoStop}%
\bibitem [{Note2()}]{Note2}%
  \BibitemOpen
  \bibinfo {note} {One may question the effectiveness of this procedure in the presence of spontaneous symmetry breaking. However, strictly speaking, in finite systems there is never spontaneous symmetry breaking. Moreover, even as the system size approaches infinity, local observables cannot distinguish between a symmetry-broken state and the large-but-finite-system-size ground state which is a superposition of all symmetry-equivalent configurations.}\BibitemShut {Stop}%
\bibitem [{\citenamefont {Impertro}\ \emph {et~al.}(2024)\citenamefont {Impertro}, \citenamefont {Karch}, \citenamefont {Wienand}, \citenamefont {Huh}, \citenamefont {Schweizer}, \citenamefont {Bloch},\ and\ \citenamefont {Aidelsburger}}]{impertroLocalReadoutControl2024}%
  \BibitemOpen
  \bibfield  {author} {\bibinfo {author} {\bibfnamefont {A.}~\bibnamefont {Impertro}}, \bibinfo {author} {\bibfnamefont {S.}~\bibnamefont {Karch}}, \bibinfo {author} {\bibfnamefont {J.~F.}\ \bibnamefont {Wienand}}, \bibinfo {author} {\bibfnamefont {S.}~\bibnamefont {Huh}}, \bibinfo {author} {\bibfnamefont {C.}~\bibnamefont {Schweizer}}, \bibinfo {author} {\bibfnamefont {I.}~\bibnamefont {Bloch}},\ and\ \bibinfo {author} {\bibfnamefont {M.}~\bibnamefont {Aidelsburger}},\ }\bibfield  {title} {\bibinfo {title} {Local {Readout} and {Control} of {Current} and {Kinetic} {Energy} {Operators} in {Optical} {Lattices}},\ }\href {https://doi.org/10.1103/PhysRevLett.133.063401} {\bibfield  {journal} {\bibinfo  {journal} {Physical Review Letters}\ }\textbf {\bibinfo {volume} {133}},\ \bibinfo {pages} {063401} (\bibinfo {year} {2024})}\BibitemShut {NoStop}%
\bibitem [{\citenamefont {Ke{\ss}ler}\ and\ \citenamefont {Marquardt}(2014)}]{kesslerSinglesiteresolvedMeasurementCurrent2014}%
  \BibitemOpen
  \bibfield  {author} {\bibinfo {author} {\bibfnamefont {S.}~\bibnamefont {Ke{\ss}ler}}\ and\ \bibinfo {author} {\bibfnamefont {F.}~\bibnamefont {Marquardt}},\ }\bibfield  {title} {\bibinfo {title} {Single-site-resolved measurement of the current statistics in optical lattices},\ }\href {https://doi.org/10.1103/PhysRevA.89.061601} {\bibfield  {journal} {\bibinfo  {journal} {Physical Review A}\ }\textbf {\bibinfo {volume} {89}},\ \bibinfo {pages} {061601} (\bibinfo {year} {2014})}\BibitemShut {NoStop}%
\bibitem [{Note3()}]{Note3}%
  \BibitemOpen
  \bibinfo {note} {Since for odd system size, the ground state is not in the total spin $S=0$ sector, we fix the dynamics to be in the total $S^z=\protect \frac {1}{2}$ sector by setting the last spin in the initial state to be in the spin-up state.}\BibitemShut {Stop}%
\bibitem [{\citenamefont {Anselmetti}\ \emph {et~al.}(2021)\citenamefont {Anselmetti}, \citenamefont {Wierichs}, \citenamefont {Gogolin},\ and\ \citenamefont {Parrish}}]{QNP2021}%
  \BibitemOpen
  \bibfield  {author} {\bibinfo {author} {\bibfnamefont {G.-L.~R.}\ \bibnamefont {Anselmetti}}, \bibinfo {author} {\bibfnamefont {D.}~\bibnamefont {Wierichs}}, \bibinfo {author} {\bibfnamefont {C.}~\bibnamefont {Gogolin}},\ and\ \bibinfo {author} {\bibfnamefont {R.~M.}\ \bibnamefont {Parrish}},\ }\bibfield  {title} {\bibinfo {title} {Local, expressive, quantum-number-preserving {VQE} ansätze for fermionic systems},\ }\href {https://doi.org/10.1088/1367-2630/ac2cb3} {\bibfield  {journal} {\bibinfo  {journal} {New Journal of Physics}\ }\textbf {\bibinfo {volume} {23}},\ \bibinfo {pages} {113010} (\bibinfo {year} {2021})}\BibitemShut {NoStop}%
\bibitem [{\citenamefont {Qu}\ \emph {et~al.}(2022)\citenamefont {Qu}, \citenamefont {Chen}, \citenamefont {Jiang}, \citenamefont {Wang},\ and\ \citenamefont {Li}}]{qu_efh_2022}%
  \BibitemOpen
  \bibfield  {author} {\bibinfo {author} {\bibfnamefont {D.-W.}\ \bibnamefont {Qu}}, \bibinfo {author} {\bibfnamefont {B.-B.}\ \bibnamefont {Chen}}, \bibinfo {author} {\bibfnamefont {H.-C.}\ \bibnamefont {Jiang}}, \bibinfo {author} {\bibfnamefont {Y.}~\bibnamefont {Wang}},\ and\ \bibinfo {author} {\bibfnamefont {W.}~\bibnamefont {Li}},\ }\bibfield  {title} {\bibinfo {title} {Spin-triplet pairing induced by near-neighbor attraction in the extended {Hubbard} model for cuprate chain},\ }\href {https://doi.org/10.1038/s42005-022-01030-x} {\bibfield  {journal} {\bibinfo  {journal} {Communications Physics}\ }\textbf {\bibinfo {volume} {5}},\ \bibinfo {pages} {1} (\bibinfo {year} {2022})}\BibitemShut {NoStop}%
\bibitem [{\citenamefont {Repellin}\ \emph {et~al.}(2020)\citenamefont {Repellin}, \citenamefont {Léonard},\ and\ \citenamefont {Goldman}}]{repellin_fractional_2020}%
  \BibitemOpen
  \bibfield  {author} {\bibinfo {author} {\bibfnamefont {C.}~\bibnamefont {Repellin}}, \bibinfo {author} {\bibfnamefont {J.}~\bibnamefont {Léonard}},\ and\ \bibinfo {author} {\bibfnamefont {N.}~\bibnamefont {Goldman}},\ }\bibfield  {title} {\bibinfo {title} {Fractional {Chern} insulators of few bosons in a box: {Hall} plateaus from center-of-mass drifts and density profiles},\ }\href {https://doi.org/10.1103/PhysRevA.102.063316} {\bibfield  {journal} {\bibinfo  {journal} {Physical Review A}\ }\textbf {\bibinfo {volume} {102}},\ \bibinfo {pages} {063316} (\bibinfo {year} {2020})}\BibitemShut {NoStop}%
\bibitem [{\citenamefont {He}\ \emph {et~al.}(2017)\citenamefont {He}, \citenamefont {Grusdt}, \citenamefont {Kaufman}, \citenamefont {Greiner},\ and\ \citenamefont {Vishwanath}}]{heRealizingAdiabaticallyPreparing2017a}%
  \BibitemOpen
  \bibfield  {author} {\bibinfo {author} {\bibfnamefont {Y.-C.}\ \bibnamefont {He}}, \bibinfo {author} {\bibfnamefont {F.}~\bibnamefont {Grusdt}}, \bibinfo {author} {\bibfnamefont {A.}~\bibnamefont {Kaufman}}, \bibinfo {author} {\bibfnamefont {M.}~\bibnamefont {Greiner}},\ and\ \bibinfo {author} {\bibfnamefont {A.}~\bibnamefont {Vishwanath}},\ }\bibfield  {title} {\bibinfo {title} {Realizing and adiabatically preparing bosonic integer and fractional quantum {{Hall}} states in optical lattices},\ }\href {https://doi.org/10.1103/PhysRevB.96.201103} {\bibfield  {journal} {\bibinfo  {journal} {Physical Review B}\ }\textbf {\bibinfo {volume} {96}},\ \bibinfo {pages} {201103} (\bibinfo {year} {2017})}\BibitemShut {NoStop}%
\bibitem [{\citenamefont {Motruk}\ and\ \citenamefont {Pollmann}(2017)}]{motrukPhaseTransitionsAdiabatic2017}%
  \BibitemOpen
  \bibfield  {author} {\bibinfo {author} {\bibfnamefont {J.}~\bibnamefont {Motruk}}\ and\ \bibinfo {author} {\bibfnamefont {F.}~\bibnamefont {Pollmann}},\ }\bibfield  {title} {\bibinfo {title} {Phase transitions and adiabatic preparation of a fractional {{Chern}} insulator in a boson cold-atom model},\ }\href {https://doi.org/10.1103/PhysRevB.96.165107} {\bibfield  {journal} {\bibinfo  {journal} {Physical Review B}\ }\textbf {\bibinfo {volume} {96}},\ \bibinfo {pages} {165107} (\bibinfo {year} {2017})}\BibitemShut {NoStop}%
\bibitem [{\citenamefont {Wang}\ \emph {et~al.}(2022)\citenamefont {Wang}, \citenamefont {Dong},\ and\ \citenamefont {Eckardt}}]{wangMeasurableSignaturesBosonic2022a}%
  \BibitemOpen
  \bibfield  {author} {\bibinfo {author} {\bibfnamefont {B.}~\bibnamefont {Wang}}, \bibinfo {author} {\bibfnamefont {X.}~\bibnamefont {Dong}},\ and\ \bibinfo {author} {\bibfnamefont {A.}~\bibnamefont {Eckardt}},\ }\bibfield  {title} {\bibinfo {title} {Measurable signatures of bosonic fractional {{Chern}} insulator states and their fractional excitations in a quantum-gas microscope},\ }\href {https://doi.org/10.21468/SciPostPhys.12.3.095} {\bibfield  {journal} {\bibinfo  {journal} {SciPost Physics}\ }\textbf {\bibinfo {volume} {12}},\ \bibinfo {pages} {095} (\bibinfo {year} {2022})}\BibitemShut {NoStop}%
\bibitem [{\citenamefont {Sørensen}\ \emph {et~al.}(2005)\citenamefont {Sørensen}, \citenamefont {Demler},\ and\ \citenamefont {Lukin}}]{FQH2005}%
  \BibitemOpen
  \bibfield  {author} {\bibinfo {author} {\bibfnamefont {A.~S.}\ \bibnamefont {Sørensen}}, \bibinfo {author} {\bibfnamefont {E.}~\bibnamefont {Demler}},\ and\ \bibinfo {author} {\bibfnamefont {M.~D.}\ \bibnamefont {Lukin}},\ }\bibfield  {title} {\bibinfo {title} {Fractional {Quantum} {Hall} {States} of {Atoms} in {Optical} {Lattices}},\ }\href {https://doi.org/10.1103/PhysRevLett.94.086803} {\bibfield  {journal} {\bibinfo  {journal} {Physical Review Letters}\ }\textbf {\bibinfo {volume} {94}},\ \bibinfo {pages} {086803} (\bibinfo {year} {2005})}\BibitemShut {NoStop}%
\bibitem [{\citenamefont {Rosson}\ \emph {et~al.}(2019)\citenamefont {Rosson}, \citenamefont {Lubasch}, \citenamefont {Kiffner},\ and\ \citenamefont {Jaksch}}]{rosson_bosonic_2019}%
  \BibitemOpen
  \bibfield  {author} {\bibinfo {author} {\bibfnamefont {P.}~\bibnamefont {Rosson}}, \bibinfo {author} {\bibfnamefont {M.}~\bibnamefont {Lubasch}}, \bibinfo {author} {\bibfnamefont {M.}~\bibnamefont {Kiffner}},\ and\ \bibinfo {author} {\bibfnamefont {D.}~\bibnamefont {Jaksch}},\ }\bibfield  {title} {\bibinfo {title} {Bosonic fractional quantum {Hall} states on a finite cylinder},\ }\href {https://doi.org/10.1103/PhysRevA.99.033603} {\bibfield  {journal} {\bibinfo  {journal} {Physical Review A}\ }\textbf {\bibinfo {volume} {99}},\ \bibinfo {pages} {033603} (\bibinfo {year} {2019})}\BibitemShut {NoStop}%
\bibitem [{\citenamefont {Pauw}\ \emph {et~al.}(2024)\citenamefont {Pauw}, \citenamefont {Palm}, \citenamefont {Schollwöck}, \citenamefont {Bohrdt}, \citenamefont {Paeckel},\ and\ \citenamefont {Grusdt}}]{pauw_detecting_2024}%
  \BibitemOpen
  \bibfield  {author} {\bibinfo {author} {\bibfnamefont {F.}~\bibnamefont {Pauw}}, \bibinfo {author} {\bibfnamefont {F.~A.}\ \bibnamefont {Palm}}, \bibinfo {author} {\bibfnamefont {U.}~\bibnamefont {Schollwöck}}, \bibinfo {author} {\bibfnamefont {A.}~\bibnamefont {Bohrdt}}, \bibinfo {author} {\bibfnamefont {S.}~\bibnamefont {Paeckel}},\ and\ \bibinfo {author} {\bibfnamefont {F.}~\bibnamefont {Grusdt}},\ }\bibfield  {title} {\bibinfo {title} {Detecting hidden order in fractional {Chern} insulators},\ }\href {https://doi.org/10.1103/PhysRevResearch.6.023180} {\bibfield  {journal} {\bibinfo  {journal} {Physical Review Research}\ }\textbf {\bibinfo {volume} {6}},\ \bibinfo {pages} {023180} (\bibinfo {year} {2024})}\BibitemShut {NoStop}%
\bibitem [{\citenamefont {Lin}\ \emph {et~al.}(2009)\citenamefont {Lin}, \citenamefont {Compton}, \citenamefont {{Jim{\'e}nez-Garc{\'i}a}}, \citenamefont {Porto},\ and\ \citenamefont {Spielman}}]{linSyntheticMagneticFields2009}%
  \BibitemOpen
  \bibfield  {author} {\bibinfo {author} {\bibfnamefont {Y.-J.}\ \bibnamefont {Lin}}, \bibinfo {author} {\bibfnamefont {R.~L.}\ \bibnamefont {Compton}}, \bibinfo {author} {\bibfnamefont {K.}~\bibnamefont {{Jim{\'e}nez-Garc{\'i}a}}}, \bibinfo {author} {\bibfnamefont {J.~V.}\ \bibnamefont {Porto}},\ and\ \bibinfo {author} {\bibfnamefont {I.~B.}\ \bibnamefont {Spielman}},\ }\bibfield  {title} {\bibinfo {title} {Synthetic magnetic fields for ultracold neutral atoms},\ }\href {https://doi.org/10.1038/nature08609} {\bibfield  {journal} {\bibinfo  {journal} {Nature}\ }\textbf {\bibinfo {volume} {462}},\ \bibinfo {pages} {628} (\bibinfo {year} {2009})}\BibitemShut {NoStop}%
\bibitem [{\citenamefont {Miyake}\ \emph {et~al.}(2013)\citenamefont {Miyake}, \citenamefont {Siviloglou}, \citenamefont {Kennedy}, \citenamefont {Burton},\ and\ \citenamefont {Ketterle}}]{miyakeRealizingHarperHamiltonian2013a}%
  \BibitemOpen
  \bibfield  {author} {\bibinfo {author} {\bibfnamefont {H.}~\bibnamefont {Miyake}}, \bibinfo {author} {\bibfnamefont {G.~A.}\ \bibnamefont {Siviloglou}}, \bibinfo {author} {\bibfnamefont {C.~J.}\ \bibnamefont {Kennedy}}, \bibinfo {author} {\bibfnamefont {W.~C.}\ \bibnamefont {Burton}},\ and\ \bibinfo {author} {\bibfnamefont {W.}~\bibnamefont {Ketterle}},\ }\bibfield  {title} {\bibinfo {title} {Realizing the {{Harper Hamiltonian}} with {{Laser-Assisted Tunneling}} in {{Optical Lattices}}},\ }\href {https://doi.org/10.1103/PhysRevLett.111.185302} {\bibfield  {journal} {\bibinfo  {journal} {Physical Review Letters}\ }\textbf {\bibinfo {volume} {111}},\ \bibinfo {pages} {185302} (\bibinfo {year} {2013})}\BibitemShut {NoStop}%
\bibitem [{Note4()}]{Note4}%
  \BibitemOpen
  \bibinfo {note} {The definition of $T$ for the HHM differs slightly from the one we used earlier to work naturally in the hard core limit, see App.~\ref {app:qtime}.}\BibitemShut {Stop}%
\bibitem [{Note5()}]{Note5}%
  \BibitemOpen
  \bibinfo {note} {This can be understood by bosonization. The triplet and singlet pairing correlations are in the form of $C_T(r)\sim \mathinner {\langle {\sin (\protect \sqrt {2}\phi _\sigma (r))\sin (\protect \sqrt {2}\phi _\sigma (0))}\rangle }$ and $C_S(r)\sim \mathinner {\langle {\cos (\protect \sqrt {2}\phi _\sigma (r))\cos (\protect \sqrt {2}\phi _\sigma (0))}\rangle }$ with $\phi _\sigma $ the spin field \cite {giamarchi2003quantum}. If the phase has a gapless spin sector as $V=0$ FHM ground state, the two pairing correlations have the same asymptotic behavior; but when the spin field $\phi _\sigma $ is pinned by the $\cos (4\phi _\sigma )$ term \cite {seidelLutherEmeryLiquidSpin2005} arising from the attractive interaction in the EFHM, the two correlations differ.}\BibitemShut {Stop}%
\bibitem [{\citenamefont {Girvin}\ and\ \citenamefont {MacDonald}(1987)}]{GMD1987}%
  \BibitemOpen
  \bibfield  {author} {\bibinfo {author} {\bibfnamefont {S.~M.}\ \bibnamefont {Girvin}}\ and\ \bibinfo {author} {\bibfnamefont {A.~H.}\ \bibnamefont {MacDonald}},\ }\bibfield  {title} {\bibinfo {title} {Off-diagonal long-range order, oblique confinement, and the fractional quantum {Hall} effect},\ }\href {https://doi.org/10.1103/PhysRevLett.58.1252} {\bibfield  {journal} {\bibinfo  {journal} {Physical Review Letters}\ }\textbf {\bibinfo {volume} {58}},\ \bibinfo {pages} {1252} (\bibinfo {year} {1987})}\BibitemShut {NoStop}%
\bibitem [{\citenamefont {Weitenberg}\ \emph {et~al.}(2026)\citenamefont {Weitenberg}, \citenamefont {Asteria}, \citenamefont {Carlsson}, \citenamefont {Bohrdt},\ and\ \citenamefont {Grusdt}}]{weitenbergProtocolsManybodyPhase2026}%
  \BibitemOpen
  \bibfield  {author} {\bibinfo {author} {\bibfnamefont {C.}~\bibnamefont {Weitenberg}}, \bibinfo {author} {\bibfnamefont {L.}~\bibnamefont {Asteria}}, \bibinfo {author} {\bibfnamefont {O.}~\bibnamefont {Carlsson}}, \bibinfo {author} {\bibfnamefont {A.}~\bibnamefont {Bohrdt}},\ and\ \bibinfo {author} {\bibfnamefont {F.}~\bibnamefont {Grusdt}},\ }\href@noop {} {\bibinfo {title} {Protocols for a many-body phase microscope: {{From}} coherences and d-wave superconductivity to {{Green}}'s functions}} (\bibinfo {year} {2026}),\ \Eprint {https://arxiv.org/abs/2602.12142} {arXiv:2602.12142 [cond-mat]} \BibitemShut {NoStop}%
\bibitem [{\citenamefont {Oszmaniec}\ and\ \citenamefont {Zimborás}(2017)}]{oszmaniec_universal_2017}%
  \BibitemOpen
  \bibfield  {author} {\bibinfo {author} {\bibfnamefont {M.}~\bibnamefont {Oszmaniec}}\ and\ \bibinfo {author} {\bibfnamefont {Z.}~\bibnamefont {Zimborás}},\ }\bibfield  {title} {\bibinfo {title} {Universal {Extensions} of {Restricted} {Classes} of {Quantum} {Operations}},\ }\href {https://doi.org/10.1103/PhysRevLett.119.220502} {\bibfield  {journal} {\bibinfo  {journal} {Physical Review Letters}\ }\textbf {\bibinfo {volume} {119}},\ \bibinfo {pages} {220502} (\bibinfo {year} {2017})}\BibitemShut {NoStop}%
\bibitem [{Note6()}]{Note6}%
  \BibitemOpen
  \bibinfo {note} {Technically, this result assumes that an $M$-mode system has a number of particles $N_p\protect \notin \{0,1,M,M-1\}$. This presents no significant obstacle, as these cases involve either exactly zero or one particle, or zero or one hole. Therefore, their state space dimensions are no larger than $M$ and can be treated efficiently classically.}\BibitemShut {Stop}%
\bibitem [{\citenamefont {Trivedi}\ \emph {et~al.}(2024)\citenamefont {Trivedi}, \citenamefont {Franco~Rubio},\ and\ \citenamefont {Cirac}}]{CiracAnalogAdvantage2024}%
  \BibitemOpen
  \bibfield  {author} {\bibinfo {author} {\bibfnamefont {R.}~\bibnamefont {Trivedi}}, \bibinfo {author} {\bibfnamefont {A.}~\bibnamefont {Franco~Rubio}},\ and\ \bibinfo {author} {\bibfnamefont {J.~I.}\ \bibnamefont {Cirac}},\ }\bibfield  {title} {\bibinfo {title} {Quantum advantage and stability to errors in analogue quantum simulators},\ }\href {https://doi.org/10.1038/s41467-024-50750-x} {\bibfield  {journal} {\bibinfo  {journal} {Nature Communications}\ }\textbf {\bibinfo {volume} {15}},\ \bibinfo {pages} {6507} (\bibinfo {year} {2024})}\BibitemShut {NoStop}%
\bibitem [{\citenamefont {Sachdev}(2011)}]{Sachdev_2011}%
  \BibitemOpen
  \bibfield  {author} {\bibinfo {author} {\bibfnamefont {S.}~\bibnamefont {Sachdev}},\ }\href@noop {} {\emph {\bibinfo {title} {Quantum Phase Transitions}}},\ \bibinfo {edition} {2nd}\ ed.\ (\bibinfo  {publisher} {Cambridge University Press},\ \bibinfo {year} {2011})\BibitemShut {NoStop}%
\bibitem [{\citenamefont {Hastings}\ and\ \citenamefont {Wen}(2005)}]{hastings_qac_2005}%
  \BibitemOpen
  \bibfield  {author} {\bibinfo {author} {\bibfnamefont {M.~B.}\ \bibnamefont {Hastings}}\ and\ \bibinfo {author} {\bibfnamefont {X.-G.}\ \bibnamefont {Wen}},\ }\bibfield  {title} {\bibinfo {title} {Quasiadiabatic continuation of quantum states: {The} stability of topological ground-state degeneracy and emergent gauge invariance},\ }\href {https://doi.org/10.1103/PhysRevB.72.045141} {\bibfield  {journal} {\bibinfo  {journal} {Physical Review B}\ }\textbf {\bibinfo {volume} {72}},\ \bibinfo {pages} {045141} (\bibinfo {year} {2005})}\BibitemShut {NoStop}%
\bibitem [{Note7()}]{Note7}%
  \BibitemOpen
  \bibinfo {note} {In this work we keep $L_x$ fixed and increase $L_y$, so the two edges, each with finite-size scaling $\sim L^{-p}$, contribute $O(L_y)$ sites, i.e. an $O(1)$ fraction of the system volume. In the isotropic 2D limit $L_x,L_y\to \infty $, the edge contribution is instead suppressed as $\sim 1/L_{x,y}$, so the finite-size error of the energy density decays slightly faster, $\epsilon _L\sim L^{-(p+1)}$. In both cases, the finite-size error remains power-law in the linear system size.}\BibitemShut {Stop}%
\bibitem [{\citenamefont {Bachmann}\ \emph {et~al.}(2019)\citenamefont {Bachmann}, \citenamefont {Roeck},\ and\ \citenamefont {Fraas}}]{bachmannAdiabaticTheoremQuantum2019}%
  \BibitemOpen
  \bibfield  {author} {\bibinfo {author} {\bibfnamefont {S.}~\bibnamefont {Bachmann}}, \bibinfo {author} {\bibfnamefont {W.~D.}\ \bibnamefont {Roeck}},\ and\ \bibinfo {author} {\bibfnamefont {M.}~\bibnamefont {Fraas}},\ }\href {https://doi.org/10.48550/arXiv.1808.09985} {\bibinfo {title} {The adiabatic theorem in a quantum many-body setting}} (\bibinfo {year} {2019}),\ \Eprint {https://arxiv.org/abs/1808.09985} {arXiv:1808.09985 [math-ph]} \BibitemShut {NoStop}%
\bibitem [{\citenamefont {Bachmann}\ \emph {et~al.}(2017)\citenamefont {Bachmann}, \citenamefont {De~Roeck},\ and\ \citenamefont {Fraas}}]{bachmannAdiabaticTheoremQuantum2017}%
  \BibitemOpen
  \bibfield  {author} {\bibinfo {author} {\bibfnamefont {S.}~\bibnamefont {Bachmann}}, \bibinfo {author} {\bibfnamefont {W.}~\bibnamefont {De~Roeck}},\ and\ \bibinfo {author} {\bibfnamefont {M.}~\bibnamefont {Fraas}},\ }\bibfield  {title} {\bibinfo {title} {Adiabatic {{Theorem}} for {{Quantum Spin Systems}}},\ }\href {https://doi.org/10.1103/PhysRevLett.119.060201} {\bibfield  {journal} {\bibinfo  {journal} {Physical Review Letters}\ }\textbf {\bibinfo {volume} {119}},\ \bibinfo {pages} {060201} (\bibinfo {year} {2017})}\BibitemShut {NoStop}%
\bibitem [{\citenamefont {Teufel}(2022)}]{teufelQuantumAdiabaticTheorem}%
  \BibitemOpen
  \bibfield  {author} {\bibinfo {author} {\bibfnamefont {S.}~\bibnamefont {Teufel}},\ }\bibfield  {title} {\bibinfo {title} {Quantum adiabatic theorem},\ }in\ \href@noop {} {\emph {\bibinfo {booktitle} {Perturbation Theory: Mathematics, Methods and Applications}}}\ (\bibinfo  {publisher} {Springer},\ \bibinfo {year} {2022})\ pp.\ \bibinfo {pages} {419--431}\BibitemShut {NoStop}%
\bibitem [{\citenamefont {Haah}\ \emph {et~al.}(2023)\citenamefont {Haah}, \citenamefont {Hastings}, \citenamefont {Kothari},\ and\ \citenamefont {Low}}]{haah_qalgo_2023}%
  \BibitemOpen
  \bibfield  {author} {\bibinfo {author} {\bibfnamefont {J.}~\bibnamefont {Haah}}, \bibinfo {author} {\bibfnamefont {M.~B.}\ \bibnamefont {Hastings}}, \bibinfo {author} {\bibfnamefont {R.}~\bibnamefont {Kothari}},\ and\ \bibinfo {author} {\bibfnamefont {G.~H.}\ \bibnamefont {Low}},\ }\bibfield  {title} {\bibinfo {title} {Quantum {{Algorithm}} for {{Simulating Real Time Evolution}} of {{Lattice Hamiltonians}}},\ }\href {https://doi.org/10.1137/18M1231511} {\bibfield  {journal} {\bibinfo  {journal} {SIAM Journal on Computing}\ }\textbf {\bibinfo {volume} {52}},\ \bibinfo {pages} {FOCS18} (\bibinfo {year} {2023})}\BibitemShut {NoStop}%
\bibitem [{\citenamefont {de~Wolf}(2023)}]{wolfQuantumComputingLecture2023}%
  \BibitemOpen
  \bibfield  {author} {\bibinfo {author} {\bibfnamefont {R.}~\bibnamefont {de~Wolf}},\ }\href@noop {} {\bibinfo {title} {Quantum {{Computing}}: {{Lecture Notes}}}} (\bibinfo {year} {2023}),\ \Eprint {https://arxiv.org/abs/1907.09415} {arXiv:1907.09415 [quant-ph]} \BibitemShut {NoStop}%
\bibitem [{\citenamefont {Wang}\ \emph {et~al.}(2021)\citenamefont {Wang}, \citenamefont {Foss-Feig},\ and\ \citenamefont {Hazzard}}]{zywang_bounding_2021}%
  \BibitemOpen
  \bibfield  {author} {\bibinfo {author} {\bibfnamefont {Z.}~\bibnamefont {Wang}}, \bibinfo {author} {\bibfnamefont {M.}~\bibnamefont {Foss-Feig}},\ and\ \bibinfo {author} {\bibfnamefont {K.~R.~A.}\ \bibnamefont {Hazzard}},\ }\bibfield  {title} {\bibinfo {title} {Bounding the finite-size error of quantum many-body dynamics simulations},\ }\href {https://doi.org/10.1103/PhysRevResearch.3.L032047} {\bibfield  {journal} {\bibinfo  {journal} {Physical Review Research}\ }\textbf {\bibinfo {volume} {3}},\ \bibinfo {pages} {L032047} (\bibinfo {year} {2021})}\BibitemShut {NoStop}%
\bibitem [{\citenamefont {Pollmann}\ \emph {et~al.}(2009)\citenamefont {Pollmann}, \citenamefont {Mukerjee}, \citenamefont {Turner},\ and\ \citenamefont {Moore}}]{pollmannTheoryFiniteEntanglementScaling2009}%
  \BibitemOpen
  \bibfield  {author} {\bibinfo {author} {\bibfnamefont {F.}~\bibnamefont {Pollmann}}, \bibinfo {author} {\bibfnamefont {S.}~\bibnamefont {Mukerjee}}, \bibinfo {author} {\bibfnamefont {A.~M.}\ \bibnamefont {Turner}},\ and\ \bibinfo {author} {\bibfnamefont {J.~E.}\ \bibnamefont {Moore}},\ }\bibfield  {title} {\bibinfo {title} {Theory of {{Finite-Entanglement Scaling}} at {{One-Dimensional Quantum Critical Points}}},\ }\href {https://doi.org/10.1103/PhysRevLett.102.255701} {\bibfield  {journal} {\bibinfo  {journal} {Physical Review Letters}\ }\textbf {\bibinfo {volume} {102}},\ \bibinfo {pages} {255701} (\bibinfo {year} {2009})}\BibitemShut {NoStop}%
\bibitem [{\citenamefont {Pirvu}\ \emph {et~al.}(2012)\citenamefont {Pirvu}, \citenamefont {Vidal}, \citenamefont {Verstraete},\ and\ \citenamefont {Tagliacozzo}}]{pirvuMatrixProductStates2012}%
  \BibitemOpen
  \bibfield  {author} {\bibinfo {author} {\bibfnamefont {B.}~\bibnamefont {Pirvu}}, \bibinfo {author} {\bibfnamefont {G.}~\bibnamefont {Vidal}}, \bibinfo {author} {\bibfnamefont {F.}~\bibnamefont {Verstraete}},\ and\ \bibinfo {author} {\bibfnamefont {L.}~\bibnamefont {Tagliacozzo}},\ }\bibfield  {title} {\bibinfo {title} {Matrix product states for critical spin chains: {{Finite-size}} versus finite-entanglement scaling},\ }\href {https://doi.org/10.1103/PhysRevB.86.075117} {\bibfield  {journal} {\bibinfo  {journal} {Physical Review B}\ }\textbf {\bibinfo {volume} {86}},\ \bibinfo {pages} {075117} (\bibinfo {year} {2012})}\BibitemShut {NoStop}%
\bibitem [{\citenamefont {Wei}\ and\ \citenamefont {Hazzard}()}]{sun_inprep}%
  \BibitemOpen
  \bibfield  {author} {\bibinfo {author} {\bibfnamefont {H.-T.}\ \bibnamefont {Wei}}\ and\ \bibinfo {author} {\bibfnamefont {K.~R.~A.}\ \bibnamefont {Hazzard}},\ }\bibinfo {title} {In preparation}\BibitemShut {NoStop}%
\bibitem [{\citenamefont {Wu}\ \emph {et~al.}(2024)\citenamefont {Wu}, \citenamefont {Rossi}, \citenamefont {Vicentini}, \citenamefont {Astrakhantsev}, \citenamefont {Becca}, \citenamefont {Cao}, \citenamefont {Carrasquilla}, \citenamefont {Ferrari}, \citenamefont {Georges}, \citenamefont {{Hibat-Allah}}, \citenamefont {Imada}, \citenamefont {L{\"a}uchli}, \citenamefont {Mazzola}, \citenamefont {Mezzacapo}, \citenamefont {Millis}, \citenamefont {Robledo~Moreno}, \citenamefont {Neupert}, \citenamefont {Nomura}, \citenamefont {Nys}, \citenamefont {Parcollet}, \citenamefont {Pohle}, \citenamefont {Romero}, \citenamefont {Schmid}, \citenamefont {Silvester}, \citenamefont {Sorella}, \citenamefont {Tocchio}, \citenamefont {Wang}, \citenamefont {White}, \citenamefont {Wietek}, \citenamefont {Yang}, \citenamefont {Yang}, \citenamefont {Zhang},\ and\ \citenamefont {Carleo}}]{wuVariationalBenchmarksQuantum2024}%
  \BibitemOpen
\bibfield  {title} {  }\bibfield  {author} {\bibinfo {author} {\bibfnamefont {D.}~\bibnamefont {Wu}}, \bibinfo {author} {\bibfnamefont {R.}~\bibnamefont {Rossi}}, \bibinfo {author} {\bibfnamefont {F.}~\bibnamefont {Vicentini}}, \bibinfo {author} {\bibfnamefont {N.}~\bibnamefont {Astrakhantsev}}, \bibinfo {author} {\bibfnamefont {F.}~\bibnamefont {Becca}}, \bibinfo {author} {\bibfnamefont {X.}~\bibnamefont {Cao}}, \bibinfo {author} {\bibfnamefont {J.}~\bibnamefont {Carrasquilla}}, \bibinfo {author} {\bibfnamefont {F.}~\bibnamefont {Ferrari}}, \bibinfo {author} {\bibfnamefont {A.}~\bibnamefont {Georges}}, \bibinfo {author} {\bibfnamefont {M.}~\bibnamefont {{Hibat-Allah}}}, \bibinfo {author} {\bibfnamefont {M.}~\bibnamefont {Imada}}, \bibinfo {author} {\bibfnamefont {A.~M.}\ \bibnamefont {L{\"a}uchli}}, \bibinfo {author} {\bibfnamefont {G.}~\bibnamefont {Mazzola}}, \bibinfo {author} {\bibfnamefont {A.}~\bibnamefont {Mezzacapo}}, \bibinfo {author} {\bibfnamefont {A.}~\bibnamefont {Millis}}, \bibinfo {author}
  {\bibfnamefont {J.}~\bibnamefont {Robledo~Moreno}}, \bibinfo {author} {\bibfnamefont {T.}~\bibnamefont {Neupert}}, \bibinfo {author} {\bibfnamefont {Y.}~\bibnamefont {Nomura}}, \bibinfo {author} {\bibfnamefont {J.}~\bibnamefont {Nys}}, \bibinfo {author} {\bibfnamefont {O.}~\bibnamefont {Parcollet}}, \bibinfo {author} {\bibfnamefont {R.}~\bibnamefont {Pohle}}, \bibinfo {author} {\bibfnamefont {I.}~\bibnamefont {Romero}}, \bibinfo {author} {\bibfnamefont {M.}~\bibnamefont {Schmid}}, \bibinfo {author} {\bibfnamefont {J.~M.}\ \bibnamefont {Silvester}}, \bibinfo {author} {\bibfnamefont {S.}~\bibnamefont {Sorella}}, \bibinfo {author} {\bibfnamefont {L.~F.}\ \bibnamefont {Tocchio}}, \bibinfo {author} {\bibfnamefont {L.}~\bibnamefont {Wang}}, \bibinfo {author} {\bibfnamefont {S.~R.}\ \bibnamefont {White}}, \bibinfo {author} {\bibfnamefont {A.}~\bibnamefont {Wietek}}, \bibinfo {author} {\bibfnamefont {Q.}~\bibnamefont {Yang}}, \bibinfo {author} {\bibfnamefont {Y.}~\bibnamefont {Yang}}, \bibinfo {author}
  {\bibfnamefont {S.}~\bibnamefont {Zhang}},\ and\ \bibinfo {author} {\bibfnamefont {G.}~\bibnamefont {Carleo}},\ }\bibfield  {title} {\bibinfo {title} {Variational benchmarks for quantum many-body problems},\ }\href {https://doi.org/10.1126/science.adg9774} {\bibfield  {journal} {\bibinfo  {journal} {Science}\ }\textbf {\bibinfo {volume} {386}},\ \bibinfo {pages} {296} (\bibinfo {year} {2024})}\BibitemShut {NoStop}%
\bibitem [{\citenamefont {McArdle}\ \emph {et~al.}(2020{\natexlab{b}})\citenamefont {McArdle}, \citenamefont {Endo}, \citenamefont {Aspuru-Guzik}, \citenamefont {Benjamin},\ and\ \citenamefont {Yuan}}]{mcardle_qchem_2020}%
  \BibitemOpen
  \bibfield  {author} {\bibinfo {author} {\bibfnamefont {S.}~\bibnamefont {McArdle}}, \bibinfo {author} {\bibfnamefont {S.}~\bibnamefont {Endo}}, \bibinfo {author} {\bibfnamefont {A.}~\bibnamefont {Aspuru-Guzik}}, \bibinfo {author} {\bibfnamefont {S.~C.}\ \bibnamefont {Benjamin}},\ and\ \bibinfo {author} {\bibfnamefont {X.}~\bibnamefont {Yuan}},\ }\bibfield  {title} {\bibinfo {title} {Quantum computational chemistry},\ }\href {https://doi.org/10.1103/RevModPhys.92.015003} {\bibfield  {journal} {\bibinfo  {journal} {Reviews of Modern Physics}\ }\textbf {\bibinfo {volume} {92}},\ \bibinfo {pages} {015003} (\bibinfo {year} {2020}{\natexlab{b}})}\BibitemShut {NoStop}%
\bibitem [{\citenamefont {Chien}\ \emph {et~al.}(2026)\citenamefont {Chien}, \citenamefont {Chiew}, \citenamefont {Harrison}, \citenamefont {Necaise}, \citenamefont {Wang}, \citenamefont {Mudassar}, \citenamefont {McLauchlan}, \citenamefont {Henderson}, \citenamefont {Scuseria}, \citenamefont {Strelchuk},\ and\ \citenamefont {Whitfield}}]{chienSimulatingFermionsDigital2026}%
  \BibitemOpen
  \bibfield  {author} {\bibinfo {author} {\bibfnamefont {R.~W.}\ \bibnamefont {Chien}}, \bibinfo {author} {\bibfnamefont {M.}~\bibnamefont {Chiew}}, \bibinfo {author} {\bibfnamefont {B.}~\bibnamefont {Harrison}}, \bibinfo {author} {\bibfnamefont {J.}~\bibnamefont {Necaise}}, \bibinfo {author} {\bibfnamefont {W.}~\bibnamefont {Wang}}, \bibinfo {author} {\bibfnamefont {M.}~\bibnamefont {Mudassar}}, \bibinfo {author} {\bibfnamefont {C.}~\bibnamefont {McLauchlan}}, \bibinfo {author} {\bibfnamefont {T.~M.}\ \bibnamefont {Henderson}}, \bibinfo {author} {\bibfnamefont {G.~E.}\ \bibnamefont {Scuseria}}, \bibinfo {author} {\bibfnamefont {S.}~\bibnamefont {Strelchuk}},\ and\ \bibinfo {author} {\bibfnamefont {J.~D.}\ \bibnamefont {Whitfield}},\ }\bibfield  {title} {\bibinfo {title} {Simulating fermions with a digital quantum computer},\ }\href {https://doi.org/10.1038/s42254-025-00914-5} {\bibfield  {journal} {\bibinfo  {journal} {Nature Reviews Physics}\ }\textbf {\bibinfo {volume} {8}},\ \bibinfo {pages} {131}
  (\bibinfo {year} {2026})}\BibitemShut {NoStop}%
\bibitem [{\citenamefont {Setia}\ \emph {et~al.}(2019)\citenamefont {Setia}, \citenamefont {Bravyi}, \citenamefont {Mezzacapo},\ and\ \citenamefont {Whitfield}}]{setiaSuperfastEncodingsFermionic2019}%
  \BibitemOpen
  \bibfield  {author} {\bibinfo {author} {\bibfnamefont {K.}~\bibnamefont {Setia}}, \bibinfo {author} {\bibfnamefont {S.}~\bibnamefont {Bravyi}}, \bibinfo {author} {\bibfnamefont {A.}~\bibnamefont {Mezzacapo}},\ and\ \bibinfo {author} {\bibfnamefont {J.~D.}\ \bibnamefont {Whitfield}},\ }\bibfield  {title} {\bibinfo {title} {Superfast encodings for fermionic quantum simulation},\ }\href {https://doi.org/10.1103/PhysRevResearch.1.033033} {\bibfield  {journal} {\bibinfo  {journal} {Physical Review Research}\ }\textbf {\bibinfo {volume} {1}},\ \bibinfo {pages} {033033} (\bibinfo {year} {2019})}\BibitemShut {NoStop}%
\bibitem [{\citenamefont {Chien}\ and\ \citenamefont {Klassen}(2022)}]{chien_optimizing_2022}%
  \BibitemOpen
  \bibfield  {author} {\bibinfo {author} {\bibfnamefont {R.~W.}\ \bibnamefont {Chien}}\ and\ \bibinfo {author} {\bibfnamefont {J.}~\bibnamefont {Klassen}},\ }\href@noop {} {\bibinfo {title} {Optimizing fermionic encodings for both {{Hamiltonian}} and hardware}} (\bibinfo {year} {2022}),\ \Eprint {https://arxiv.org/abs/2210.05652} {arXiv:2210.05652 [quant-ph]} \BibitemShut {NoStop}%
\bibitem [{\citenamefont {Kirby}\ \emph {et~al.}(2022)\citenamefont {Kirby}, \citenamefont {Fuller}, \citenamefont {Hadfield},\ and\ \citenamefont {Mezzacapo}}]{kirbySecondQuantizedFermionicOperators2022}%
  \BibitemOpen
  \bibfield  {author} {\bibinfo {author} {\bibfnamefont {W.}~\bibnamefont {Kirby}}, \bibinfo {author} {\bibfnamefont {B.}~\bibnamefont {Fuller}}, \bibinfo {author} {\bibfnamefont {C.}~\bibnamefont {Hadfield}},\ and\ \bibinfo {author} {\bibfnamefont {A.}~\bibnamefont {Mezzacapo}},\ }\bibfield  {title} {\bibinfo {title} {Second-{{Quantized Fermionic Operators}} with {{Polylogarithmic Qubit}} and {{Gate Complexity}}},\ }\href {https://doi.org/10.1103/PRXQuantum.3.020351} {\bibfield  {journal} {\bibinfo  {journal} {PRX Quantum}\ }\textbf {\bibinfo {volume} {3}},\ \bibinfo {pages} {020351} (\bibinfo {year} {2022})}\BibitemShut {NoStop}%
\bibitem [{\citenamefont {Algaba}\ \emph {et~al.}(2024)\citenamefont {Algaba}, \citenamefont {Sriluckshmy}, \citenamefont {Leib},\ and\ \citenamefont {IV}}]{algabaLowdepthSimulationsFermionic2024}%
  \BibitemOpen
  \bibfield  {author} {\bibinfo {author} {\bibfnamefont {M.~G.}\ \bibnamefont {Algaba}}, \bibinfo {author} {\bibfnamefont {P.~V.}\ \bibnamefont {Sriluckshmy}}, \bibinfo {author} {\bibfnamefont {M.}~\bibnamefont {Leib}},\ and\ \bibinfo {author} {\bibfnamefont {F.~{\v S}.}\ \bibnamefont {IV}},\ }\bibfield  {title} {\bibinfo {title} {Low-depth simulations of fermionic systems on square-grid quantum hardware},\ }\href {https://doi.org/10.22331/q-2024-04-30-1327} {\bibfield  {journal} {\bibinfo  {journal} {Quantum}\ }\textbf {\bibinfo {volume} {8}},\ \bibinfo {pages} {1327} (\bibinfo {year} {2024})}\BibitemShut {NoStop}%
\bibitem [{\citenamefont {Nigmatullin}\ \emph {et~al.}(2025)\citenamefont {Nigmatullin}, \citenamefont {H{\'e}mery}, \citenamefont {Ghanem}, \citenamefont {Moses}, \citenamefont {Gresh}, \citenamefont {Siegfried}, \citenamefont {Mills}, \citenamefont {Gatterman}, \citenamefont {Hewitt}, \citenamefont {Granet},\ and\ \citenamefont {Dreyer}}]{nigmatullinExperimentalDemonstrationBreakeven2025}%
  \BibitemOpen
  \bibfield  {author} {\bibinfo {author} {\bibfnamefont {R.}~\bibnamefont {Nigmatullin}}, \bibinfo {author} {\bibfnamefont {K.}~\bibnamefont {H{\'e}mery}}, \bibinfo {author} {\bibfnamefont {K.}~\bibnamefont {Ghanem}}, \bibinfo {author} {\bibfnamefont {S.}~\bibnamefont {Moses}}, \bibinfo {author} {\bibfnamefont {D.}~\bibnamefont {Gresh}}, \bibinfo {author} {\bibfnamefont {P.}~\bibnamefont {Siegfried}}, \bibinfo {author} {\bibfnamefont {M.}~\bibnamefont {Mills}}, \bibinfo {author} {\bibfnamefont {T.}~\bibnamefont {Gatterman}}, \bibinfo {author} {\bibfnamefont {N.}~\bibnamefont {Hewitt}}, \bibinfo {author} {\bibfnamefont {E.}~\bibnamefont {Granet}},\ and\ \bibinfo {author} {\bibfnamefont {H.}~\bibnamefont {Dreyer}},\ }\bibfield  {title} {\bibinfo {title} {Experimental demonstration of breakeven for a compact fermionic encoding},\ }\href {https://doi.org/10.1038/s41567-025-02931-8} {\bibfield  {journal} {\bibinfo  {journal} {Nature Physics}\ }\textbf {\bibinfo {volume} {21}},\ \bibinfo {pages} {1319} (\bibinfo
  {year} {2025})}\BibitemShut {NoStop}%
\bibitem [{\citenamefont {Evered}\ \emph {et~al.}(2025)\citenamefont {Evered}, \citenamefont {Kalinowski}, \citenamefont {Geim}, \citenamefont {Manovitz}, \citenamefont {Bluvstein}, \citenamefont {Li}, \citenamefont {Maskara}, \citenamefont {Zhou}, \citenamefont {Ebadi}, \citenamefont {Xu}, \citenamefont {Campo}, \citenamefont {Cain}, \citenamefont {Ostermann}, \citenamefont {Yelin}, \citenamefont {Sachdev}, \citenamefont {Greiner}, \citenamefont {Vuleti{\'c}},\ and\ \citenamefont {Lukin}}]{everedProbingKitaevHoneycomb2025}%
  \BibitemOpen
  \bibfield  {author} {\bibinfo {author} {\bibfnamefont {S.~J.}\ \bibnamefont {Evered}}, \bibinfo {author} {\bibfnamefont {M.}~\bibnamefont {Kalinowski}}, \bibinfo {author} {\bibfnamefont {A.~A.}\ \bibnamefont {Geim}}, \bibinfo {author} {\bibfnamefont {T.}~\bibnamefont {Manovitz}}, \bibinfo {author} {\bibfnamefont {D.}~\bibnamefont {Bluvstein}}, \bibinfo {author} {\bibfnamefont {S.~H.}\ \bibnamefont {Li}}, \bibinfo {author} {\bibfnamefont {N.}~\bibnamefont {Maskara}}, \bibinfo {author} {\bibfnamefont {H.}~\bibnamefont {Zhou}}, \bibinfo {author} {\bibfnamefont {S.}~\bibnamefont {Ebadi}}, \bibinfo {author} {\bibfnamefont {M.}~\bibnamefont {Xu}}, \bibinfo {author} {\bibfnamefont {J.}~\bibnamefont {Campo}}, \bibinfo {author} {\bibfnamefont {M.}~\bibnamefont {Cain}}, \bibinfo {author} {\bibfnamefont {S.}~\bibnamefont {Ostermann}}, \bibinfo {author} {\bibfnamefont {S.~F.}\ \bibnamefont {Yelin}}, \bibinfo {author} {\bibfnamefont {S.}~\bibnamefont {Sachdev}}, \bibinfo {author} {\bibfnamefont {M.}~\bibnamefont
  {Greiner}}, \bibinfo {author} {\bibfnamefont {V.}~\bibnamefont {Vuleti{\'c}}},\ and\ \bibinfo {author} {\bibfnamefont {M.~D.}\ \bibnamefont {Lukin}},\ }\bibfield  {title} {\bibinfo {title} {Probing the {{Kitaev}} honeycomb model on a neutral-atom quantum computer},\ }\href {https://doi.org/10.1038/s41586-025-09475-0} {\bibfield  {journal} {\bibinfo  {journal} {Nature}\ }\textbf {\bibinfo {volume} {645}},\ \bibinfo {pages} {341} (\bibinfo {year} {2025})}\BibitemShut {NoStop}%
\bibitem [{\citenamefont {Granet}\ \emph {et~al.}(2025)\citenamefont {Granet}, \citenamefont {Lin}, \citenamefont {H{\'e}mery}, \citenamefont {Haghshenas}, \citenamefont {{Andres-Martinez}}, \citenamefont {Stephen}, \citenamefont {Ransford}, \citenamefont {Arkinstall}, \citenamefont {Allman}, \citenamefont {Campora}, \citenamefont {Cooper}, \citenamefont {Delaney}, \citenamefont {Dreiling}, \citenamefont {Estey}, \citenamefont {Figgatt}, \citenamefont {Foltz}, \citenamefont {Gaebler}, \citenamefont {Hall}, \citenamefont {Husain}, \citenamefont {Isanaka}, \citenamefont {Kennedy}, \citenamefont {Kotibhaskar}, \citenamefont {Madjarov}, \citenamefont {Mills}, \citenamefont {Milne}, \citenamefont {Park}, \citenamefont {Reed}, \citenamefont {Neyenhuis}, \citenamefont {Bohnet}, \citenamefont {{Foss-Feig}}, \citenamefont {Potter}, \citenamefont {Nigmatullin}, \citenamefont {Iqbal},\ and\ \citenamefont {Dreyer}}]{granetSuperconductingPairingCorrelations2025}%
  \BibitemOpen
  \bibfield  {author} {\bibinfo {author} {\bibfnamefont {E.}~\bibnamefont {Granet}}, \bibinfo {author} {\bibfnamefont {S.-H.}\ \bibnamefont {Lin}}, \bibinfo {author} {\bibfnamefont {K.}~\bibnamefont {H{\'e}mery}}, \bibinfo {author} {\bibfnamefont {R.}~\bibnamefont {Haghshenas}}, \bibinfo {author} {\bibfnamefont {P.}~\bibnamefont {{Andres-Martinez}}}, \bibinfo {author} {\bibfnamefont {D.~T.}\ \bibnamefont {Stephen}}, \bibinfo {author} {\bibfnamefont {A.}~\bibnamefont {Ransford}}, \bibinfo {author} {\bibfnamefont {J.}~\bibnamefont {Arkinstall}}, \bibinfo {author} {\bibfnamefont {M.~S.}\ \bibnamefont {Allman}}, \bibinfo {author} {\bibfnamefont {P.}~\bibnamefont {Campora}}, \bibinfo {author} {\bibfnamefont {S.~F.}\ \bibnamefont {Cooper}}, \bibinfo {author} {\bibfnamefont {R.~D.}\ \bibnamefont {Delaney}}, \bibinfo {author} {\bibfnamefont {J.~M.}\ \bibnamefont {Dreiling}}, \bibinfo {author} {\bibfnamefont {B.}~\bibnamefont {Estey}}, \bibinfo {author} {\bibfnamefont {C.}~\bibnamefont {Figgatt}}, \bibinfo {author}
  {\bibfnamefont {C.}~\bibnamefont {Foltz}}, \bibinfo {author} {\bibfnamefont {J.~P.}\ \bibnamefont {Gaebler}}, \bibinfo {author} {\bibfnamefont {A.}~\bibnamefont {Hall}}, \bibinfo {author} {\bibfnamefont {A.}~\bibnamefont {Husain}}, \bibinfo {author} {\bibfnamefont {A.}~\bibnamefont {Isanaka}}, \bibinfo {author} {\bibfnamefont {C.~J.}\ \bibnamefont {Kennedy}}, \bibinfo {author} {\bibfnamefont {N.}~\bibnamefont {Kotibhaskar}}, \bibinfo {author} {\bibfnamefont {I.~S.}\ \bibnamefont {Madjarov}}, \bibinfo {author} {\bibfnamefont {M.}~\bibnamefont {Mills}}, \bibinfo {author} {\bibfnamefont {A.~R.}\ \bibnamefont {Milne}}, \bibinfo {author} {\bibfnamefont {A.~J.}\ \bibnamefont {Park}}, \bibinfo {author} {\bibfnamefont {A.~P.}\ \bibnamefont {Reed}}, \bibinfo {author} {\bibfnamefont {B.}~\bibnamefont {Neyenhuis}}, \bibinfo {author} {\bibfnamefont {J.~G.}\ \bibnamefont {Bohnet}}, \bibinfo {author} {\bibfnamefont {M.}~\bibnamefont {{Foss-Feig}}}, \bibinfo {author} {\bibfnamefont {A.~C.}\ \bibnamefont {Potter}},
  \bibinfo {author} {\bibfnamefont {R.}~\bibnamefont {Nigmatullin}}, \bibinfo {author} {\bibfnamefont {M.}~\bibnamefont {Iqbal}},\ and\ \bibinfo {author} {\bibfnamefont {H.}~\bibnamefont {Dreyer}},\ }\href@noop {} {\bibinfo {title} {Superconducting pairing correlations on a trapped-ion quantum computer}} (\bibinfo {year} {2025}),\ \Eprint {https://arxiv.org/abs/2511.02125} {arXiv:2511.02125 [quant-ph]} \BibitemShut {NoStop}%
\bibitem [{\citenamefont {Guaita}(2025)}]{guaita_locality_2024}%
  \BibitemOpen
  \bibfield  {author} {\bibinfo {author} {\bibfnamefont {T.}~\bibnamefont {Guaita}},\ }\bibfield  {title} {\bibinfo {title} {On the locality of qubit encodings of local fermionic modes},\ }\href {https://doi.org/10.22331/q-2025-02-25-1644} {\bibfield  {journal} {\bibinfo  {journal} {Quantum}\ }\textbf {\bibinfo {volume} {9}},\ \bibinfo {pages} {1644} (\bibinfo {year} {2025})}\BibitemShut {NoStop}%
\bibitem [{\citenamefont {Tantivasadakarn}\ \emph {et~al.}(2024)\citenamefont {Tantivasadakarn}, \citenamefont {Thorngren}, \citenamefont {Vishwanath},\ and\ \citenamefont {Verresen}}]{tantivasadakarnLongRangeEntanglementMeasuring2024}%
  \BibitemOpen
  \bibfield  {author} {\bibinfo {author} {\bibfnamefont {N.}~\bibnamefont {Tantivasadakarn}}, \bibinfo {author} {\bibfnamefont {R.}~\bibnamefont {Thorngren}}, \bibinfo {author} {\bibfnamefont {A.}~\bibnamefont {Vishwanath}},\ and\ \bibinfo {author} {\bibfnamefont {R.}~\bibnamefont {Verresen}},\ }\bibfield  {title} {\bibinfo {title} {Long-{{Range Entanglement}} from {{Measuring Symmetry-Protected Topological Phases}}},\ }\href {https://doi.org/10.1103/PhysRevX.14.021040} {\bibfield  {journal} {\bibinfo  {journal} {Physical Review X}\ }\textbf {\bibinfo {volume} {14}},\ \bibinfo {pages} {021040} (\bibinfo {year} {2024})}\BibitemShut {NoStop}%
\bibitem [{\citenamefont {Schuckert}\ \emph {et~al.}(2024)\citenamefont {Schuckert}, \citenamefont {Crane}, \citenamefont {Gorshkov}, \citenamefont {Hafezi},\ and\ \citenamefont {Gullans}}]{schuckertFermionqubitFaulttolerantQuantum2024a}%
  \BibitemOpen
  \bibfield  {author} {\bibinfo {author} {\bibfnamefont {A.}~\bibnamefont {Schuckert}}, \bibinfo {author} {\bibfnamefont {E.}~\bibnamefont {Crane}}, \bibinfo {author} {\bibfnamefont {A.~V.}\ \bibnamefont {Gorshkov}}, \bibinfo {author} {\bibfnamefont {M.}~\bibnamefont {Hafezi}},\ and\ \bibinfo {author} {\bibfnamefont {M.~J.}\ \bibnamefont {Gullans}},\ }\href@noop {} {\bibinfo {title} {Fermion-qubit fault-tolerant quantum computing}} (\bibinfo {year} {2024}),\ \Eprint {https://arxiv.org/abs/2411.08955} {arXiv:2411.08955 [quant-ph]} \BibitemShut {NoStop}%
\bibitem [{\citenamefont {Ott}\ \emph {et~al.}(2024)\citenamefont {Ott}, \citenamefont {{Gonz{\'a}lez-Cuadra}}, \citenamefont {Zache}, \citenamefont {Zoller}, \citenamefont {Kaufman},\ and\ \citenamefont {Pichler}}]{ottErrorcorrectedFermionicQuantum2024}%
  \BibitemOpen
  \bibfield  {author} {\bibinfo {author} {\bibfnamefont {R.}~\bibnamefont {Ott}}, \bibinfo {author} {\bibfnamefont {D.}~\bibnamefont {{Gonz{\'a}lez-Cuadra}}}, \bibinfo {author} {\bibfnamefont {T.~V.}\ \bibnamefont {Zache}}, \bibinfo {author} {\bibfnamefont {P.}~\bibnamefont {Zoller}}, \bibinfo {author} {\bibfnamefont {A.~M.}\ \bibnamefont {Kaufman}},\ and\ \bibinfo {author} {\bibfnamefont {H.}~\bibnamefont {Pichler}},\ }\href@noop {} {\bibinfo {title} {Error-corrected fermionic quantum processors with neutral atoms}} (\bibinfo {year} {2024}),\ \Eprint {https://arxiv.org/abs/2412.16081} {arXiv:2412.16081 [quant-ph]} \BibitemShut {NoStop}%
\bibitem [{\citenamefont {Chen}\ \emph {et~al.}(2024)\citenamefont {Chen}, \citenamefont {Gorshkov},\ and\ \citenamefont {Xu}}]{chenErrorcorrectingCodesFermionic2024}%
  \BibitemOpen
  \bibfield  {author} {\bibinfo {author} {\bibfnamefont {Y.-A.}\ \bibnamefont {Chen}}, \bibinfo {author} {\bibfnamefont {A.~V.}\ \bibnamefont {Gorshkov}},\ and\ \bibinfo {author} {\bibfnamefont {Y.}~\bibnamefont {Xu}},\ }\bibfield  {title} {\bibinfo {title} {Error-correcting codes for fermionic quantum simulation},\ }\href {https://doi.org/10.21468/SciPostPhys.16.1.033} {\bibfield  {journal} {\bibinfo  {journal} {SciPost Physics}\ }\textbf {\bibinfo {volume} {16}},\ \bibinfo {pages} {033} (\bibinfo {year} {2024})}\BibitemShut {NoStop}%
\bibitem [{\citenamefont {Chalopin}\ \emph {et~al.}(2025)\citenamefont {Chalopin}, \citenamefont {Bojovi{\'c}}, \citenamefont {Bourgund}, \citenamefont {Wang}, \citenamefont {Franz}, \citenamefont {Bloch},\ and\ \citenamefont {Hilker}}]{chalopinOpticalSuperlatticeEngineering2025a}%
  \BibitemOpen
  \bibfield  {author} {\bibinfo {author} {\bibfnamefont {T.}~\bibnamefont {Chalopin}}, \bibinfo {author} {\bibfnamefont {P.}~\bibnamefont {Bojovi{\'c}}}, \bibinfo {author} {\bibfnamefont {D.}~\bibnamefont {Bourgund}}, \bibinfo {author} {\bibfnamefont {S.}~\bibnamefont {Wang}}, \bibinfo {author} {\bibfnamefont {T.}~\bibnamefont {Franz}}, \bibinfo {author} {\bibfnamefont {I.}~\bibnamefont {Bloch}},\ and\ \bibinfo {author} {\bibfnamefont {T.}~\bibnamefont {Hilker}},\ }\bibfield  {title} {\bibinfo {title} {Optical {{Superlattice}} for {{Engineering Hubbard Couplings}} in {{Quantum Simulation}}},\ }\href {https://doi.org/10.1103/PhysRevLett.134.053402} {\bibfield  {journal} {\bibinfo  {journal} {Physical Review Letters}\ }\textbf {\bibinfo {volume} {134}},\ \bibinfo {pages} {053402} (\bibinfo {year} {2025})}\BibitemShut {NoStop}%
\bibitem [{\citenamefont {Osborne}(2007)}]{osborne_2007}%
  \BibitemOpen
  \bibfield  {author} {\bibinfo {author} {\bibfnamefont {T.~J.}\ \bibnamefont {Osborne}},\ }\bibfield  {title} {\bibinfo {title} {Simulating adiabatic evolution of gapped spin systems},\ }\href {https://doi.org/10.1103/PhysRevA.75.032321} {\bibfield  {journal} {\bibinfo  {journal} {Physical Review A}\ }\textbf {\bibinfo {volume} {75}},\ \bibinfo {pages} {032321} (\bibinfo {year} {2007})}\BibitemShut {NoStop}%
\bibitem [{\citenamefont {Dalzell}\ and\ \citenamefont {Brand{\~a}o}(2019)}]{dalzellLocallyAccurateMPS2019}%
  \BibitemOpen
  \bibfield  {author} {\bibinfo {author} {\bibfnamefont {A.~M.}\ \bibnamefont {Dalzell}}\ and\ \bibinfo {author} {\bibfnamefont {F.~G. S.~L.}\ \bibnamefont {Brand{\~a}o}},\ }\bibfield  {title} {\bibinfo {title} {Locally accurate {{MPS}} approximations for ground states of one-dimensional gapped local {{Hamiltonians}}},\ }\href {https://doi.org/10.22331/q-2019-09-23-187} {\bibfield  {journal} {\bibinfo  {journal} {Quantum}\ }\textbf {\bibinfo {volume} {3}},\ \bibinfo {pages} {187} (\bibinfo {year} {2019})}\BibitemShut {NoStop}%
\bibitem [{\citenamefont {Arad}\ \emph {et~al.}(2013)\citenamefont {Arad}, \citenamefont {Kitaev}, \citenamefont {Landau},\ and\ \citenamefont {Vazirani}}]{AKLV2013}%
  \BibitemOpen
  \bibfield  {author} {\bibinfo {author} {\bibfnamefont {I.}~\bibnamefont {Arad}}, \bibinfo {author} {\bibfnamefont {A.}~\bibnamefont {Kitaev}}, \bibinfo {author} {\bibfnamefont {Z.}~\bibnamefont {Landau}},\ and\ \bibinfo {author} {\bibfnamefont {U.}~\bibnamefont {Vazirani}},\ }\href@noop {} {\bibinfo {title} {An area law and sub-exponential algorithm for {{1D}} systems}} (\bibinfo {year} {2013}),\ \Eprint {https://arxiv.org/abs/1301.1162} {arXiv:1301.1162 [quant-ph]} \BibitemShut {NoStop}%
\bibitem [{\citenamefont {Hu}\ \emph {et~al.}(2025)\citenamefont {Hu}, \citenamefont {Gomez}, \citenamefont {Chen}, \citenamefont {Trowbridge}, \citenamefont {Goldschmidt}, \citenamefont {Manchester}, \citenamefont {Chong}, \citenamefont {Jaffe},\ and\ \citenamefont {Yelin}}]{huUniversalDynamicsGlobally2025}%
  \BibitemOpen
  \bibfield  {author} {\bibinfo {author} {\bibfnamefont {H.-Y.}\ \bibnamefont {Hu}}, \bibinfo {author} {\bibfnamefont {A.~M.}\ \bibnamefont {Gomez}}, \bibinfo {author} {\bibfnamefont {L.}~\bibnamefont {Chen}}, \bibinfo {author} {\bibfnamefont {A.}~\bibnamefont {Trowbridge}}, \bibinfo {author} {\bibfnamefont {A.~J.}\ \bibnamefont {Goldschmidt}}, \bibinfo {author} {\bibfnamefont {Z.}~\bibnamefont {Manchester}}, \bibinfo {author} {\bibfnamefont {F.~T.}\ \bibnamefont {Chong}}, \bibinfo {author} {\bibfnamefont {A.}~\bibnamefont {Jaffe}},\ and\ \bibinfo {author} {\bibfnamefont {S.~F.}\ \bibnamefont {Yelin}},\ }\href@noop {} {\bibinfo {title} {Universal {{Dynamics}} with {{Globally Controlled Analog Quantum Simulators}}}} (\bibinfo {year} {2025}),\ \Eprint {https://arxiv.org/abs/2508.19075} {arXiv:2508.19075 [quant-ph]} \BibitemShut {NoStop}%
\bibitem [{\citenamefont {Weitenberg}\ \emph {et~al.}(2011)\citenamefont {Weitenberg}, \citenamefont {Endres}, \citenamefont {Sherson}, \citenamefont {Cheneau}, \citenamefont {Schau{\ss}}, \citenamefont {Fukuhara}, \citenamefont {Bloch},\ and\ \citenamefont {Kuhr}}]{weitenbergSinglespinAddressingAtomic2011}%
  \BibitemOpen
  \bibfield  {author} {\bibinfo {author} {\bibfnamefont {C.}~\bibnamefont {Weitenberg}}, \bibinfo {author} {\bibfnamefont {M.}~\bibnamefont {Endres}}, \bibinfo {author} {\bibfnamefont {J.~F.}\ \bibnamefont {Sherson}}, \bibinfo {author} {\bibfnamefont {M.}~\bibnamefont {Cheneau}}, \bibinfo {author} {\bibfnamefont {P.}~\bibnamefont {Schau{\ss}}}, \bibinfo {author} {\bibfnamefont {T.}~\bibnamefont {Fukuhara}}, \bibinfo {author} {\bibfnamefont {I.}~\bibnamefont {Bloch}},\ and\ \bibinfo {author} {\bibfnamefont {S.}~\bibnamefont {Kuhr}},\ }\bibfield  {title} {\bibinfo {title} {Single-spin addressing in an atomic {{Mott}} insulator},\ }\href {https://doi.org/10.1038/nature09827} {\bibfield  {journal} {\bibinfo  {journal} {Nature}\ }\textbf {\bibinfo {volume} {471}},\ \bibinfo {pages} {319} (\bibinfo {year} {2011})}\BibitemShut {NoStop}%
\bibitem [{\citenamefont {Trubko}\ \emph {et~al.}(2017)\citenamefont {Trubko}, \citenamefont {Gregoire}, \citenamefont {Holmgren},\ and\ \citenamefont {Cronin}}]{trubkoPotassiumTuneoutwavelengthMeasurement2017}%
  \BibitemOpen
  \bibfield  {author} {\bibinfo {author} {\bibfnamefont {R.}~\bibnamefont {Trubko}}, \bibinfo {author} {\bibfnamefont {M.~D.}\ \bibnamefont {Gregoire}}, \bibinfo {author} {\bibfnamefont {W.~F.}\ \bibnamefont {Holmgren}},\ and\ \bibinfo {author} {\bibfnamefont {A.~D.}\ \bibnamefont {Cronin}},\ }\bibfield  {title} {\bibinfo {title} {Potassium tune-out-wavelength measurement using atom interferometry and a multipass optical cavity},\ }\href {https://doi.org/10.1103/PhysRevA.95.052507} {\bibfield  {journal} {\bibinfo  {journal} {Physical Review A}\ }\textbf {\bibinfo {volume} {95}},\ \bibinfo {pages} {052507} (\bibinfo {year} {2017})}\BibitemShut {NoStop}%
\bibitem [{\citenamefont {Naldesi}\ \emph {et~al.}(2023)\citenamefont {Naldesi}, \citenamefont {Elben}, \citenamefont {Minguzzi}, \citenamefont {Clément}, \citenamefont {Zoller},\ and\ \citenamefont {Vermersch}}]{naldesi_fermionic_2023}%
  \BibitemOpen
  \bibfield  {author} {\bibinfo {author} {\bibfnamefont {P.}~\bibnamefont {Naldesi}}, \bibinfo {author} {\bibfnamefont {A.}~\bibnamefont {Elben}}, \bibinfo {author} {\bibfnamefont {A.}~\bibnamefont {Minguzzi}}, \bibinfo {author} {\bibfnamefont {D.}~\bibnamefont {Clément}}, \bibinfo {author} {\bibfnamefont {P.}~\bibnamefont {Zoller}},\ and\ \bibinfo {author} {\bibfnamefont {B.}~\bibnamefont {Vermersch}},\ }\bibfield  {title} {\bibinfo {title} {Fermionic correlation functions from randomized measurements in programmable atomic quantum devices},\ }\href {https://doi.org/10.1103/PhysRevLett.131.060601} {\bibfield  {journal} {\bibinfo  {journal} {Physical Review Letters}\ }\textbf {\bibinfo {volume} {131}},\ \bibinfo {pages} {060601} (\bibinfo {year} {2023})}\BibitemShut {NoStop}%
\bibitem [{\citenamefont {Tran}\ \emph {et~al.}(2023)\citenamefont {Tran}, \citenamefont {Mark}, \citenamefont {Ho},\ and\ \citenamefont {Choi}}]{tran_measuring_2023}%
  \BibitemOpen
  \bibfield  {author} {\bibinfo {author} {\bibfnamefont {M.~C.}\ \bibnamefont {Tran}}, \bibinfo {author} {\bibfnamefont {D.~K.}\ \bibnamefont {Mark}}, \bibinfo {author} {\bibfnamefont {W.~W.}\ \bibnamefont {Ho}},\ and\ \bibinfo {author} {\bibfnamefont {S.}~\bibnamefont {Choi}},\ }\bibfield  {title} {\bibinfo {title} {Measuring {Arbitrary} {Physical} {Properties} in {Analog} {Quantum} {Simulation}},\ }\href {https://doi.org/10.1103/PhysRevX.13.011049} {\bibfield  {journal} {\bibinfo  {journal} {Physical Review X}\ }\textbf {\bibinfo {volume} {13}},\ \bibinfo {pages} {011049} (\bibinfo {year} {2023})}\BibitemShut {NoStop}%
\bibitem [{\citenamefont {Bourgund}\ \emph {et~al.}(2025)\citenamefont {Bourgund}, \citenamefont {Chalopin}, \citenamefont {Bojovi{\'c}}, \citenamefont {Schl{\"o}mer}, \citenamefont {Wang}, \citenamefont {Franz}, \citenamefont {Hirthe}, \citenamefont {Bohrdt}, \citenamefont {Grusdt}, \citenamefont {Bloch},\ and\ \citenamefont {Hilker}}]{bourgundFormationIndividualStripes2025a}%
  \BibitemOpen
  \bibfield  {author} {\bibinfo {author} {\bibfnamefont {D.}~\bibnamefont {Bourgund}}, \bibinfo {author} {\bibfnamefont {T.}~\bibnamefont {Chalopin}}, \bibinfo {author} {\bibfnamefont {P.}~\bibnamefont {Bojovi{\'c}}}, \bibinfo {author} {\bibfnamefont {H.}~\bibnamefont {Schl{\"o}mer}}, \bibinfo {author} {\bibfnamefont {S.}~\bibnamefont {Wang}}, \bibinfo {author} {\bibfnamefont {T.}~\bibnamefont {Franz}}, \bibinfo {author} {\bibfnamefont {S.}~\bibnamefont {Hirthe}}, \bibinfo {author} {\bibfnamefont {A.}~\bibnamefont {Bohrdt}}, \bibinfo {author} {\bibfnamefont {F.}~\bibnamefont {Grusdt}}, \bibinfo {author} {\bibfnamefont {I.}~\bibnamefont {Bloch}},\ and\ \bibinfo {author} {\bibfnamefont {T.~A.}\ \bibnamefont {Hilker}},\ }\bibfield  {title} {\bibinfo {title} {Formation of individual stripes in a mixed-dimensional cold-atom {{Fermi}}--{{Hubbard}} system},\ }\href {https://doi.org/10.1038/s41586-024-08270-7} {\bibfield  {journal} {\bibinfo  {journal} {Nature}\ }\textbf {\bibinfo {volume} {637}},\ \bibinfo {pages}
  {57} (\bibinfo {year} {2025})}\BibitemShut {NoStop}%
\bibitem [{\citenamefont {Hirthe}\ \emph {et~al.}(2023)\citenamefont {Hirthe}, \citenamefont {Chalopin}, \citenamefont {Bourgund}, \citenamefont {Bojovi{\'c}}, \citenamefont {Bohrdt}, \citenamefont {Demler}, \citenamefont {Grusdt}, \citenamefont {Bloch},\ and\ \citenamefont {Hilker}}]{hirtheMagneticallyMediatedHole2023}%
  \BibitemOpen
  \bibfield  {author} {\bibinfo {author} {\bibfnamefont {S.}~\bibnamefont {Hirthe}}, \bibinfo {author} {\bibfnamefont {T.}~\bibnamefont {Chalopin}}, \bibinfo {author} {\bibfnamefont {D.}~\bibnamefont {Bourgund}}, \bibinfo {author} {\bibfnamefont {P.}~\bibnamefont {Bojovi{\'c}}}, \bibinfo {author} {\bibfnamefont {A.}~\bibnamefont {Bohrdt}}, \bibinfo {author} {\bibfnamefont {E.}~\bibnamefont {Demler}}, \bibinfo {author} {\bibfnamefont {F.}~\bibnamefont {Grusdt}}, \bibinfo {author} {\bibfnamefont {I.}~\bibnamefont {Bloch}},\ and\ \bibinfo {author} {\bibfnamefont {T.~A.}\ \bibnamefont {Hilker}},\ }\bibfield  {title} {\bibinfo {title} {Magnetically mediated hole pairing in fermionic ladders of ultracold atoms},\ }\href {https://doi.org/10.1038/s41586-022-05437-y} {\bibfield  {journal} {\bibinfo  {journal} {Nature}\ }\textbf {\bibinfo {volume} {613}},\ \bibinfo {pages} {463} (\bibinfo {year} {2023})}\BibitemShut {NoStop}%
\bibitem [{\citenamefont {Yang}\ \emph {et~al.}(2021)\citenamefont {Yang}, \citenamefont {Liu}, \citenamefont {Mongkolkiattichai},\ and\ \citenamefont {Schauss}}]{yangSiteResolvedImagingUltracold2021}%
  \BibitemOpen
  \bibfield  {author} {\bibinfo {author} {\bibfnamefont {J.}~\bibnamefont {Yang}}, \bibinfo {author} {\bibfnamefont {L.}~\bibnamefont {Liu}}, \bibinfo {author} {\bibfnamefont {J.}~\bibnamefont {Mongkolkiattichai}},\ and\ \bibinfo {author} {\bibfnamefont {P.}~\bibnamefont {Schauss}},\ }\bibfield  {title} {\bibinfo {title} {Site-{{Resolved Imaging}} of {{Ultracold Fermions}} in a {{Triangular-Lattice Quantum Gas Microscope}}},\ }\href {https://doi.org/10.1103/PRXQuantum.2.020344} {\bibfield  {journal} {\bibinfo  {journal} {PRX Quantum}\ }\textbf {\bibinfo {volume} {2}},\ \bibinfo {pages} {020344} (\bibinfo {year} {2021})}\BibitemShut {NoStop}%
\bibitem [{\citenamefont {Fukui}\ \emph {et~al.}(2005)\citenamefont {Fukui}, \citenamefont {Hatsugai},\ and\ \citenamefont {Suzuki}}]{fukuiChernNumbersDiscretized2005}%
  \BibitemOpen
  \bibfield  {author} {\bibinfo {author} {\bibfnamefont {T.}~\bibnamefont {Fukui}}, \bibinfo {author} {\bibfnamefont {Y.}~\bibnamefont {Hatsugai}},\ and\ \bibinfo {author} {\bibfnamefont {H.}~\bibnamefont {Suzuki}},\ }\bibfield  {title} {\bibinfo {title} {Chern {{Numbers}} in {{Discretized Brillouin Zone}}: {{Efficient Method}} of {{Computing}} ({{Spin}}) {{Hall Conductances}}},\ }\href {https://doi.org/10.1143/JPSJ.74.1674} {\bibfield  {journal} {\bibinfo  {journal} {Journal of the Physical Society of Japan}\ }\textbf {\bibinfo {volume} {74}},\ \bibinfo {pages} {1674} (\bibinfo {year} {2005})}\BibitemShut {NoStop}%
\bibitem [{\citenamefont {Giamarchi}(2003)}]{giamarchi2003quantum}%
  \BibitemOpen
  \bibfield  {author} {\bibinfo {author} {\bibfnamefont {T.}~\bibnamefont {Giamarchi}},\ }\href@noop {} {\emph {\bibinfo {title} {Quantum physics in one dimension}}},\ Vol.\ \bibinfo {volume} {121}\ (\bibinfo  {publisher} {Clarendon press},\ \bibinfo {year} {2003})\BibitemShut {NoStop}%
\bibitem [{\citenamefont {Seidel}\ and\ \citenamefont {Lee}(2005)}]{seidelLutherEmeryLiquidSpin2005}%
  \BibitemOpen
  \bibfield  {author} {\bibinfo {author} {\bibfnamefont {A.}~\bibnamefont {Seidel}}\ and\ \bibinfo {author} {\bibfnamefont {D.-H.}\ \bibnamefont {Lee}},\ }\bibfield  {title} {\bibinfo {title} {The {{Luther-Emery}} liquid: {{Spin}} gap and anomalous flux period},\ }\href {https://doi.org/10.1103/PhysRevB.71.045113} {\bibfield  {journal} {\bibinfo  {journal} {Physical Review B}\ }\textbf {\bibinfo {volume} {71}},\ \bibinfo {pages} {045113} (\bibinfo {year} {2005})}\BibitemShut {NoStop}%
\end{thebibliography}%

\end{document}